\documentclass[aps,prd,twocolumn,superscriptaddress,nofootinbib,nobibnotes,groupedaddress]{revtex4}

\usepackage{tikz}
\usetikzlibrary{matrix}
\usepackage{latexsym}
\usepackage{epsfig}
\usepackage{graphicx}
\usepackage{epstopdf}
\usepackage{amssymb}
\usepackage{amsmath}
\usepackage{dcolumn}
\usepackage{bm}
\usepackage{color}
\usepackage{comment}
\usepackage{subcaption}  
\usepackage{float}
\usepackage{amsfonts}
\usepackage[titletoc]{appendix}
\usepackage{hyperref}
\usepackage{cleveref}
\usepackage{soul}
\usepackage{cancel}
\usepackage{caption}
\usepackage{ragged2e}
\everymath{\displaystyle}

\usepackage{array}
\usepackage{ctable}
\usepackage{multirow}
\usepackage{siunitx}
\usepackage{tabularx}
\usepackage{booktabs}
\usepackage{float}

\def\be{\begin{equation}}
\def\ee{\end{equation}}
\def\bea{\begin{eqnarray}}
\def\eea{\end{eqnarray}}

\def\mnras{MNRAS}

\definecolor{vividviolet}{rgb}{0.62, 0.0, 1.0}
\definecolor{amaranth}{rgb}{0.9, 0.17, 0.31}
\definecolor{palatinateblue}{rgb}{0.15, 0.23, 0.89}
\definecolor{brightpink}{rgb}{1.0, 0.0, 0.5}
\definecolor{cornflowerblue}{rgb}{0.39, 0.58, 0.93}
\definecolor{deepcarminepink}{rgb}{0.94, 0.19, 0.22}
\definecolor{radicalred}{rgb}{1.0, 0.21, 0.37}

\hypersetup{
    linktoc=all,
    colorlinks,
    linkcolor={palatinateblue},
    citecolor={brightpink},
    urlcolor={amaranth},
}

\DeclareUnicodeCharacter{2212}{\ensuremath{-}}

\begin{document}

\title{Comparing Bondi and Novikov-Thorne  accretion disk luminosity around  regular black holes}
\author{Salvatore Capozziello}
\email{capozziello@na.infn.it}
\affiliation{Dipartimento di Fisica ”E. Pancini”, Università di Napoli “Federico II”, Via Cinthia,  Napoli, Italy.}
\affiliation{Istituto Nazionale di Fisica Nucleare (INFN), Sezione di Napoli, Napoli, Italy.}
\affiliation{Scuola Superiore Meridionale (SSM), Largo San Marcellino 10, 80138 Napoli, Italy.}
\affiliation{Research Center of Astrophysics and Cosmology, Khazar University, Baku, AZ1096, 41 Mehseti Street, Azerbaijan.}
\author{Serena Gambino}
\email{s.gambino@ssmeridionale.it}
\affiliation{Scuola Superiore Meridionale (SSM), Largo San Marcellino 10, 80138 Napoli, Italy.}
\affiliation{Istituto Nazionale di Fisica Nucleare, Sezione di Napoli, Napoli, Italy.}
\author{Orlando Luongo}
\email{orlando.luongo@unicam.it}
\affiliation{School of Science and Technology, University of Camerino, Via Madonna delle Carceri, Camerino, 62032, Italy.}
\affiliation{Istituto Nazionale di Fisica Nucleare (INFN), Sezione di Perugia, Perugia, 06123, Italy.}
\affiliation{INAF - Osservatorio Astronomico di Brera, Milano, Italy,}
\affiliation{Al-Farabi Kazakh National University, Al-Farabi av. 71, 050040 Almaty, Kazakhstan.}
\begin{abstract}
We consider and compare the Bondi and Novikov-Thorne accretion mechanisms in spherically symmetric regular black holes. To do so, we model the dark matter distribution adopting two main approaches. The first takes into account a cosmologically-inspired dark fluid, whose pressure turns out to be constant, whereas the density and equation of state depend on the radial coordinate. The second case employs an exponential dark matter energy distribution that describes the dark matter cloud in the accretion process. Accordingly, we quantify the mass accreted by regular solutions with the above information on the energy-momentum tensor and their luminosity and thermal properties, derived from the background models.
Numerical simulations of the accretion processes are then reported, comparing regular black holes with Schwarzschild and Schwarzschild-de Sitter metrics. In so doing, we consider the Hayward, Bardeen, Dymnikova and Fang-Wang solutions, where the addiction of a further parameter determines regularity. Implications and physical consequences of using the Novikov-Thorne and Bondi accretion backgrounds are thus critically discussed in varying the free parameters of each spacetime.
\end{abstract}

\maketitle

%
%
\section{Introduction}
\label{Introduction}

Black holes (BHs) are natural solutions to general relativity (GR), exhibiting horizons and thermodynamic properties that have sparked significant speculations, especially after the first BH accretion image \cite{EHT}, and the LIGO gravitational wave detection \cite{LIGO}. In particular, thermodynamic properties of BHs are a very active research area after the Bekenstein seminal paper  \cite{Bekenstein:1973}.

Recently, a renewed interest in the study of regular BHs (RBHs) gained more attention since, unlike traditional BHs, RBHs bypass the Penrose singularity theorems \cite{Hawking:1970, Hawking:1996}, ensuring finite geometric invariants everywhere and avoiding singularities, while still exhibiting horizons similar to classic BH solutions \cite{Lorentz-Euclid}.

They are candidates to describe compact objects, strong gravitational regimes, as well as associated effects such as quasi-periodic oscillations \cite{Boshkayev:2023}, quasi-normal modes \cite{Flachi:2012,Fernando:2012}, non-linear electrodynamic environments \cite{Bronnikov:2001,Fan-Wang:2016}, quantum corrections \cite{Belfiglio:2024,Zhang:2018qdk} and so on, see e.g. \cite{Sharif:2010}.

The first inspired mathematical solution\footnote{The origin of RBHs can be traced back to the pioneering work by Sakharov and Gliner \cite{Sakharov:1996,Gliner:1996}, who firstly introduced the idea of a vacuum-like medium in a de Sitter metric. There, the concept of a smooth behavior of the core, placed at the center, has been explored, in analogy to de Sitter space, culminating in possible de Sitter generalizations, see e.g. \cite{Vertogradov:2024,Zhuk:1995}. } of RBH was proposed by Bardeen in Ref.\cite{Bardeen:1968}, making use of a topological charge, $q$, and was followed by notable models as the Dymnikova RBH \cite{Dymnikova:2004}, Hayward solution \cite{Hayward:2005} and others \cite{Fan-Wang:2016,Bronnikov:2005,Babichev:2020}.

Accordingly, the study of compact objects has been reformulated, including the concept of mimickers \cite{Olmo:2023,Rosa:2022,Tsukamoto:2014,Belissarova:2020}, namely those objects that mime the BH properties being non-necessarily real BHs.

In this respect, observations have led to study the interaction of matter in the proximity of BHs, compact objects and slightly mimickers.

In particular, the formation of accretion disks contributing to the mass growth or the generation of relativistic jets and winds appear crucial both in presence of absence of dark matter \cite{Molla:2025yoh, Capozziello:2023rfv}.

Early studies of accretion, such as the Bondi work in Newtonian gravity \cite{Bondi:1952} and the Michel extension to GR \cite{Michel:1972}, explored accretion around spherically symmetric objects such as the Schwarzschild BH. Later studies included various BH solutions and exotic fluids, such as phantom energy \cite{Babichev:2004}, and extended to RBHs \cite{Debnath:2015, Bahamonde:2015}.

Typically, choosing different spacetimes leads to remarkable differences in the accretion disks. Examples of different behaviors consist in effects due to tangential pressure \cite{Boshkayev:2022}, effects related to the kind of dark matter distribution \cite{Pepe:2012}, consequences related to metrics exhibiting quadrupole \cite{Boshkayev:2021}, acceleration properties \cite{Ashoorioon:2024,Bukhari:2023}, and so forth, see e.g. \cite{Guzman:2011,Pugliese:2022}.

More in detail, we naively interpret at least two main models to characterize accretion regions:
\begin{itemize}
    \item[-] The Novikov-Thorne model, in which the thin accretion disk lies in the equatorial plane, assumes that particles in the disk move in nearly circular orbits. This model allows to evaluate the differential luminosity emitted by the disk and the efficiency of mass conversion into radiation \cite{Kurmanov:2024}.
    \item[-] The Bondi model, distinct from the above, assumes spherical symmetry and isotropy of the disk. Here, the energy and angular momentum of the particle in the accretion disk are conserved. The model also allows the evaluation of the accretion disk luminosity, which is linked to the Eddington luminosity limit.
\end{itemize}

Hence, motivated by the above models, in this paper we consider the accretion mechanism in spherical symmetry for some RBH solutions, with the aim of distinguishing those solutions from standard BHs. To characterize the dark matter distribution, we pose two distinct problems: In the first case, we single out the well-established dark fluid, surrounding the gravitational fields induced by the RBH solutions. A dark fluid is a compelling model mimicking the effects of a cosmological constant, but turning out to be quite different from it. The fluid is characterized by a constant pressure, quite more evident in laboratory on Earth, with a varying density and an equation of state (EoS), $w$ proportional to the inverse of  density itself. The fluid has several applications in cosmology \cite{Arbey:2006,Wang:2017} and has been assumed as a possible way out to heal the cosmological constant problem, see e.g. \cite{Carturan:2002}. Hence, its importance in relativistic astrophysics is recently acquiring great relevance. Accordingly, we study the effect of a dark fluid with constant negative pressure and build on the formalism for relativistic spherical accretion \cite{Debnath:2015}. Instead, for the second approach, we identify the energy density profile and the exponential profile depending only on the radial coordinate, already seen in different work to describe a dark matter cloud in accretion processes \cite{Boshkayev:2020,Boshkayev:2022}. We thus apply such recipes to RBHs. Specifically, we examine the Bardeen and Dymnikova BH solutions, which are characterized by topological charges, as well as the Fan-Wang and Hayward BH solutions, the latter being associated with vacuum energy. For each of these cases, we quantify the mass accreted by these objects and their associated luminosity and thermal properties. We show that the theoretical formalism commonly used to derive the accretion conservation equations remains consistent when using the exponential density profile, while, when using the dark fluid, the procedure requires some correction from the classical method given the particular EoS. More precisely, our results suggest that for a dark fluid, the critical point, which is typically of this accretion process, deviates from the standard relativistic formulation, where the polytropic EoS is commonly involved. Thus, physical differences with respect to the standard polytropic case are reported in detail, as well as a direct comparison with the Schwarzschild and Schwarzschild-de Sitter solutions. Moreover, we compare the results obtained in the Bondi accretion with those found in the Novikov-Thorne model, showing the main departures between the two background theories and highlighting the expected differences.

The paper is structured as follows: Sec. \ref{SEC:accreting_fluids} introduces the two fluids that we aim to investigate in this work. Sec. \ref{SEC:BONDI_accretion} presents the general formalism for Bondi spherical accretion around compact objects using a spherically symmetric metric and derives key relations from conservation laws for the dark fluid case. Sec. \ref{SEC:NT_accretion} reports the formalism used to investigate the Novikov-Thorne accretion model and derives the key relations for both fluid cases. Sec. \ref{SEC:Regular_metrics} discusses the RBH solutions used in this study. Sec. \ref{SEC:BONDI_results} presents numerical simulations of Bondi accretion, comparing RBHs with Schwarzschild and Schwarzschild-de Sitter BHs, first for the dark fluid case and then for the exponential profile. In Sec. \ref{SEC:NT_results} we perform a similar analysis but for the Novikov-Thorne model by modeling the two different accreting fluids for all the RBH solutions considered and incorporating the comparison with Schwarzschild and Schwarzschild-de Sitter BHs. Finally, Sec. \ref{SEC:Discussion} explores the implications of our findings in relation to observations and future research. Throughout the text, we adopt units where $G = c = \hbar = 1$.

%
%
\section{The accreting fluids}
\label{SEC:accreting_fluids}
We here focus on the two typologies of fluids involved here. In particular, we consider the case of \emph{dark fluid} accretion plus the study of \emph{exponential density profile}. The first picture is cosmologically important \cite{Hipolito-Ricaldi:2010,Xu:2011}. It refers to a fully-degenerate model compared with the standard $\Lambda$CDM model and can be arguable from scalar fields considerations \cite{Arbey:2006}, having relevant consequences on the cosmological constant problem \cite{Carturan:2002}, inflation \cite{Wang:2017,Arbey:2020} and late-time cosmological evolution. Conversely, the exponential distribution has been initially proposed in Ref. \cite{Freeman:1970} and by Sofue in Ref. \cite{Sofue:2013a,Sofue:2013b} to study the rotational curves of galaxies. It was also severely investigated in the frameworks of accretion disks \cite{Boshkayev:2020,Kurmanov:2021}. Modeling dark matter with this profile has remarkable advantages \cite{Sofue:2020}, especially if compared with other profiles, exhibiting pathologies in central regions or discontinuities \cite{Boshkayev:2022}. Below, we start with the dark fluid case and then, we describe the exponential distribution.

\subsection{The dark fluid accretion}
\label{SEC:dark_fluid}
Here we derive the equations needed for the dark fluid accretion process represented by the following EoS for the pressure,
\begin{equation}
\label{equation of state}
P = w \rho = \text{const},
\end{equation}
where $w=w(r)$ is the state parameter or barotropic factor\footnote{Usually, a barotropic factor is well-established if the EoS does not depend on the temperature and entropy. Since this appears our case, we may confuse the terms state parameter and barotropic factors as it happens in cosmology.} depending only on $r$.

This EoS can be used to describe a model associated with the dark fluid \cite{Arbey:2009}.

\subsection{The exponential density profile}
\label{SEC:exponential_profile}
The other relevant case is offered by the exponential density profile, first introduced in Ref. \cite{Sofue:2013a,Sofue:2013b} and extensively used in Refs. \cite{Boshkayev:2022,deSalas:2020,Boshkayev:2020,Jusufi:2020,DAgostino:2022,Kurmanov:2021}
that appears directly associated with the density of dark matter\footnote{Alternative perspectives, not explored in this work, include the Navarro-Frenk-White profile \cite{Navarro:1995}, the phenomenological Burkert profile \cite{Burkert:1995}, their extensions \cite{Moore:1997}, the Einasto profile \cite{Graham:2005} and several others, see e.g. \cite{Boshkayev:2020vrg,Hernquist:1990}.}.

The density is thus defined as
\begin{equation}
\label{exponential_prof}
    \rho(r) = \rho_0 e^{-r/r_0} \,,
\end{equation}
where $\rho_0$ is a characteristic density, whereas $r_0$ is the core radius.

In the next sections, we will adopt these two cases to study two accretion models and it will be explain in detail how their behavior affects the type of accretion but also the different results we can get.

%
%
\section{Bondi spherical symmetric accretion and relativistic fluid formalism}
\label{SEC:BONDI_accretion}
%
%
\subsection{General formalism}
\label{SEC:BONDI_general_formalism}
In the context of Bondi accretion, we start by considering a generic spherically-symmetric spacetime
\begin{equation}
\label{general_metric} ds^{2} = -A(r) dt^2 + \frac{dr^2}{B(r)} + r^2(d\theta^2 + \sin^2\theta\, d\phi^2),
\end{equation}
where the two functions, $ A(r)\neq B(r)$, represent positive definite quantities that depend only on the radial coordinate $r$.

The accretion process around a BH requires the definition of an energy-momentum tensor and, accordingly, of the kind of fluids around it. For a perfect fluid, we have
\begin{equation}
\label{stress_energy_tensor}
T_{\mu\nu} = (P + \rho) u_{\mu} u_{\nu} + P g_{\mu\nu},
\end{equation}
where $ \rho $ is the energy density, $ P $ the pressure and $ u^\mu = u^\mu(r) $ is the four-velocity vector, which is reduced within a spherical symmetry to
\begin{equation}
\label{four_velocity_components}
u^{\mu} = \frac{dx^{\mu}}{d\tau} = (u^t, u^r, 0, 0),
\end{equation}
with $ \tau $  the proper time and $ u^\theta = 0 $ and $ u^\phi = 0 $.

From the four-velocity normalization, we may express $ u^t $ in terms of $ u^r $ as
\begin{equation}\label{normalization_condition}
u_{\mu} u^{\mu} = -A(r) (u^t)^2 +\frac{1}{B(r)} (u^r)^2 = -1,
\end{equation}
and, immediately, we obtain
\begin{equation}
\label{expression_for_the_velocity}
u^t = \pm \sqrt{\frac{(u^r)^2 + B(r)}{A(r) B(r)}}.
\end{equation}
To ensure causality, we choose from the two solutions the one with a positive sign to ensure causality, which corresponds to a backward condition. In this way, we enforce a condition in which the fluid is \emph{flowing into the BH rather than moving out of it}.

For simplicity, it appears convenient to hereafter redefine the radial velocity as $ u^r = u $ and, thus, to take off the superscript. Depending on the sign of $u$, we describe two different physical processes,

\begin{itemize}
    \item[-] $u>0$ corresponds to an outgoing flow,
    \item[-] $u<0$ represents an inflow, characteristic of accretion.
\end{itemize}
To characterize the relativistic fluid accretion, we set $ \theta = \pi/2 $ to focus on the equatorial plane, where the metric determinant simplifies to $\sqrt{-g} = r^2 \sqrt{A(r)B(r)^{-1}}$.From this, we can define:

\begin{itemize}

\item[-] The conservation of the energy-momentum tensor, $ T^{\mu\nu}_{~~;\mu} =
\frac{1}{\sqrt{-g}} \partial_r \left( \sqrt{-g} T^{r\nu} \right) = 0$ that leads to the continuity equation,
\begin{equation}
\label{1}
(P + \rho) u r^2 \frac{A(r)}{B(r)} \sqrt{u^2 + B(r)} = \mathcal{C}_1,
\end{equation}
where $ \mathcal{C}_1 $ is an integration constant.

\item[-] The Euler equation consisting in the projection of the conservation equation onto the four-velocity vector, $u_\mu$, $u_\mu T^{\mu\nu}\! _{;\nu} = u^\mu \partial_\mu \rho + (P + \rho) \nabla_\mu u^\mu = 0$, that reads
\begin{equation}
\label{Eueler_differential_form}
\frac{\rho'}{P + \rho} + \frac{u'}{u} + \frac{A'}{2A} + \frac{B'}{2B} + \frac{2}{r} = 0.
\end{equation}
From the above relation, upon integration, immediately the definition of velocity $u$ can be found.

Please note that in this and in the above and following relations, $u^t$ is irrelevant, while $u$ becomes the quantity to handle later. We thus have
\begin{equation}
\label{2}
u r^2 \sqrt{\frac{A(r)}{B(r)}} \exp\left( \int \frac{d\tilde{\rho}}{P + \tilde{\rho}} \right) = -\mathcal{C}_2,
\end{equation}
where $ \mathcal{C}_2 $ is an integration constant, chosen to be negative, $\mathcal C_2<0$, to ensure an inward flow with $ u < 0 $.

\item[-] Finally, the mass flux conservation $J^{\mu}$ is given by:
\begin{equation}
\label{mass_flux_conservation}
J^{\mu}\! _{;\mu} = (\rho u^\mu)_{;\mu} = \frac{1}{\sqrt{-g}} \frac{d}{dr} \left( J^r \sqrt{-g} \right) = 0,
\end{equation}
and, after the integration, we have
\begin{equation}
\label{3}
\rho u r^2 \sqrt{\frac{A(r)}{B(r)}} = \mathcal{C}_3,
\end{equation}
with $ \mathcal{C}_3 $ another constant. In the above, we are deriving the density that will be useful later.
\end{itemize}
By combining these results, we obtain two important relations. From Eqs. \eqref{1} and \eqref{2}, we find
\begin{equation}
\label{4}
(P + \rho) \sqrt{u^2 + B(r)} \sqrt{\frac{A(r)}{B(r)}} \exp \left( - \int \frac{d\tilde{\rho}}{P + \tilde{\rho}} \right) = \mathcal{C},
\end{equation}
where $ \mathcal{C} $ is a constant derived from $ \mathcal{C}_1 $ and $ \mathcal{C}_2 $.
Additionally, combining Eqs. \eqref{1} and \eqref{3}, we derive the Bernoulli equation:
\begin{equation}
\label{5}
\frac{P + \rho}{\rho} \sqrt{\frac{A(r)}{B(r)}} \sqrt{u^2 + B(r)} = \mathcal{C}_4,
\end{equation}
with $\mathcal{C}_4$ being a constant determined by $\mathcal{C}_1$ and $\mathcal{C}_3$. Eq. \eqref{5} is often used to derive the equation for locating the critical point.
In our analysis, as discussed in Subsec. \ref{SEC:BONDI_df_variables}, we adopt an alternative approach to construct the critical point for the case of dark fluid EoS. However, this method will be later applied to the second case under investigation, namely the exponential density profile in Subsec. \ref{SEC:BONDI_exp_variables}.

%
%

\subsection{Bondi accretion with dark fluid equation of state}
\label{SEC:BONDI_df_variables}
Using the integral relations derived in Sec. \ref{SEC:BONDI_general_formalism}, we can express the unknown variables in terms of metric functions and integration constants. This method, outlined in Ref. \cite{Debnath:2015,Bahamonde:2015}, allows us to express the mass accretion rate $\dot{M}$ in terms of these variables, as pioneering discussed in Ref. \cite{Babichev:2004}.

Isolating the radial velocity $ u $ in Eq. \eqref{4} and solving the integral by Eq. \eqref{equation of state}, we compute the radial velocity,
\begin{equation}
\label{velocity_expression}
u = \pm \sqrt{\frac{B(r)}{A(r)}\mathcal{C}^2 - B(r)}\,.
\end{equation}
Here, the negative sign is selected in order to model the accretion process. The energy density $ \rho $ can then be expressed using Eq. \eqref{3} and, so, substituting the velocity from Eq. \eqref{velocity_expression}, we obtain
\begin{equation}
\label{density_expression}
\rho = -\frac{\mathcal{C}_3}{r^2}\frac{\sqrt{B(r)/A(r)}}{\sqrt{\frac{B(r)}{A(r)}\mathcal{C}^2 - B(r)}}\,.
\end{equation}
Both Eqs. \eqref{velocity_expression} and \eqref{density_expression} depend on some integration constants, that can be singled out as later we will show in the text, see Subsec. \ref{SEC:BONDI_df_results}. From their magnitudes, it is thus possible to fix reasonable bounds over the observable quantities that we are going to constrain.

Further, within the spherically symmetric accretion,
\begin{itemize}
    \item[-] a \textit{critical point} can be identified where the fluid transits from a subsonic to a supersonic regime,
    \item[-] the aforementioned transition occurs when the radial velocity equals the speed of sound, a phenomenon dubbed \textit{sonic point},
    \item[-]  in the relativistic extension of the Bondi accretion, a critical point is also present but the fluid can here become supersonic prior to reaching this critical point, indicating that the critical point does not always coincide with the sonic point.
\end{itemize}
In our specific case, by virtue of Eq.  \eqref{equation of state}, the speed of sound is given by
\begin{equation}
\label{sound_speed}
c_s^2 = \frac{\partial P}{\partial \rho} = 0\,, \end{equation}
namely, the dark fluid enables no sound perturbations throughout the entire evolution of the fluid. This fluid is, therefore, particularly relevant since it suggests a matter-like component that can be mimicked by exotic scalar fields \cite{Arbey:2020,MadrizAguilar:2020} or simply by \emph{ad hoc} barotropic fluids \cite{Aviles:2014}.

Accordingly, the speed of sound vanishes at all scales, thus implying that we need to extend the usual analysis proposed in previous works through alternative evaluation of the fluid properties, as it will be clarified later in the following sections.

\subsubsection{Mass accretion rate and disk luminosity}
\label{SEC:BONDI_df_accretion_rate}
Let us now discuss the formula for Bondi accretion, starting from the mass accretion rate $\dot{M}$, which is derived from the conservation of mass flux. We integrate the mass flux over a two-dimensional surface \cite{Debnath:2015} and obtain the following equation:

\begin{equation}
\label{accretion_rate}
\dot{M} = - \int d\theta d\phi \sqrt{-g} T^r\! _0\,.
\end{equation}
Focusing on the dark-fluid case, we derive the following general expression:
\begin{equation}
    \label{Mdot_dark_fluid}
    \dot{M}_{\rm dark\ fluid} = -4\pi r^2(P +\rho) \frac{u\sqrt{u^2+B(r)}}{A(r)B(r)}\,.
\end{equation}
This equation forms the basis of the first method, whose reasoning will be elaborated later. Notably, it allows us to recover a specific subcase of the accretion rate equation, used in prior studies. By applying Eq. \eqref{1} to Eq. \eqref{Mdot_dark_fluid}, we find:
\begin{equation}
    \label{Mdot_sottocaso1}
    \dot{M}_{\rm special} = 4\pi \mathcal{C}_2 M^2 (P_\infty + \rho_\infty)\,,
\end{equation}
where $ P_\infty $ and $ \rho_\infty $ represent the pressure and density at infinity, i.e., where the variables reach their equilibrium values. Assuming constant pressure, this further simplifies to:
\begin{equation}
    \label{Mdot_sottocaso2}
    \dot{M}_{\rm special} = 4\pi \mathcal{C}_2 M^2 (P + \rho(r))\,.
\end{equation}
This accretion rate describes a scenario governed by a cosmological constant $\Lambda$ where the density, $\rho$, remains constant. This becomes evident if we assume the velocity in Eq. \eqref{velocity_expression} to be constant, $u = \text{const}$. Under this assumption, the density expression in Eq. \eqref{density_expression} also remains constant, resulting in $\rho = \text{const.}$. Consequently, in the specific case of Eq. \eqref{accretion_rate}, the mass accretion rate, $\dot{M}$, is constant and reflects a universe dominated by a cosmological constant, satisfying the condition $\rho = \rho_\infty = \text{const}$.

%
%
\subsubsection{Critical points analysis}
\label{SEC:BONDI_df_critical_points}
Despite the above limitation, related to the precise form of the dark fluid EoS, a critical point still exists at $r=r_{\textit{crit}}$. Even though, at $r_{\textit{crit}}$, the radial velocity does not equate the sound speed, the relativistic nature of this fluid allows us to identify a critical point that depends on the geometry and properties of the specific spacetime under exam, rather than on the pressure $P$, which instead turns out to be a constant across the entire accretion process.

As done in previous works \cite{Debnath:2015,Babichev:2004}, we can redefine the sound speed with this new variable:
\begin{equation}
\label{new_variable_V}
V^2 = \frac{d \ln(P + \rho)}{d \ln \rho} - 1\,.
\end{equation}
However, the latter does not work well with an EoS with constant pressure and can even lead to discontinuities as we will discuss in Subsec. \ref{SEC:V2_problem}. Hence, in lieu of using Eq. \eqref{new_variable_V} to derive the conventional equation for the critical point, following for example the procedure reported in Ref. \cite{Bahamonde:2015}), we notice that,

\begin{itemize}
    \item[-] $V^2$ remains positive only near the inner region.
    \item[-] It turns to negative values as we move outward.
\end{itemize}

From the second condition, we argue that by construction $V^2$ cannot be positive everywhere, regardless the EoS, as this does not hold in our case.
Indeed, this stands in contrasts with other EoS, such as a polytropic one or the one chosen by the work mentioned above, where an EoS of the type $P(r)=w\rho(r)$, with $w=const$ is chosen.

Since $V^2$ has to remain positive for physical consistency, we identify a small region where $V^2$ takes negative values, requiring a redefinition. By analyzing its behavior, we find a point—unrelated to any sonic transition of the fluid—where $V^2$ changes sign, transitioning from negative to positive. Therefore, we can still refer to this as a \emph{critical point}, as it marks the region where $V^2$ requires redefinition, which we implement as follows:
\begin{align}
\label{new_critical_point1}
&\, V^2 > 0,\quad r < r_{\text{crit}}\,,\\
\label{new_critical_point2} &\, V^2 < 0,\quad r > r_{\text{crit}} \rightarrow V^2 \rightarrow -V^2\,.
\end{align}

At this stage, we can now wonder whether the condition $V^2=0$ can help to constrain some of the unbounded integration constants.

In Subsec. \ref{SEC:BONDI_df_results}, we will report a detailed choice for these quantities.

%
%
\subsection{Bondi accretion with exponential density profile}
\label{SEC:BONDI_exp_variables}
Let us now discuss the second fluid under investigation: a fluid with an exponential density profile. We now derive the radial velocity and pressure profiles. Specifically, starting from Eq. \eqref{3}, we express the velocity explicitly by solving w.r.t. $u$:
\begin{equation}
\label{exponential_velocity_expression} u = \frac{\mathcal{C}_3}{\rho r^2} \sqrt{\frac{B(r)}{A(r)}}.
\end{equation}
This expression for the velocity is then substituted into Eq. \eqref{1} to determine the pressure profile, yielding
\begin{equation}
\label{exponential_pressure_expression} P = \left[ \frac{\mathcal{C}_3 / \mathcal{C}_1}{ \sqrt{\mathcal{C}_3^2 / \rho^2 r^4 + A(r)} } - 1 \right] \rho \,.
\end{equation}

\subsubsection{Mass accretion rate and disk luminosity}
\label{SEC:BONDI_exp_accretion_rate}
For the second case, which involves accretion modeled by an exponential density profile, we start from Eq. \eqref{Mdot_sottocaso1} and generalize it for any relation of the form $ P = P(\rho) $, which includes our scenario. Thus, we can write:
\begin{align}
    \label{accretion_exponential}
    \dot{M}_{\rm exponential} &= 4\pi \mathcal{C}_2 M^2 (P_\infty + \rho_\infty) \nonumber \\
    &= 4\pi \mathcal{C}_2 M^2 (P(r) + \rho(r))\,.
\end{align}

Now, to describe the luminosity, we use the expression of Eddington luminosity. This limit indicates the maximum luminosity above which the radiation pressure exceeds the gravitational attraction, halting the accretion process \cite{Frank:2002}. It is derived by equating the gravitational force acting on a proton to the radiation pressure force exerted by a luminous source. Since radiation interacts primarily with electrons via Thomson scattering, the electrostatic coupling between electrons and protons ensures that the entire plasma feels the effect.

The result is the following expression:
\begin{equation}
\label{Eddington_lum}
L_{\rm Edd} = \frac{4\pi GM m_p}{\sigma_T},
\end{equation}
where $ m_p $ is the mass of the proton, and $ \sigma_T $ is the Thomson cross-section.

To account for the efficiency of converting accreted mass into radiation, we introduce an efficiency parameter $ \eta_{\rm eff} $, typically around 0.1. The final expression for the luminosity is:
\begin{equation}
\label{bondi_luminosity}
L = \eta_{\rm eff} \dot{M}\,,
\end{equation}
with $\dot{M}$ given by Eq. \eqref{Mdot_dark_fluid} for the first case and Eq. \eqref{accretion_exponential} for the second case.

\subsubsection{Critical points analysis}
\label{SEC:BONDI_exp_critical_points}
In order to highlight the main differences with respect to the first case, let us perform a critical point analysis to determine a \emph{critical or sonic  point, where the accreting fluid transits from subsonic to supersonic flow}.

Notably, as observed in the relativistic case, this transition can sometimes occur before reaching the critical point, i.e., the reason to  refer to this as  \emph{critical point}.

Here, since the sound speed is non-zero, it can be easily derived directly from Eq. \eqref{exponential_pressure_expression}, as
\begin{align}
\label{exponential_sound_speed}
    &\, c_s^2 =  \frac{\mathcal{C}_3/\mathcal{C}_1}{\left[\mathcal{C}_3^2/\rho ^2 r^4 + A(r)\right]^{3/2}} \left(\frac{2\mathcal{C}_3^2}{\rho ^2 r^4} + A(r)\right) -1 \,.
\end{align}
To obtain the critical point expression, following Ref. \cite{Michel:1972}, we require a combination of all the other conservation laws.

To do so, we can take the logarithmic derivatives of Eq. \eqref{3}-\eqref{5}, combining them and using the variable $V^2$ introduced in Eq. \eqref{new_variable_V}.

We then obtain the following equation for the critical point $r_c$
\begin{align}
\label{critical_point_eq}
   & \left[ (V^2 - 1) - \frac{u^2}{u^2 + B} \right] \frac{du}{u} +
   \left[ (V^2 - 1) \left( \frac{A'}{2A} - \frac{B'}{2B} \right) \right. \nonumber \\
   & \left. + \frac{2}{r} V^2 - \frac{B'}{2(u^2 + B)} \right] dr = 0\,.
\end{align}
The critical point is determined by setting both of the bracketed terms to zero simultaneously, leading to the following system of equations
\begin{align}
\label{crit_point_conditions}
    & V_c^2 = \frac{u_c^2}{u_c^2 + B(r_c)} \,, \\
    & (V_c^2 - 1) \left( \frac{A'(r_c)}{2A(r_c)} - \frac{B'(r_c)}{2B(r_c)} \right) + \frac{2}{r_c} V_c^2 - \frac{B'(r_c)}{u_c^2 + B(r_c)} = 0 \,.
\end{align}
By separating the two variables, $ u^2 $ and $ V^2 $, we obtain the following final set of decoupled equations:
\begin{align}
    &\label{variables_crit_point_vc2} V_c^2 = \frac{1}{1 + \frac{4A(r_c)}{A'(r_c) r_c}} \,, \\
    &\label{variables_crit_point_uc2} u_c^2 = \frac{B(r_c) A'(r_c) r_c}{4 A(r_c)} \,.
\end{align}
These equations, are related, together with the expressions from Eqs. \eqref{new_variable_V}-\eqref{exponential_velocity_expression}, to the study of the critical point $r_c$ and consequently, to constrain one of the integration constants, namely $\mathcal{C}_3$.

Indeed, by manipulating these relations, as will be demonstrated in Subsec. \ref{SEC:BONDI_exp_rc}, and by equating the expressions at the critical point with their general counterparts, we can derive at least one constraint on the integration constants, specifically on $ \mathcal{C}_3 $.

%
%
\section{Novikov-Thorne accretion model and relativistic fluid formalism}
\label{SEC:NT_accretion}
We now introduce the second accretion model considered in this work: the Novikov-Thorne accretion model. In this framework, the disk is modeled as a thin, optically thick structure within a relativistic setting. A steady-state assumption guarantees that the rest-mass accretion rate, $\dot{M}$, remains constant throughout the disk.

We adopt the same metric as in Eq. \eqref{general_metric}. The mass of the accreting fluid is defined by
\begin{equation}
\label{NT_mass} M_{\rm fluid}(r) = 4\pi \int_{r_{\rm b}}^{r} d\tilde{r} \rho(\tilde{r}) \tilde{r}^2,
\end{equation}
 where $\rho(\tilde{r})$ is left unspecified at this stage. Here, $r_{\rm b}$ denotes a threshold radius, which needs to be greater than the event horizon, and marks the inner boundary of the fluid envelope. The total mass of the system, i.e. the BH plus the fluid envelope, is then given by
\begin{align}
                M(r) &=  M_{\rm BH},\ \hspace{2cm} \ r_{\rm g} \leq r \leq r_{\rm b}   \,,\nonumber \\
\label{NT_mass_distribution} M(r) &=  M_{\rm BH} + M_{\rm fluid}(r),\quad \ r_{\rm b} \leq r \leq r_{\rm s}\,, \nonumber \\
                M(r)  &=  M_{\rm BH} + M_{\rm fluid}(r_{\rm s}),\quad  r_{\rm s} \leq r,
\end{align}
with $r_{\rm s}$ representing the outer radius of the accreting envelope and $r_g = 2,M_{\rm BH}$.

An essential relation in our formulation is the one connecting pressure and density. Derived from the Tolman-Oppenheimer-Volkoff (TOV) equations, see, e.g., \cite{TOV:1939} and \cite{Peirani:2008}, the TOV equation—obtained via the conservation of the stress-energy tensor $T^{\mu r}_{\quad;\mu}=0$—leads to
\begin{equation}
\label{TOV_eq} \frac{dP(r)}{dr} = -\frac{(P(r) + \rho(r))}{2 A(r)} \frac{d A(r)}{dr}.
\end{equation}
In this approach the explicit forms of the pressure and density are obtained by specifying the metric functions, which are assumed to be known for each solution.

The flux radiated by the disk—due to particles on circular geodesics in the equatorial plane $\theta = \pi/2$—is derived from conservation laws for energy, angular momentum, and mass. It is given by
\begin{equation}
\label{NT_flux}
\mathcal{F}(r) = \frac{\dot{M}}{4\pi\sqrt{-g}}\frac{d\Omega/dr}{\left(E-\Omega L\right)^2}
\int_{r_{\rm i}}^{r}\left(E-\Omega L\right)\frac{dL}{d\tilde{r}}d\tilde{r},
\end{equation}
where $\Omega$ denotes the orbital angular velocity, and $E$ and $L$ are the specific energy and angular momentum of test particles in the disk. Here, $\sqrt{-g}$ is the square root of the absolute value of the determinant of the metric, evaluated in the equatorial plane. (For a diagonal metric, $\sqrt{-g} = \sqrt{g_{rr},g_{tt},g_{\phi\phi}}$.) The radius $r_{\rm i}$ indicates the innermost stable circular orbit (ISCO); the derivatives $d\Omega/dr$ and $dL/d\tilde{r}$ are taken with respect to the radial coordinate.

For completeness, the orbital parameters are expressed as follows:
\begin{align}
& \Omega = \frac{d\phi}{dt} = \sqrt{-\frac{\partial_r g_{\rm tt}}{\partial_r g_{\phi\phi}}}\,, \\
& E = u_{\rm t}\,, \\
& L = -u_{\phi} = - \Omega u^{\rm t}g_{\phi\phi}\,, \\
& \dot{t} = u^{\rm t} = \frac{1}{\sqrt{g_{\rm tt} + \Omega^2 g_{\phi\phi}}}\,.
\end{align}
It is useful to relate the locally emitted flux to the luminosity measured by an observer at infinity, as $\mathcal{F}(r)$ is not directly observable. In differential form, the luminosity takes the following form
\begin{equation}
\label{differential_luminosity} \frac{d\mathcal{L}_{\infty}}{d\ln r} = 4\pi r\sqrt{-g}\mathcal{F}(r).
\end{equation}
%
%

\subsection{Novikov-Thorne Accretion with Dark Fluid Equation of State}
\label{SEC:NT_df_variables}
Let us consider the first scenario, which includes the dark fluid EoS, relation \eqref{equation of state}. By doing so, Eq. \eqref{TOV_eq} will simplify and leads to the the following expression:
\begin{equation}
    \label{NT_dark_fluid_pressure}
    P = - \rho \,.
\end{equation}
Since the derivative of $ P $ is zero, because the pressure is a constant and can be seen from Eq. \eqref{equation of state}, we can automatically say that the density is also constant in this case, as is the parameter $ w $. Thereofre, we will have that the envelope of dark fluid that creates in the accretion disk, will hve a constant density and pressure for all of the extension from $ r_{\rm b} $ up to $ r_{\rm s} $.

Under these assumptions, if we take these condtions and use them to derive the mass distribution from Eq. \eqref{NT_mass}, it leads to an unphysical rsult having a mass that icncrease and increse as the radius incrase ($\propto r^3$). Therefore, it is more suitable to choose a constant mass distribution for the dark fluid $ M_{\rm dark \ fluid}=const$ that extends from $ r_{\rm b} $ up to $ r_{\rm s} $ in such a way that
\begin{align}
                M(r) &=  M_{\rm BH},\ \hspace{2cm} \ r_{\rm g} \leq r \leq r_{\rm b}   \,,\nonumber \\
\label{NT_df_mass_distr} M(r) &=  M_{\rm BH} + M_{\rm dark \ fluid},\quad  r_{\rm b} \leq r \leq r_{\rm s}\,, \nonumber \\
                M(r)  &=  M_{\rm BH} +M_{\rm dark \ fluid},\quad  r_{\rm s} \leq r.
\end{align}
%
%
%
\subsection{Novikov-Thorne Accretion with Exponential Density Profile}
\label{SEC:NT_exp_variables}
In this second case, we take a fluid envelope modeled by the exponential density profile in \eqref{exponential_prof}. In this study, since the metric functions are already known, we retain Eq. \eqref{TOV_eq} without modification. However, particular attention must be given to the numerical integration of the TOV equation across the envelope boundaries at $r_{\rm b}$ and $r_{\rm s}$.

At the inner boundary, the matching between the vacuum region and the fluid envelope is imposed by selecting a test value for the density:
\begin{equation}
    \label{density_rb_exp}
    \rho(r=r_{\rm b}) = \rho_0 e^{-r_{\rm b}/ r_0} = \rho_{\rm b}\,,
\end{equation}
and by setting the pressure equal to its Newtonian limit,
\begin{equation}
    \label{pressure_rb_exp}
    P(r=r_{\rm b}) =  P_{\rm b}\,.
\end{equation}
The Newtonian solution for the pressure profile, derived by combining Eq. \eqref{exponential_prof} with the mass distribution of Eq. \eqref{NT_mass_distribution}, reads
\begin{align}
\label{pressure_newt_limit}
P(r) = &\ 8 \pi r_0^2 \rho_0^2 \Bigg\{ -e^{-2r/r_0}\left( \frac{1}{4} + \frac{r_0}{r} \right) - \text{Ei}( -2r/r_0)\nonumber \\
    &\  + \Bigg[\frac{M}{8 \pi r_0^3 \rho_0} + e^{-r_b/r_0} \left( 1 + \frac{r_b}{r_0} + \frac{r_b^2}{2 r_0^2} \right) \Bigg] \nonumber \\
    &\ \times \left[ \frac{e^{-r/r_0}}{r/r_0} + \text{Ei}( -r/r_0)\right] \Bigg\} \,,
\end{align}
where $\text{Ei}(r)=-\int_{-r}^{\infty}dt \frac{e^{-t}}{t}$ is the exponential integral.
 (up to 1.5) as discussed in \cite{Boshkayev:2020}, which effectively regulates the envelope’s thickness.

Following the procedure of the reference above, a relativistic correction is applied to the boundary pressure $P_{\rm b}$ at $r = r_{\rm b}$. By varying $P_{\rm b}$ by up to a factor of 1.5, one effectively regulates the width—or thickness—of the fluid envelope, thereby ensuring continuity at the interface $r = r_{\rm b}$. All other quantities, such as the disk luminosity, retain the forms derived previously.
%
%
\section{Spherically symmetric regular metrics}
\label{SEC:Regular_metrics}

In this section, we examine in detail how fluid accretion is modeled using various RBHs. They are a class of BH solutions that do not exhibit essential singularities. They have been proposed as a means to circumvent the singularity theorems by Penrose and Hawking \cite{Hawking:1996}, who stated that singularities are inevitable in GR. However, these conclusions arise from a purely classical description of the theory. Since singularities are non-physical entities, a class of regular solutions can be developed. Bardeen \cite{Bardeen:1968} was the first to propose a static, spherically symmetric regular metric, where the BH is characterized by a mass term and a charge parameter that ensures the regularity of the solution. Similarly, the Hayward metric \cite{Hayward:2005} incorporates a damping factor that maintains regularity at the center and exhibits features that allow for modeling various processes, including accretion.

\subsection{Black hole classes}

In this section, we classify the metrics studied in this work. The first class comprises singular spherically symmetric solutions, which serve as reference models for comparison with RBH solutions. This comparison allows us to highlight the differences between singular and regular metrics. The subsequent sections discuss different RBH classes, categorized based on the physical mechanisms that ensure regularity.

The first class of RBHs involves solutions incorporating vacuum energy effects, such as a cosmological constant, which alter the spacetime structure, particularly at large distances, and modify the nature of the horizons.

The second class, instead, contains topological charges, which regularize the BH solutions. They are topological magnetic or electric charges. Another solution is placed between these groups as it does not fit strictly into these classifications. Below, we provide a detailed description of each RBH solution.

All of the BHs considered in this section employs a single shape function in the metric, such that, according to the metric in \eqref{general_metric}, it simplifies as $A(r) = B(r)$.

\subsection{Singular black holes}
%
%
\subsubsection{The Schwarzschild solution}
To better understand how RBHs differ from standard BHs, we consider also the standard Schwarzschild solution. It describes the gravitational field outside a static, spherically symmetric mass and here the function $A(r)$, is given by:
\begin{equation}
\label{func_schwarz} A(r)=1 - \frac{2M}{r}\,,
\end{equation}
where M is the BH mass and presents a singularity at $r=2M$, known as the event horizon.

The comparison of the behavior of RBHs with this classical solution allows for a major understanding of the accretion of fluids around RBHs.
%
%
\subsubsection{The Schwarzschild-de Sitter solution}
The Schwarzschild-de Sitter BH is a solution that includes the cosmological constant $\Lambda$, representing a BH embedded in an expanding universe. The expansion is driven by the presence of this vacuum energy. The corresponding shape function is:
\begin{equation}
    A(r)=1 -   \frac{2 M}{r} - \frac{\Lambda}{3} r^2\,,
\end{equation}
where $M$ is the mass of the BH, and $\Lambda$ is the cosmological constant. The term $2 M/r$ corresponds to the usual Schwarzschild term, while the term with the vacuum energy $\Lambda$ modifies the spacetime geometry at large distances, introducing a cosmological horizon, in addition to the physical event horizon.
\subsection{Regular black holes}
%
%
\subsubsection{The Hayward solution}
The first RBH solution we discuss is the Hayward solution \cite{Hayward:2005}. It introduces vacuum energy by modifying the BH mass with a regularizing term that prevents singularities. The shape function for this solution is given by:
\begin{equation}
    \label{Hayward_func} A(r)=1-\frac{2Mr^2}{r^3+2a^2}\,,
\end{equation}
where $a$ is a scale parameter that influences the BH core. This solution is spherically symmetric and effectively introduces a finite energy density at the BH center, mimicking the vacuum energy and removing the central singularity.
%
%
\subsubsection{The Bardeen solution}
The first RBH solution of the second class that we explore is also one of the earliest RBH models called Bardeen RBH \cite{Bardeen:1968}. It modifies the Schwarzschild BH mass and introduces a magnetic monopole charge $q_B$ to prevent the formation of a singularity. The shape function is expressed as:
\begin{equation}
   \label{Bardeen_func} A(r)=1-\frac{2Mr^2}{(r^2+q_B^2)^{3/2}}\,.
\end{equation}
The charge $q_B$ modifies the geometry of the BH, ensuring regularity at the core. The Bardeen solution is foundational in demonstrating how topological charges can be used to create a RBH.
%
%
\subsubsection{The Dymnikova solution}
The Dymnikova solution \cite{Dymnikova:2004}, based upon the works by \cite{Bronnikov:2001,Berej:2006}, is another charged RBH but this time incorporating an electric charge $q_D$ associated with the topology of spacetime. The shape function is:
\begin{equation}
   \label{Dymnikova_func} A(r)=1-\frac{4M}{\pi r}\left[\arctan\left(\frac{r}{r_{\rm D}}\right)-\frac{r \ r_{\rm D}}{(r^2+r_{\rm D}^2)}\right]\,.
\end{equation}
Here $r_{\rm D}=\frac{\pi q_D^2}{8M}$ is a characteristic length scale. This solution shows how electric topological charges can regularize BHs, similarly to the magnetic charge of Bardeen BH.
%
%
\subsubsection{The Fan-Wang solution}
The Fan-Wang BH \cite{Fan-Wang:2016} cannot be classified as the other above solutions. It introduces a charge parameter which modifies the geometry in a distinct way. The shape function is:
\begin{equation}
   \label{FW_func}  A(r)=1-\frac{2Mr^2}{(r+l_{\rm FW })^{3}}\,,
\end{equation}
where $l_{\rm FW}\leq 8/27$ \cite{Kurmanov:2024} represents a characteristic charge parameter. This parameter alters the dynamics by influencing the accretion processes around the BH and introducing a peculiar structure.

\subsection{Variables expressions for Bondi accretion}
\label{SEC:BONDI_variable_RBH_expression}
The expressions for the variables in Subsec. \ref{SEC:BONDI_df_variables} and Subsec. \ref{SEC:BONDI_exp_variables} are thereby simplified and this simplification arises because, as we said, in these cases the metric functions satisfy $A(r) = B(r)$.

For the dark fluid case, the velocity in Eq. \eqref{velocity_expression} simplifies to:
\begin{equation}
    u = -\sqrt{\mathcal{C}^2 - A(r)}\,,
\end{equation}
while the energy density in Eq. \eqref{density_expression} becomes:
\begin{equation}
    \rho = -\frac{\mathcal{C}_3}{r^2 \sqrt{\mathcal{C}^2 - A(r)}}\,,
\end{equation}
and the mass accretion rate in Eq. \eqref{accretion_rate} simplifies to:
\begin{equation}
    \dot{M} = \frac{4\pi r^2 \mathcal{C}}{A(r)^2} \left( P\sqrt{\mathcal{C}^2 - A(r)} - \frac{\mathcal{C}_3}{r^2} \right)\,.
\end{equation}
In contrast, for the exponential density profile, the expressions simplify as follows. The radial velocity is given by:
\begin{equation}
    u = \frac{\mathcal{C}_3}{\rho r^2}\,,
\end{equation}
while the pressure profile in Eq. \eqref{exponential_pressure_expression}, as well as the accretion rate and luminosity, remain unchanged.

%
%
\section{The analysis of Bondi accretion process}
\label{SEC:BONDI_results}
In this section, we investigate the accretion processes around various RBH solutions. We begin by examining the behavior of a dark fluid and plotting the relevant variables for each solution. Following this recipe, we compare the results obtained for the dark fluid with those for the classical Schwarzschild and Schwarzschild-de Sitter solutions to highlight differences and similarities among RBHs and their singular counterparts.

Afterwards, we turn our attention into a second scenario, where the exponential density profile is adopted to describe the fluid distribution in the accretion disk. Using the same methodology as before, we analyze and compare the accretion behaviors.

By remarking the differences between dark fluid and exponential profile, we intend to clarify which model better drives the  spherically symmetric accretion dynamics and whether there are or not notable similarities in their behaviors.

\subsection{The dark fluid accretion process}
\label{SEC:BONDI_df_results}

Using the relations established before, we analyze the behavior of the chosen dark fluid to characterize the accretion around RBHs. The constants of motion are chosen by us after preliminary tests, while the critical radius $r_{\text{crit}}$ and the event horizon radius $r_{\text{EH}}$ are determined numerically. The event horizon is found by imposing the condition $ A(r) = 0 $ or equivalently $ B(r) = 0 $, since they are equal in this work. The results are summarized in Tab. \ref{TAB:const_val_1st_method}.

For all the solutions, the BH mass is set to $M = 1$ and we define this scaled radius $x = r/M$ to plot all the variables. Since each solution we test is different, their event horizons will also be different, so we evaluate them all numerically and start plotting from the corresponding event horizon radius for each solution $x_{\text{EH}}=r_{\text{EH}}/M$.

\begin{figure}[!h]
    \begin{subfigure}{0.5\textwidth}
        \includegraphics[width=\linewidth]{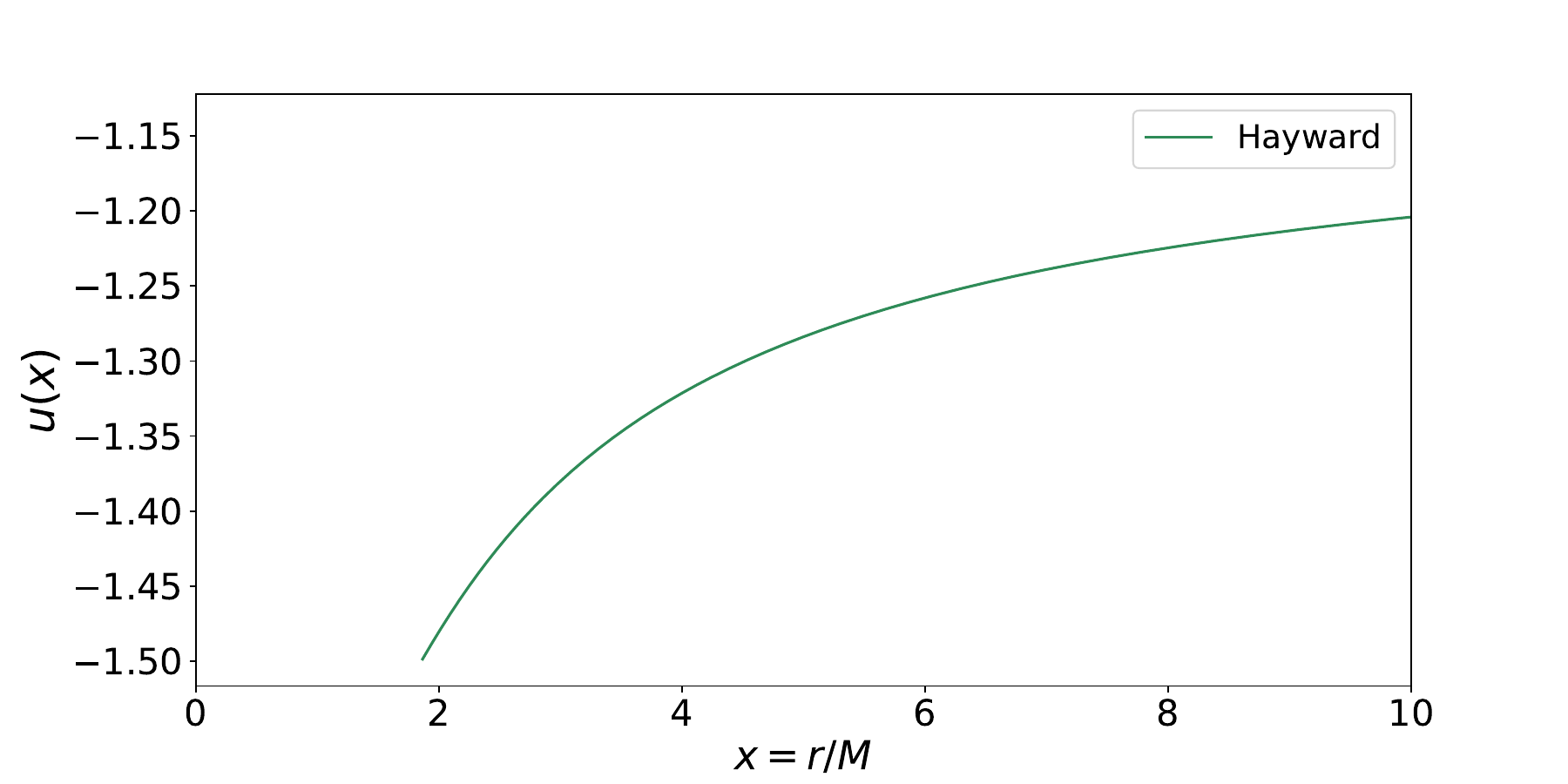}
    \end{subfigure}
    \hfill
    \begin{subfigure}{0.5\textwidth}
        \includegraphics[width=\linewidth]{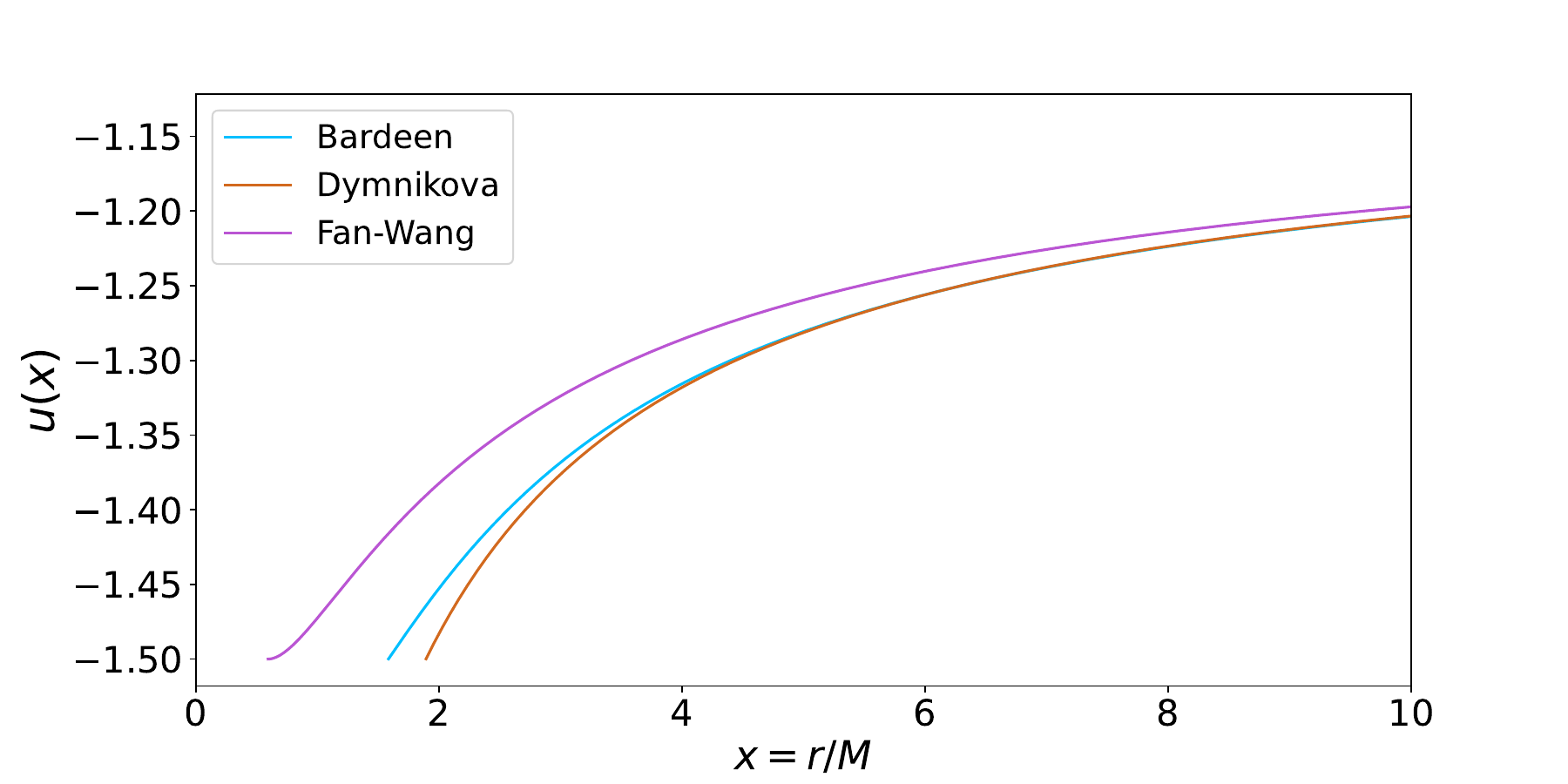}
    \end{subfigure}
       \caption{\justifying Radial velocity profiles $ u $ as a function of $ x = r/M $ for different RBH solutions.
    {\bf Left:} Behavior of vacuum solution. {\bf Right:} Behavior of topologically charged and Fan-Wang solutions.}
    \label{fig:u}
\end{figure}

After some tests, the pressure value has been set to be negative. We choose $P=-0.75$.
%
%
We start with the radial inflow velocity profiles shown in Fig.~\ref{fig:u}.
In the top panel, we analyze the BH characterized by vacuum energy, the Hayward BH. The solution exhibit a small range of variation for the velocity values and it shows a a sharp decline in its velocity profile as it approaches the event horizon.

As we move outwards from the inner region of the accretion disk, the Hayward solution shows a continuous decrease in velocity with increasing $x$, indicating a gradual slowing of the accretion flow as the fluid moves further away from the event horizon.

The bottom panel shows the radial velocity profiles for topologically charged BHs, including the Bardeen, Dymnikova and Fan-Wang BHs. All BHs show smooth velocity profiles, similar to the Hayward solution, with a steady decrease in velocity as $x$ increases. The magnetic charge introduces only minor deviations from the vacuum energy solutions, and despite variations in the radii of the event horizons for these metrics, the overall growth behavior remains similar.

\begin{figure}[!h]
    \begin{subfigure}{0.5\textwidth}
        \includegraphics[width=\linewidth]{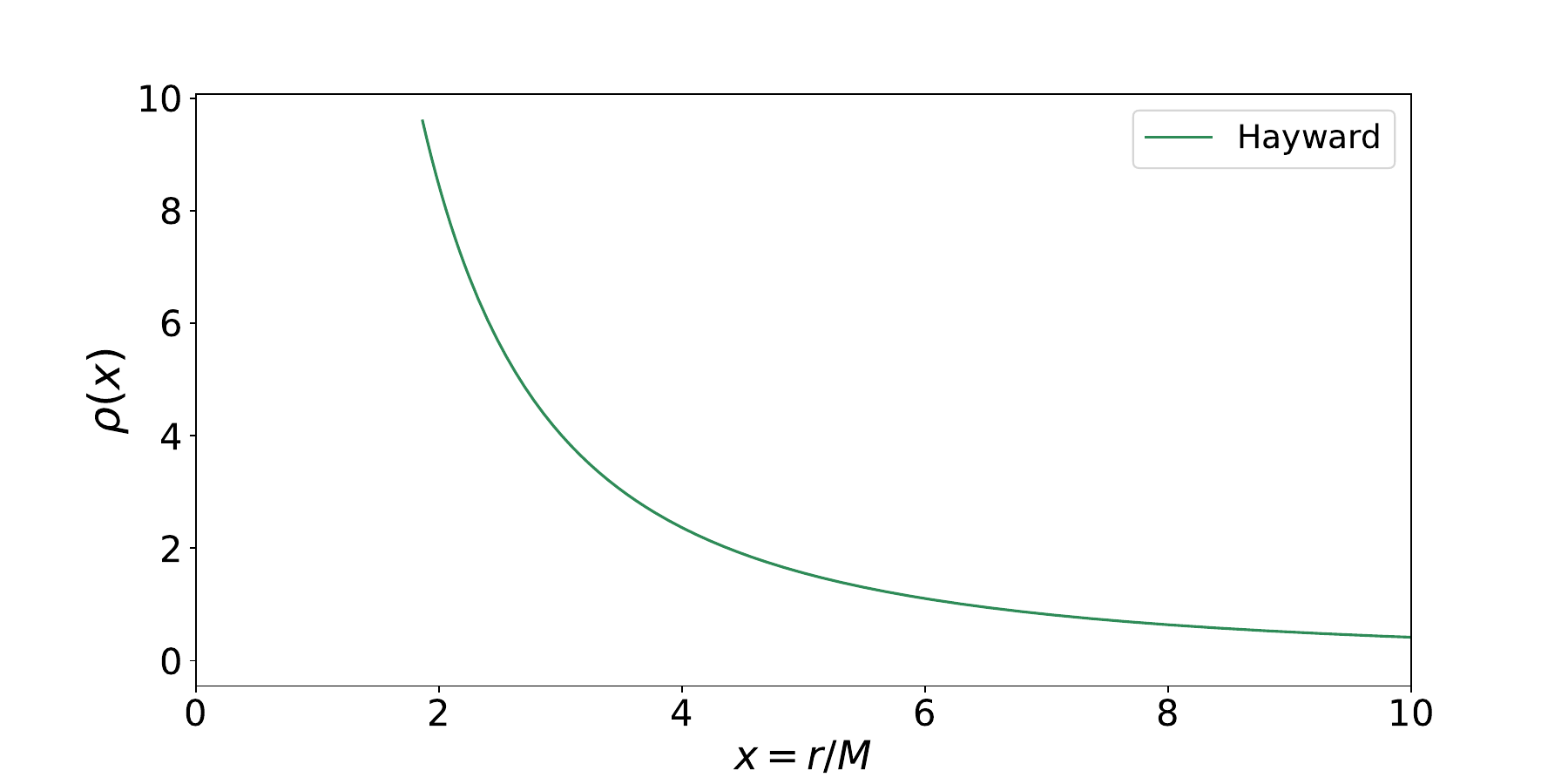}
    \end{subfigure}
    \hfill
    \begin{subfigure}{0.5\textwidth}
        \includegraphics[width=\linewidth]{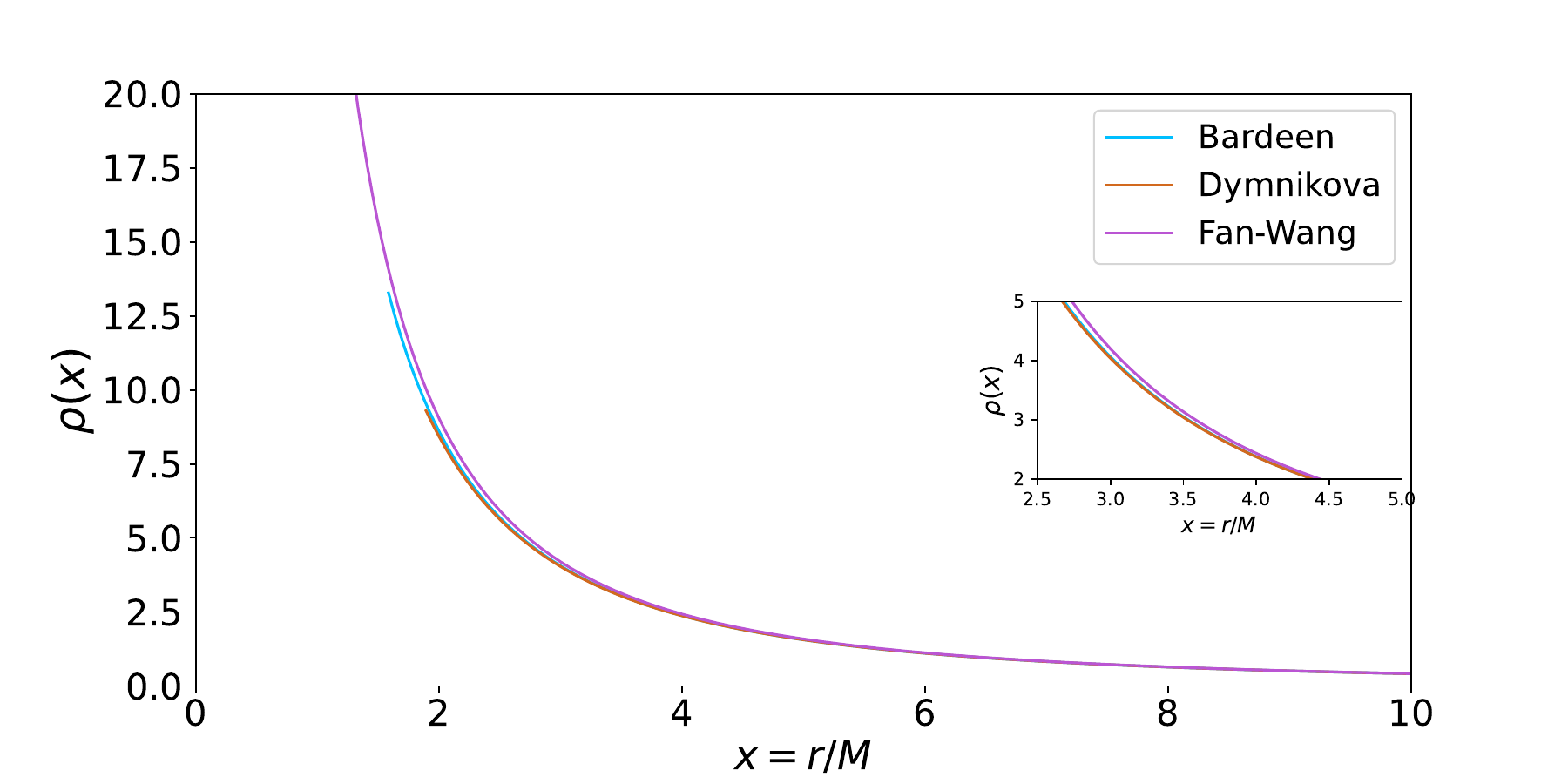}
    \end{subfigure}
      \caption{\justifying Energy density profile $ \rho $ as a function of $ x = r/M $. {\bf Top:} Behavior of vacuum solution. {\bf Bottom:} Behavior of topologically charged and Fan-Wang solutions.}
    \label{fig:rho}
\end{figure}

We only highlight how the Fan-Wang solution gives smaller values for the velocity at small radii, but approaches the other solutions as we move away from the inner region of the accretion disk.
%
%
%
Next, we examine the energy density, starting with the top panel of Fig.~\ref{fig:rho}.
For the vacuum energy solution, we have the Hayward BH that shows a rapid increase in energy density at small values of $x$. The final value towards which the energy density grows at small $x$ depends on the constant $\mathcal{C}$, and these values have been carefully chosen to give consistent results for all the solutions plotted. As we move away from the accretion disk, the profile becomes smooth and asymptotically approaches zero.

\begin{figure}[!h]
    \begin{subfigure}{0.5\textwidth}
        \includegraphics[width=\linewidth]{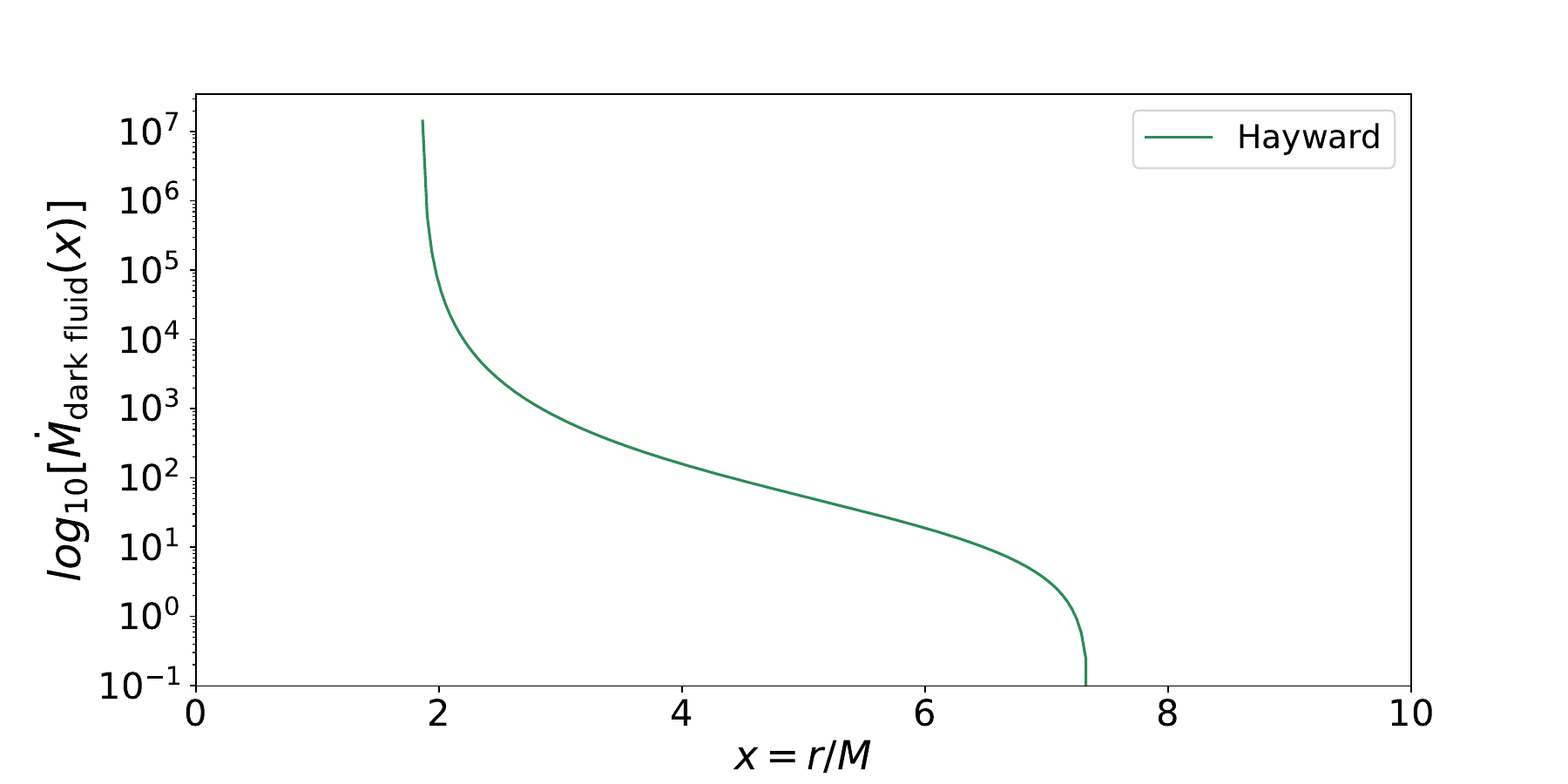}
    \end{subfigure}
    \hfill
    \begin{subfigure}{0.5\textwidth}
        \includegraphics[width=\linewidth]{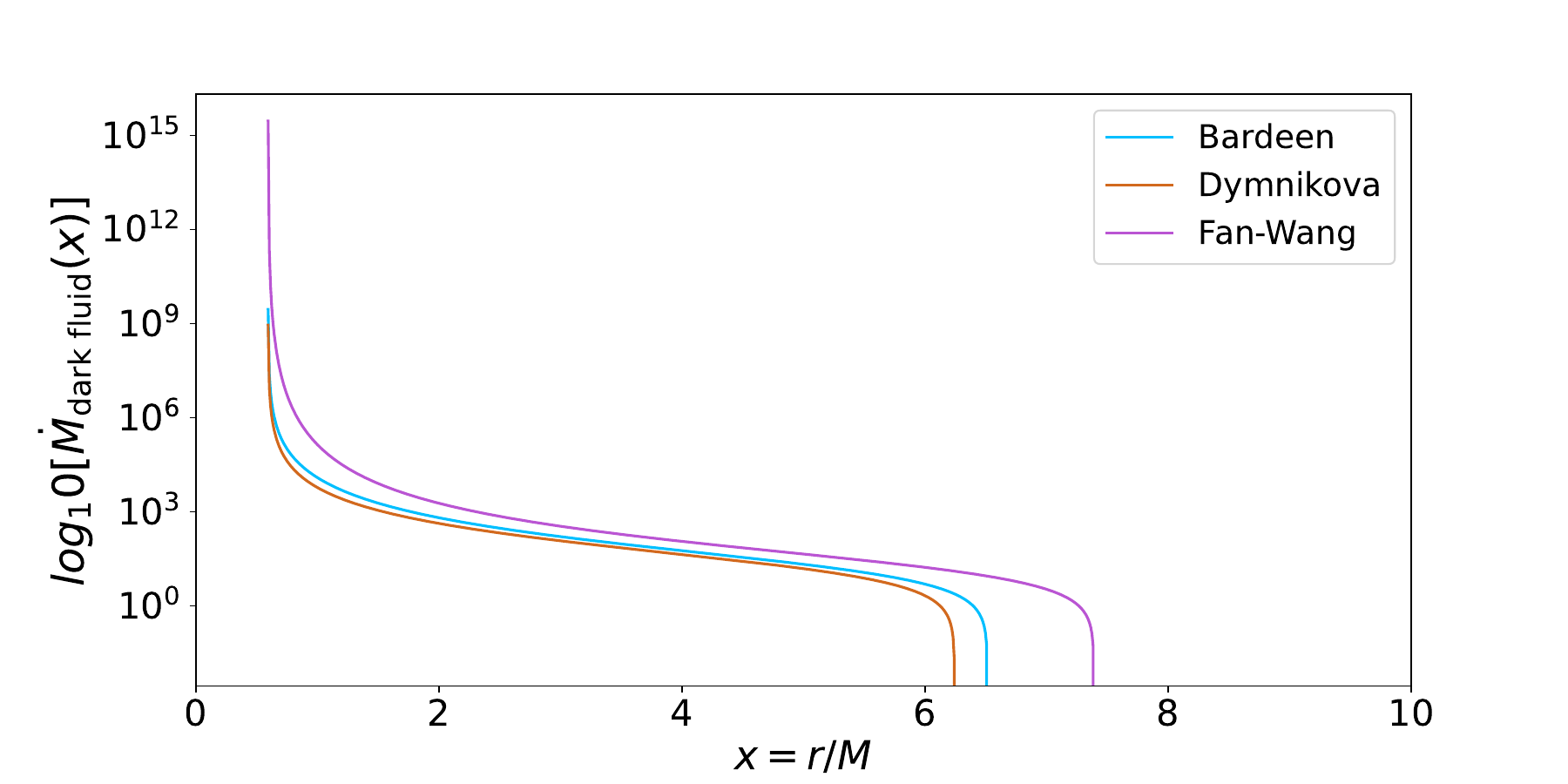}
    \end{subfigure}
      \caption{\justifying Mass accretion rate as a function of $ x = r/M $ on a logarithmic scale. \textbf{Top:} Behavior of vacuum solution. \textbf{Bottom:} Behavior of topologically charged and Fan-Wang solutions.}
    \label{fig:Mdot}
\end{figure}

In the lower panel of Fig.~\ref{fig:rho}, all three topologically charged solutions show behavior similar to that of the vacuum solutions. The energy density decreases smoothly at larger $x$, while, in the inner region of the accretion disk, a growth is observed starting at different values of $x$ and with different slopes, depending on the specific solution. The behavior of the three solutions almost overlaps, with the differences becoming more pronounced at small $x$, and the Fan-Wang solution reaching the highest values for the density in this region.
%
%
The mass accretion rate, $ \dot{M} $, is plotted in Fig.~\ref{fig:Mdot}.
The accretion rate is plotted on a logarithmic scale to capture its wide range of variation. The plot reveals a "collapse," characterized by an almost vertical drop in the accreted mass at large $x$, which is an artifact of the logarithmic scale. In general, the curves smoothly approach zero as expected for the standard trend.

All RBH solutions exhibit an increase in the accretion rate as $x$ decreases, i.e., as we approach the BH, with varying slopes for different solutions. Regardless of the solution, $\dot{M}$ asymptotically approaches zero as we move further away from the BH.

In the top panel, the vacuum energy solution shows a faster growth, approaching an almost vertical rise. This rapid growth is the reason for using a logarithmic scale to plot $\dot{M}$.

In the bottom panel of Fig.~\ref{fig:Mdot}, the topologically charged solutions exhibit similar trends. The Bardeen and Dymnikova solutions almost overlap at small $x$, as observed for the density, but diverge at larger $x$, decreasing at different rates. Conversely, the Fan-Wang solution displays higher values of the accretion rate throughout the entire range considered, reaching zero later than the other two solutions.

Both the mass accretion rate and the luminosity asymptotically approach zero as we move away from the BH, exhibiting similar behavior to the vacuum solutions.

\begin{figure}[!h]
    \begin{subfigure}{0.5\textwidth}
        \includegraphics[width=\linewidth]{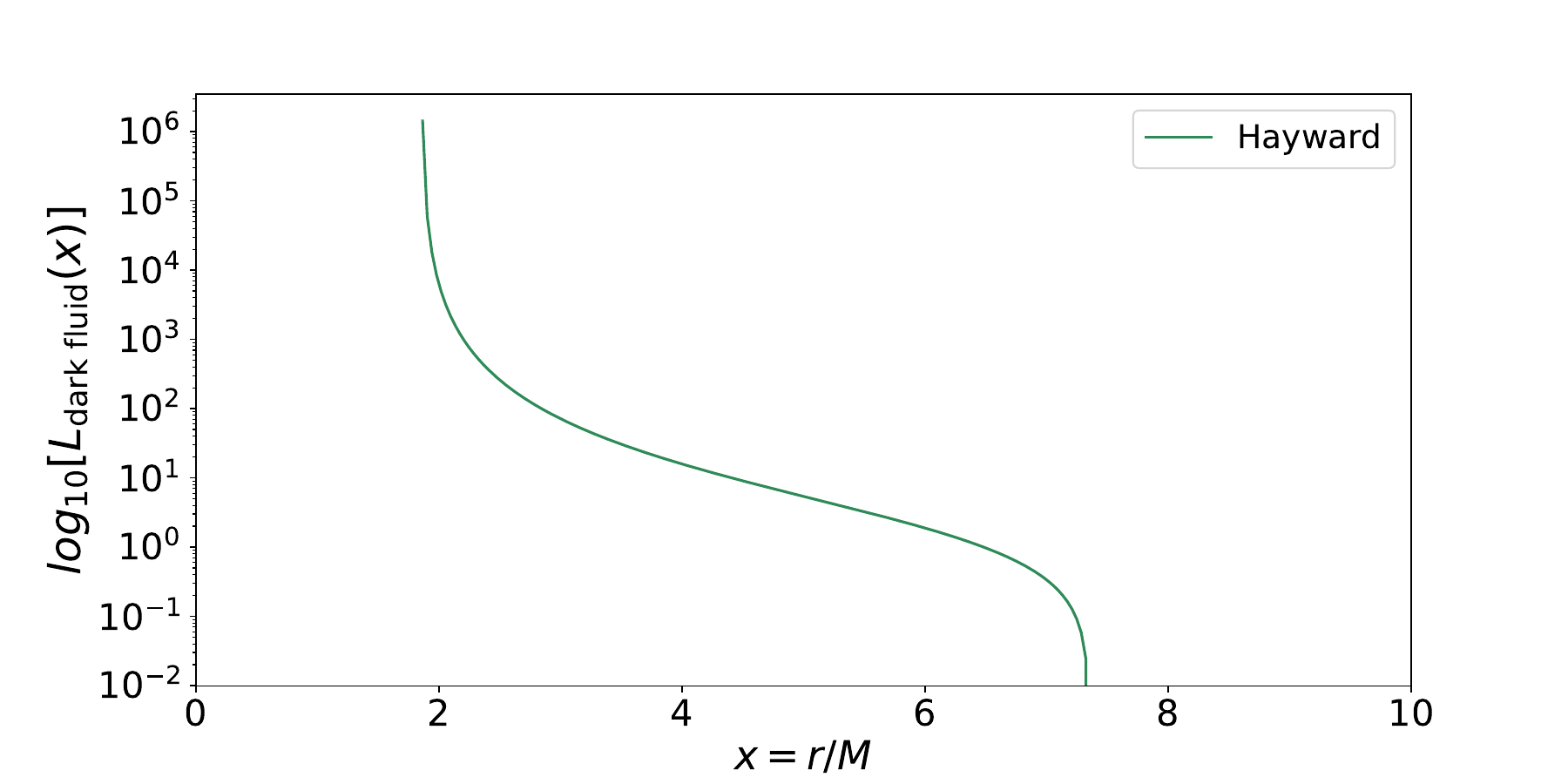}
    \end{subfigure}
    \hfill
    \begin{subfigure}{0.5\textwidth}
        \includegraphics[width=\linewidth]{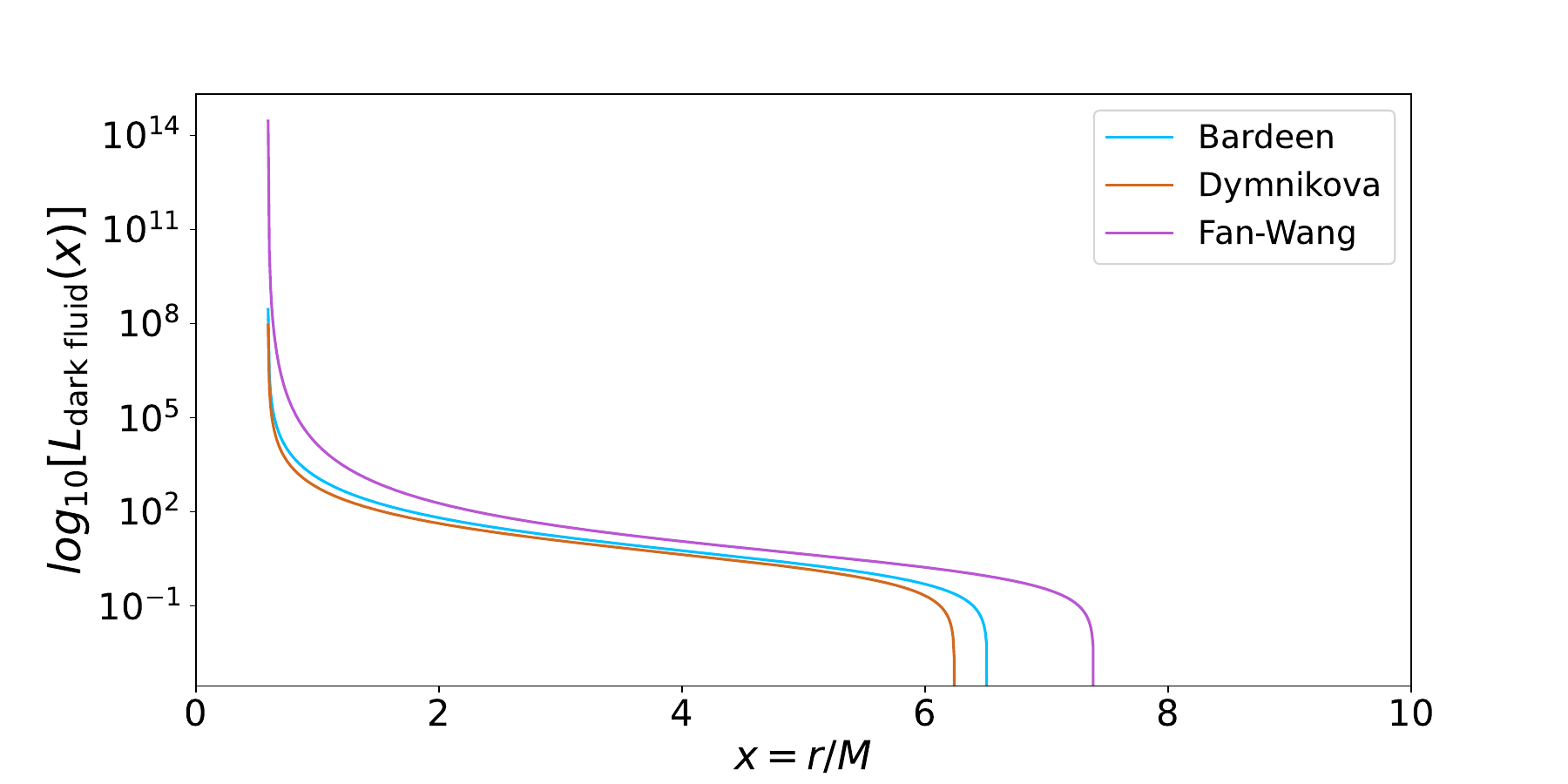}
    \end{subfigure}
      \caption{\justifying Logarithm of Bondi luminosity as a function of $x = r/M $. {\bf Top:} Behavior of vacuum solution. {\bf Bottom:} Behavior of topologically charged and Fan-Wang solutions.}
    \label{fig:L}
\end{figure}

%
%
The behavior observed for the accretion rate, modeled using a conversion efficiency coefficient $\eta_{\text{eff}} \sim 0.1 $, is also reflected in the Bondi luminosity profiles shown in Fig.~\ref{fig:L}.
It is important to note that the main difference is that the luminosity $L$ is scaled by an order of magnitude relative to the mass accretion rate $\dot{M}$, due to the effect of the efficiency parameter $\eta_{\text{eff}}$, which explains why $L$ has smaller values but the same trend of $\dot{M}$.
%
%
\subsubsection{The role of variable $V^2$}
\label{SEC:V2_problem}
Let us now examine the variable $ V^2 $, introduced in Eq. \eqref{new_variable_V}, over several RBH solutions. Although the variable is zero for most of the considered range of $ x $, we identify a region near a "critical point" around which the variable is negative for $ r > r_{\rm crit} $ and takes positive values for $ r < r_{\rm crit} $. Hence, $ V^2 $ should be redefined using the conditions \eqref{new_critical_point1} and \eqref{new_critical_point2}.

To better illustrate the behavior of $V^2 $ in the different solutions, we separate the positive and negative regions of it. This allows a clearer identification of the critical point and highlights how the dark fluid EoS affects $V^2 $, causing it to behave inconsistently with physical expectations. The observed discontinuity in the behavior of $ V^2 $, especially around the critical point, suggests that the chosen EoS is the cause of this irregularity. Specifically, we used a constant pressure, which, although unusual in this context, is further complicated by the fact that the pressure is also negative, as discussed earlier. This unusual choice is likely to contribute to the discontinuous behavior.

\begin{figure}[!h]
    \begin{subfigure}{0.5\textwidth}
        \includegraphics[width=\linewidth]{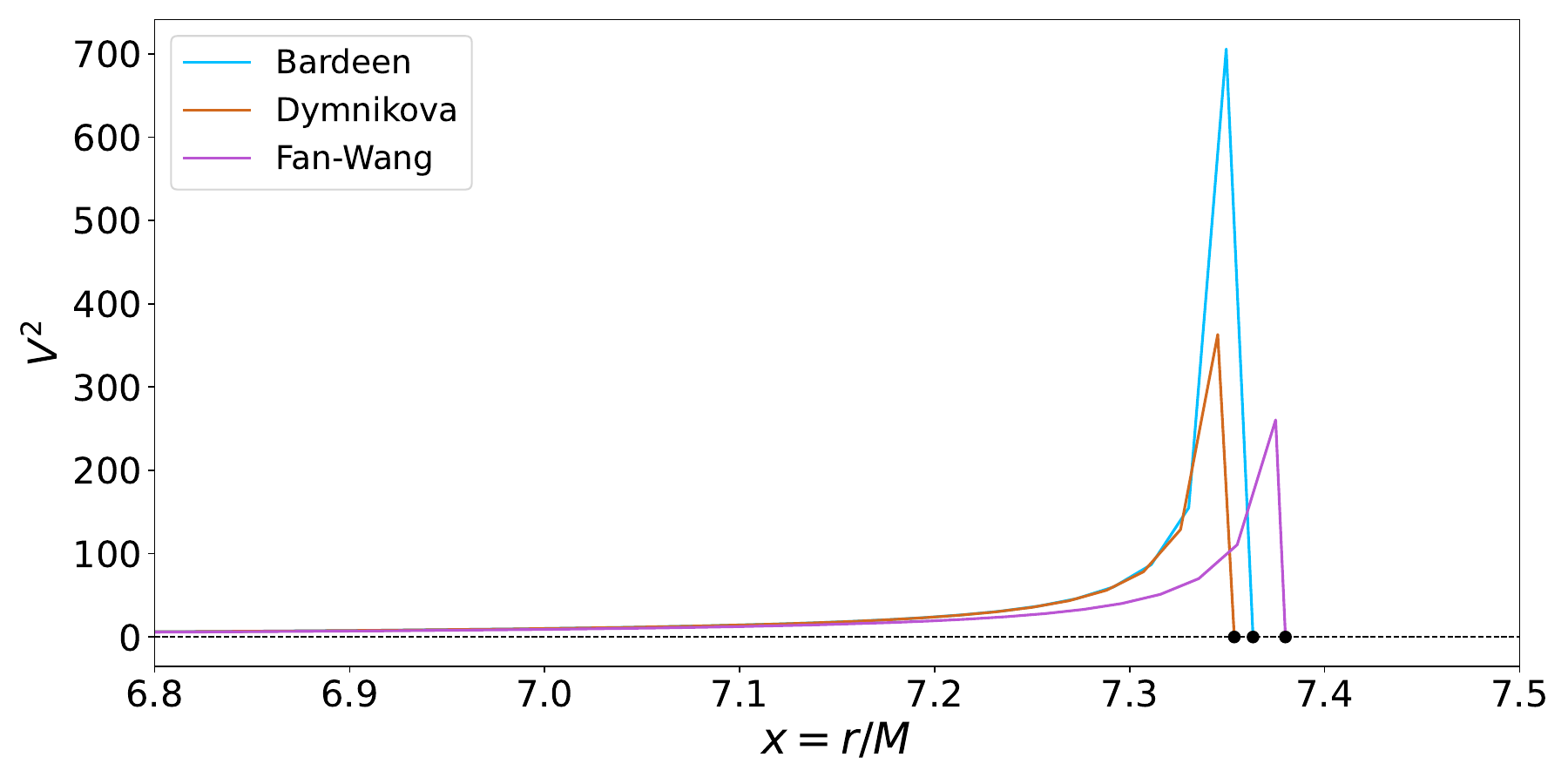}
    \end{subfigure}
    \hfill
    \begin{subfigure}{0.5\textwidth}
        \includegraphics[width=\linewidth]{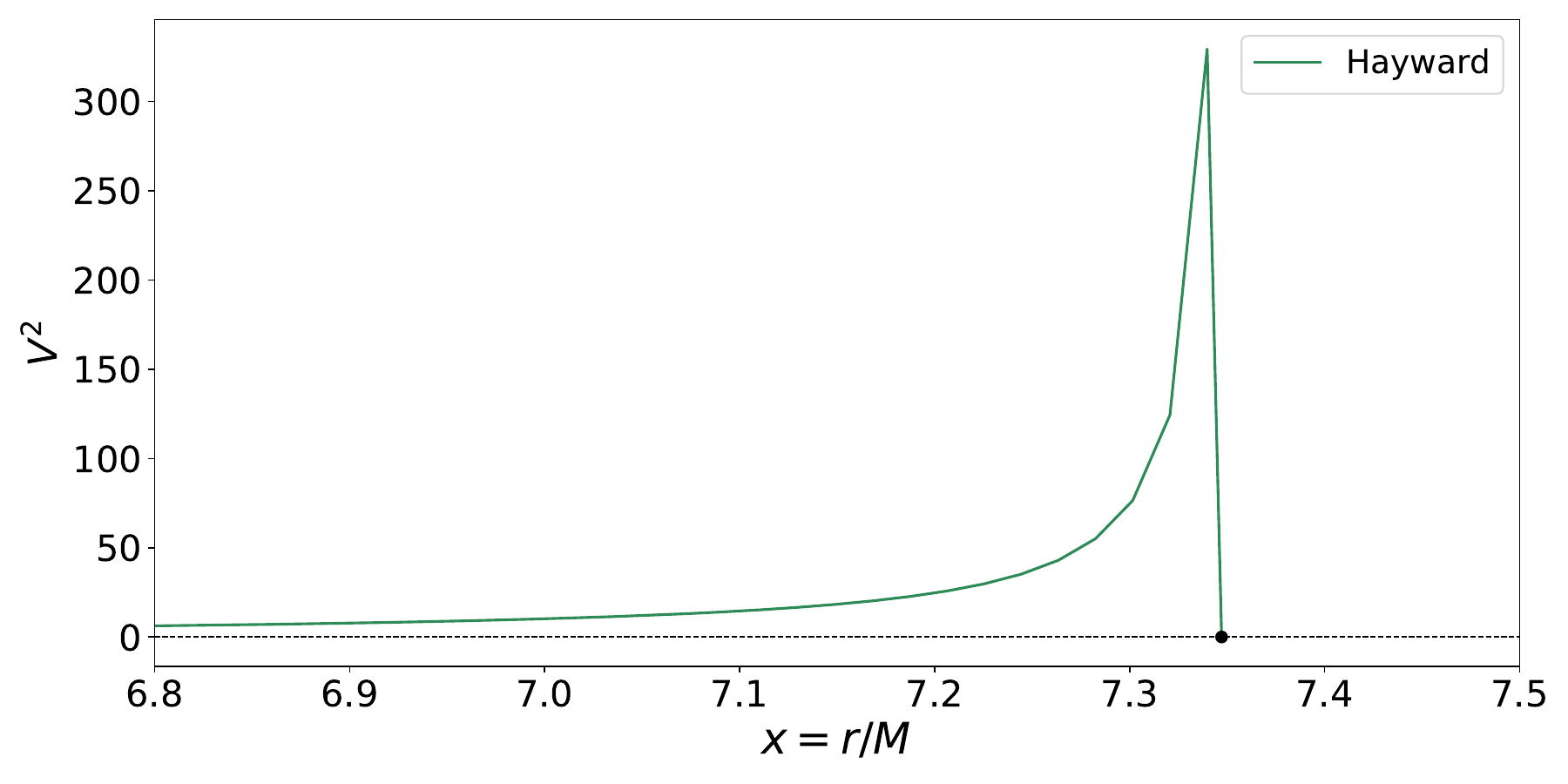}
    \end{subfigure}
      \caption{\justifying Positive range of values of $V^2$ as a function of $ x = r/M $. {\bf Top:} Behavior of vacuum solution. {\bf Bottom:} Behavior of topologically charged and Fan-Wang solutions.}
    \label{fig:V2_positive}
\end{figure}

\begin{figure}[!h]
    \begin{subfigure}{0.5\textwidth}
        \includegraphics[width=\linewidth]{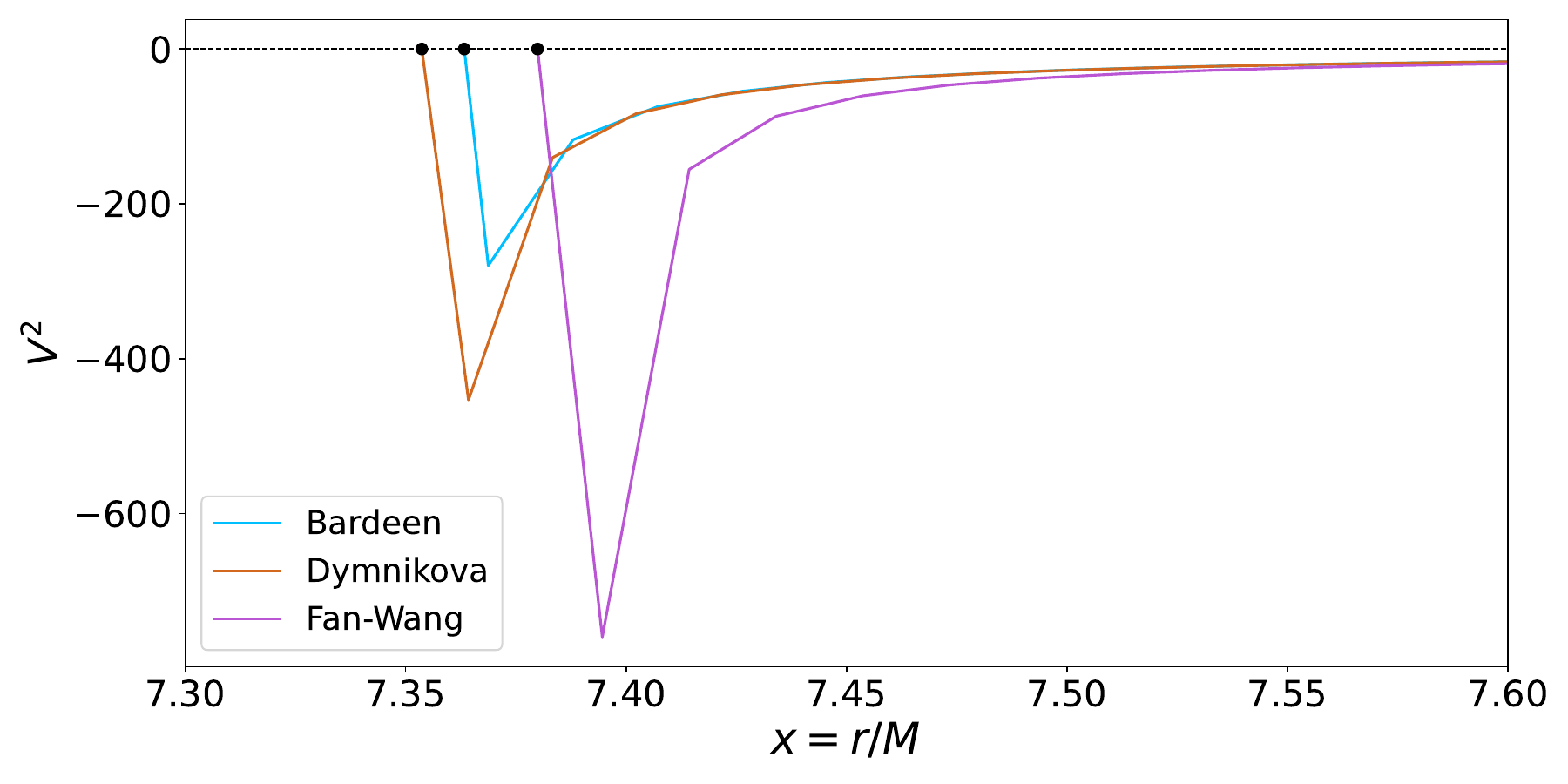}
    \end{subfigure}
    \hfill
    \begin{subfigure}{0.5\textwidth}
        \includegraphics[width=\linewidth]{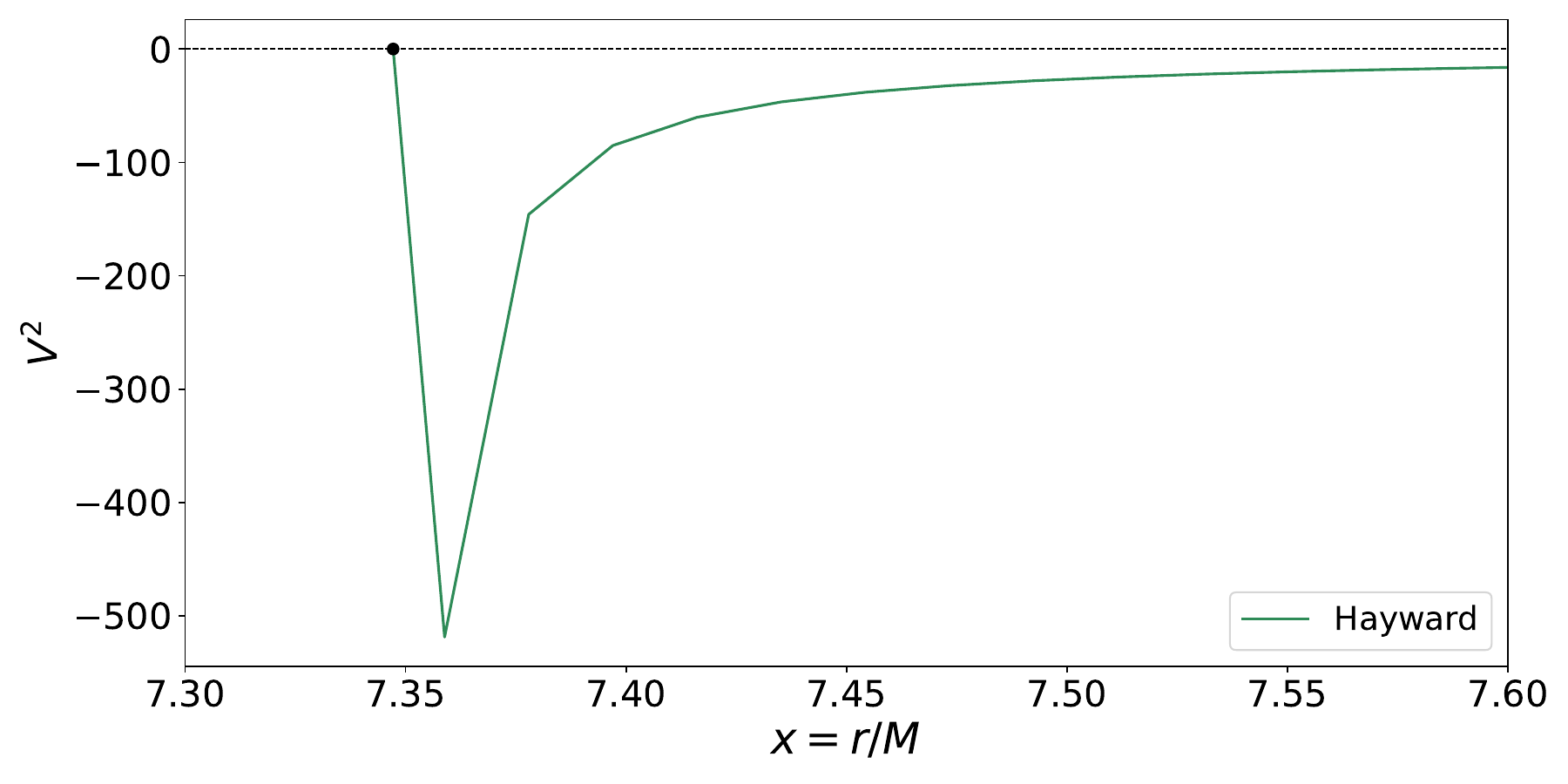}
    \end{subfigure}
      \caption{\justifying Negative range of values of $V^2$ as a function of $ x = r/M $. {\bf Top:} Behavior of vacuum solution. {\bf Bottom:} Behavior of topologically charged and Fan-Wang solutions.}
    \label{fig:V2_negative}
\end{figure}

To prompt this, we can notice that in Fig.~\ref{fig:V2_positive}, we display the variation of the positive values of $V^2$ as a function of $x=r/M$, while Fig.~\ref{fig:V2_negative} shows the corresponding negative values. It is important to note that the different solutions exhibit distinct maximum values of $V^2$, which is caused by the specific forms and parameters of the RBH solutions under exams.
We begin with Fig.~\ref{fig:V2_positive}, where the top panel illustrates the Hayward solution, which displays a critical point at a similar $x$ value to those found in the charged solutions. For the charged and Fan-Wang solutions, we observe that the critical points are located at nearly the same $x$, with the critical point of the Fan-Wang solution being only slightly larger than the others.

In Fig.~\ref{fig:V2_negative}, we present the behavior of the negative values of $V^2$, which complements the previous plot, though with differences in the maximum values of $V^2$ reached by the different solutions. In general, the solutions pass from a region where $V^2 < 0$ to large $x$, so we need to rename the variable for this region $-V^2$ to a region smaller than $x$ for $r< r_{\text crit}$ where $V^2 > 0$.
\begin{table}[!h]
    \centering
    \renewcommand{\arraystretch}{1.3} 
    \begin{tabular}{|l|c|c|c|c|}
     \hline
        \hline
    \multicolumn{5}{|c|}{\textbf{1\textsuperscript{st} case: dark fluid EoS}} \\ \hline\hline
        \textbf{Solution} & $ \mathcal{C} $  & $ \mathcal{C}_3 $ & $ r_{\text{crit}} $ & $ r_{\text{EH}} $  \\
        \hline\hline
        Bardeen                  &  1.5 &  50  & 7.36          &  1.58       \\
        \hline
        Hayward                  &  1.5 &  50  & 7.36          &  1.85       \\
        \hline
        Fan-Wang                 &  1.5 &  50  & 7.39          &  0.59       \\
        \hline
        Dymnikova                &  1.5 &  50  & 7.36          &  1.89       \\
        \hline
        Schwarzschild            &  1.5 &  50  & 7.36          &  2          \\
        \hline
        Schwarzschild-de Sitter  &  1.5 &  50  & 7.34          &   2 - 53.74 \\
        \hline
        \hline
    \end{tabular}

    \caption{\justifying In the table, we show the values of the integration constants and the numerical results for the critical radius $r_{\text{crit}}$ and the event horizon $r_{\text{EH}}$ in the case of dark fluid. Note that the Schwarzschild-de Sitter solution has two event horizons: a physical one and a cosmological one. Constants used: $ M=1 $, $ a=0.5 $, $ q_B=0.65 $, $ q_D=0.452 $, $ \tilde{\Lambda}=0.5 $, $ \Lambda=0.001 $, $ l_{FW}=8/27 $, $ P=-0.75 $.}
    \label{TAB:const_val_1st_method}
\end{table}
%
%
\subsubsection{Comparing Schwarzschild and Schwarzschild-de Sitter solutions}
\label{SEC:BONDI_df_comparison}
In this section, we compare the trends of the RBH solutions with two limiting cases: the Schwarzschild and Schwarzschild-de Sitter BHs. This comparison aims to show how the RBH solutions differ from these conventional models in the context of spherically symmetric accretion. All relevant constants are listed in Tab. \ref{TAB:const_val_1st_method}.

\begin{figure}[h!]
    \begin{subfigure}{0.5\textwidth}
    \includegraphics[width=\linewidth]{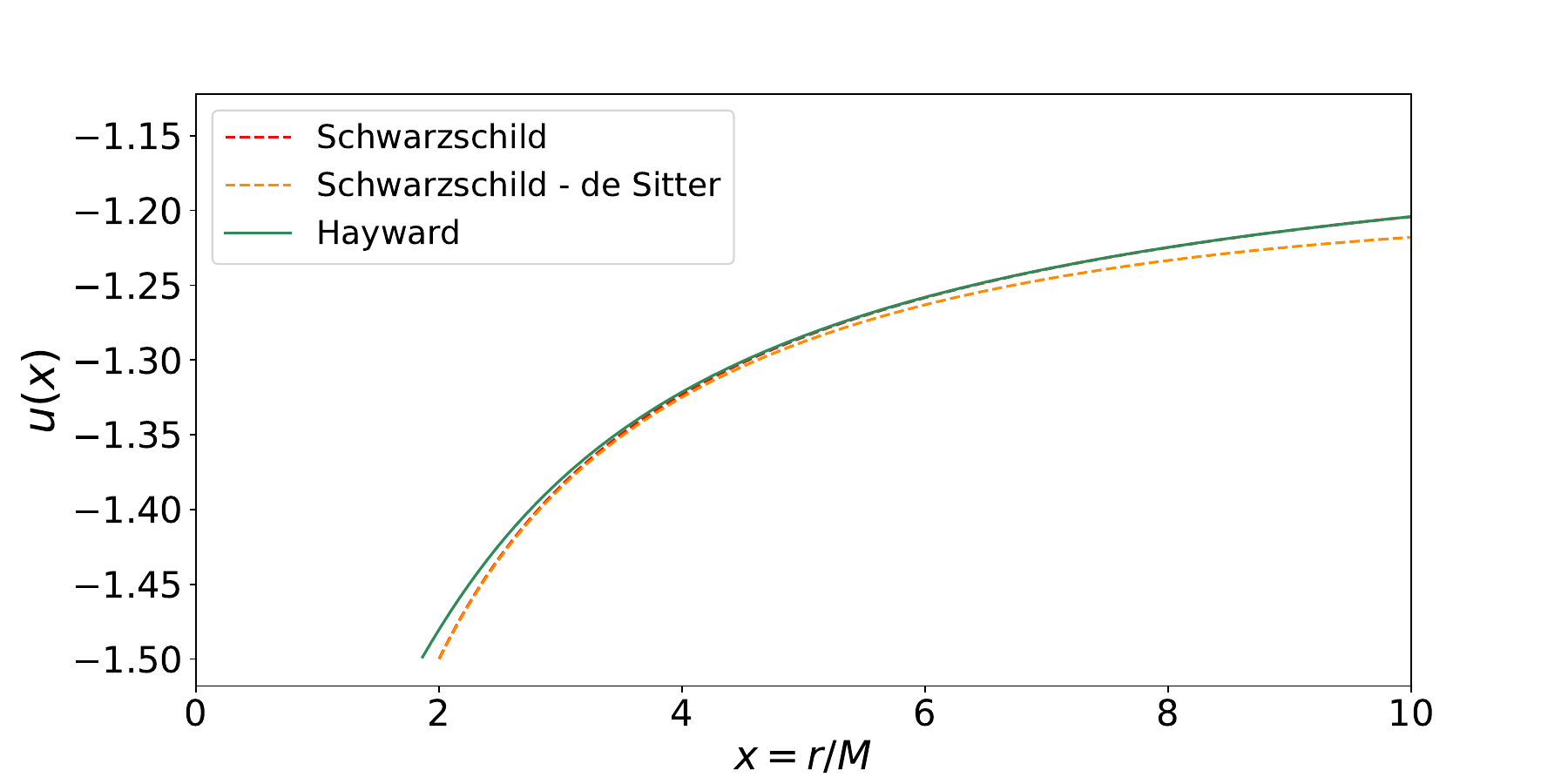}
    \end{subfigure}
\hfill
    \begin{subfigure}{0.5\textwidth}

    \includegraphics[width=\linewidth]{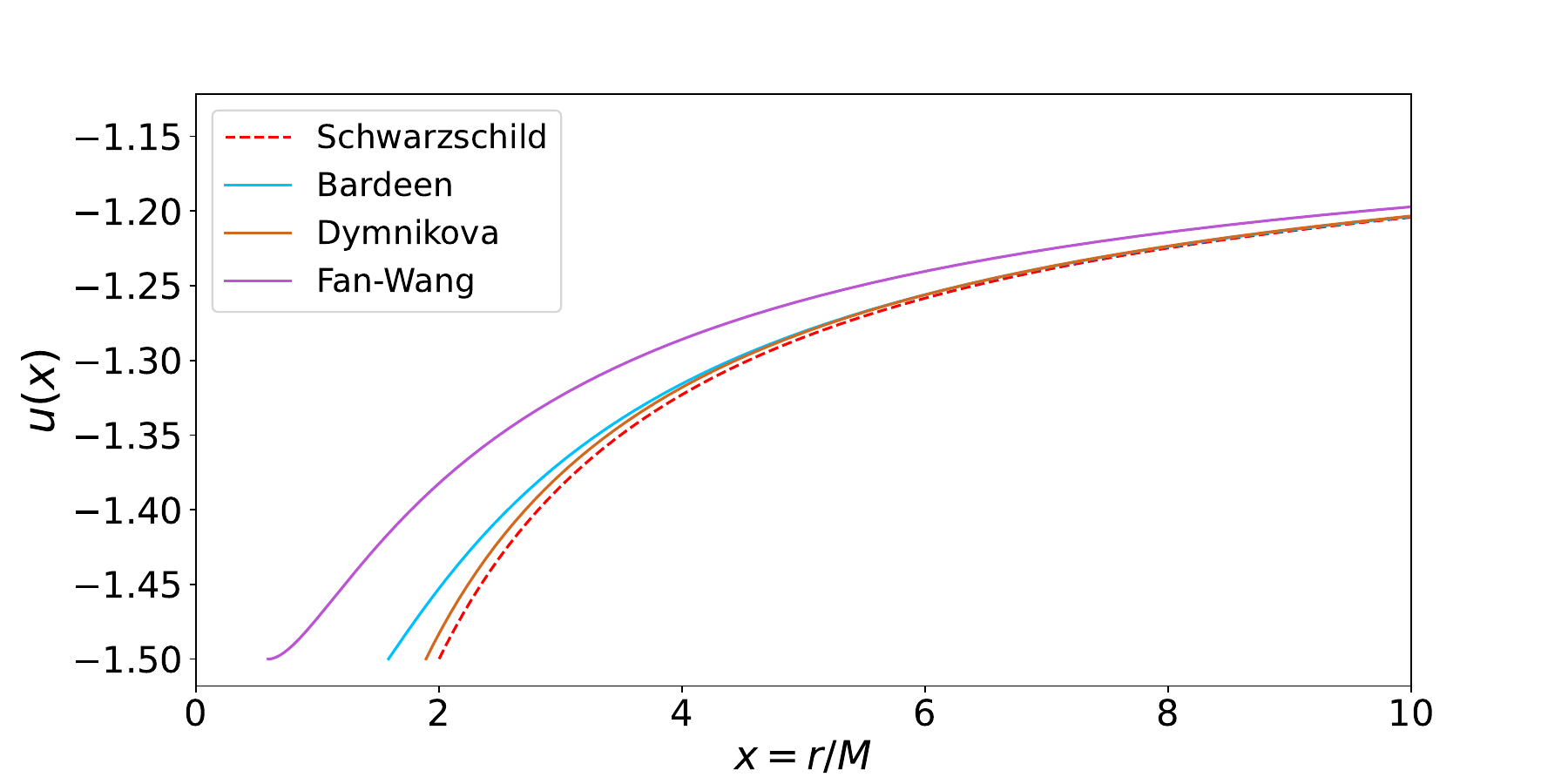}
    \end{subfigure}
  \caption{\justifying Radial velocity profiles $u$ as a function of $x = r/M$ for different RBH solutions. {\bf Top:} Behavior of the vacuum solution compared with Schwarzschild and Schwarzschild-de Sitter solutions. {\bf Bottom:} behavior of the topologically charged and Fan-Wang solutions compared with the Schwarzschild solution.}
\label{fig:u_confront}
\end{figure}

We start with the radial velocity profiles shown in Fig.~\ref{fig:u_confront}.
In the top panel, the behavior of the Hayward BH solution closely resembles that of the Schwarzschild solution, with both of these solutions nearly overlapping in the middle region of $x$. However, the Hayward solution begins to deviate slightly from both the Schwarzschild and Schwarzschild-de Sitter solutions in the inner region of the accretion disk. As we move outward, the Hayward solution approaches zero at a slightly faster rate compared to the other solutions.

The lower panel, which refers to the topologically charged solutions, shows how these solutions approach the Schwarzschild one at large $ x $. However, in the inner region they deviate slightly, though not significantly, except for the Dymnkova solution, which almost overlaps even at small $x$.

\begin{figure}[h!]
\begin{subfigure}{0.5\textwidth}
\includegraphics[width=\linewidth]{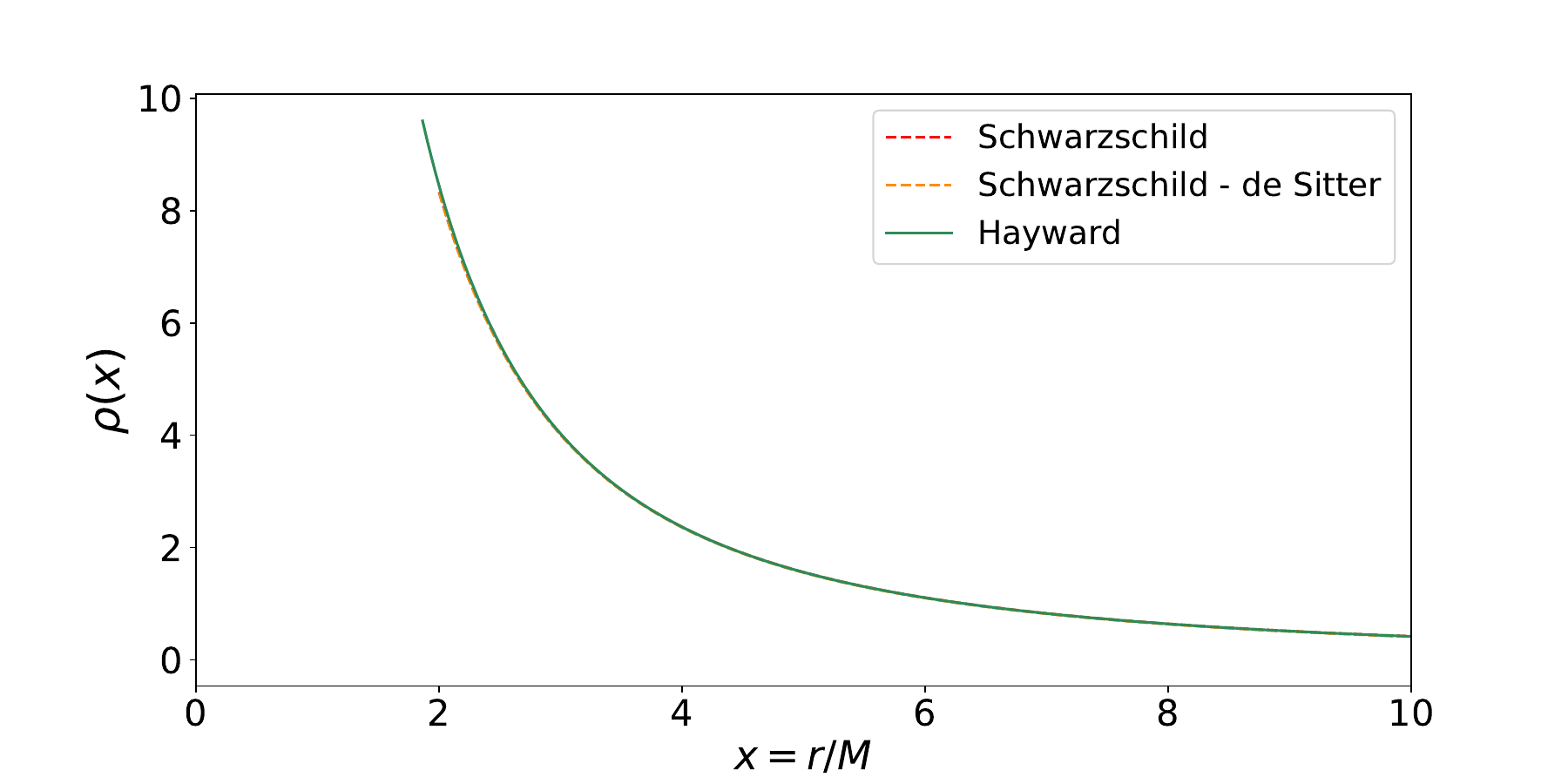}
\end{subfigure}
\hfill
\begin{subfigure}{0.5\textwidth}
\includegraphics[width=\linewidth]{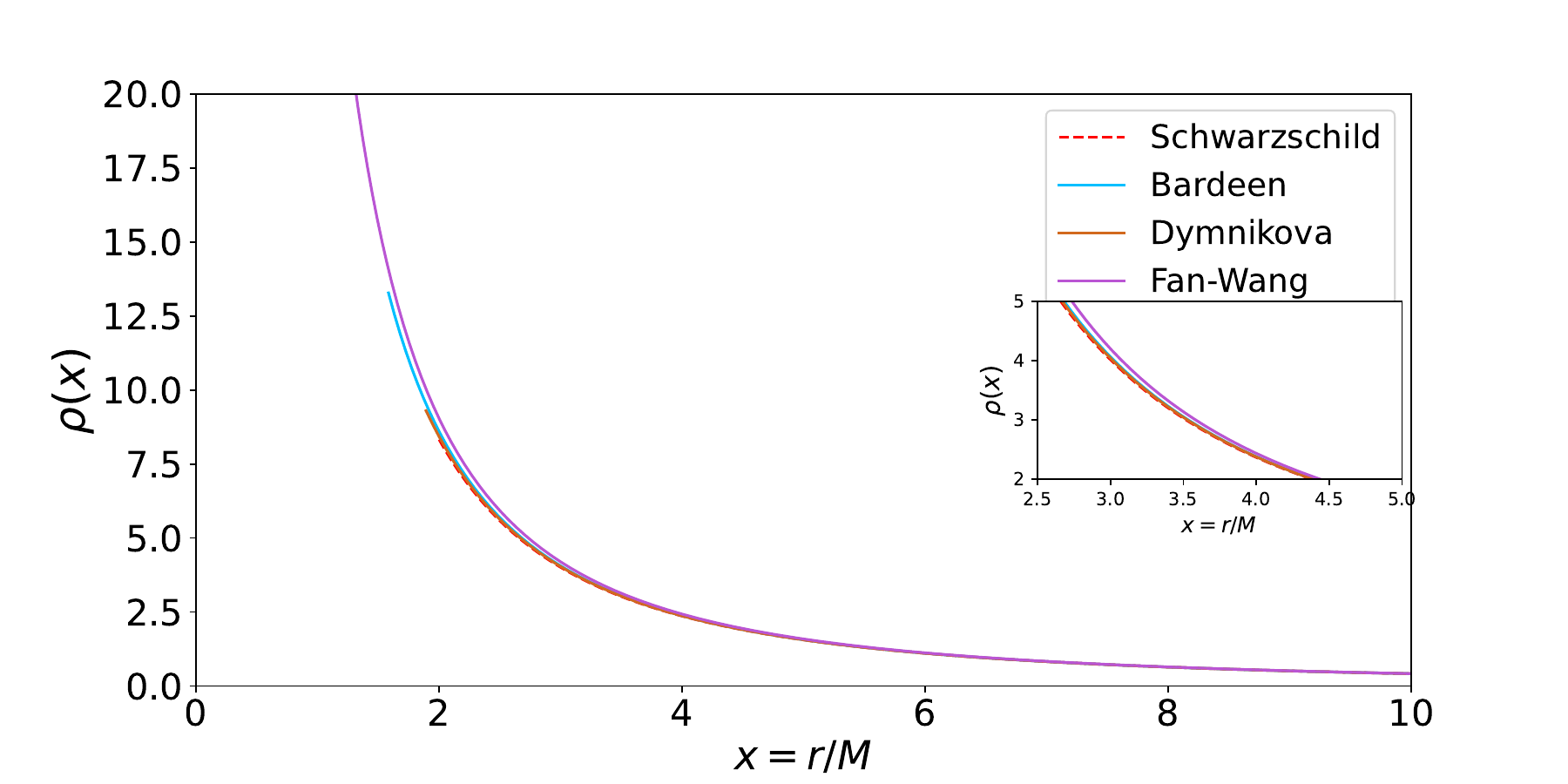}
\end{subfigure}
  \caption{\justifying Energy density profiles $\rho$ as a function of $x = r/M$ for different RBH solutions. {\bf Top:} Behavior of the vacuum solution compared with the Schwarzschild and Schwarzschild-de Sitter solutions. {\bf Bottom:} behavior of the topologically charged and Fan-Wang solutions compared with the Schwarzschild solution.}
\label{fig:rho_confront}
\end{figure}

A similar analysis applies to the energy density, as shown in Fig.~\ref{fig:rho_confront}.
In the top panel, the Hayward, Schwarzschild-de Sitter and Schwarzschild solutions show a complete overlap in the range considered, with their behavior being almost indistinguishable. In contrast, the bottom panel shows that the Bardeen, Dymnikova, Fan-Wang and Schwarzschild solutions overlap almost completely over the whole range of $x$.

All solutions eventually approach zero as we move away from the inner region of the accretion disk.

\begin{figure}[h!]
\begin{subfigure}{0.5\textwidth}
\includegraphics[width=\linewidth]{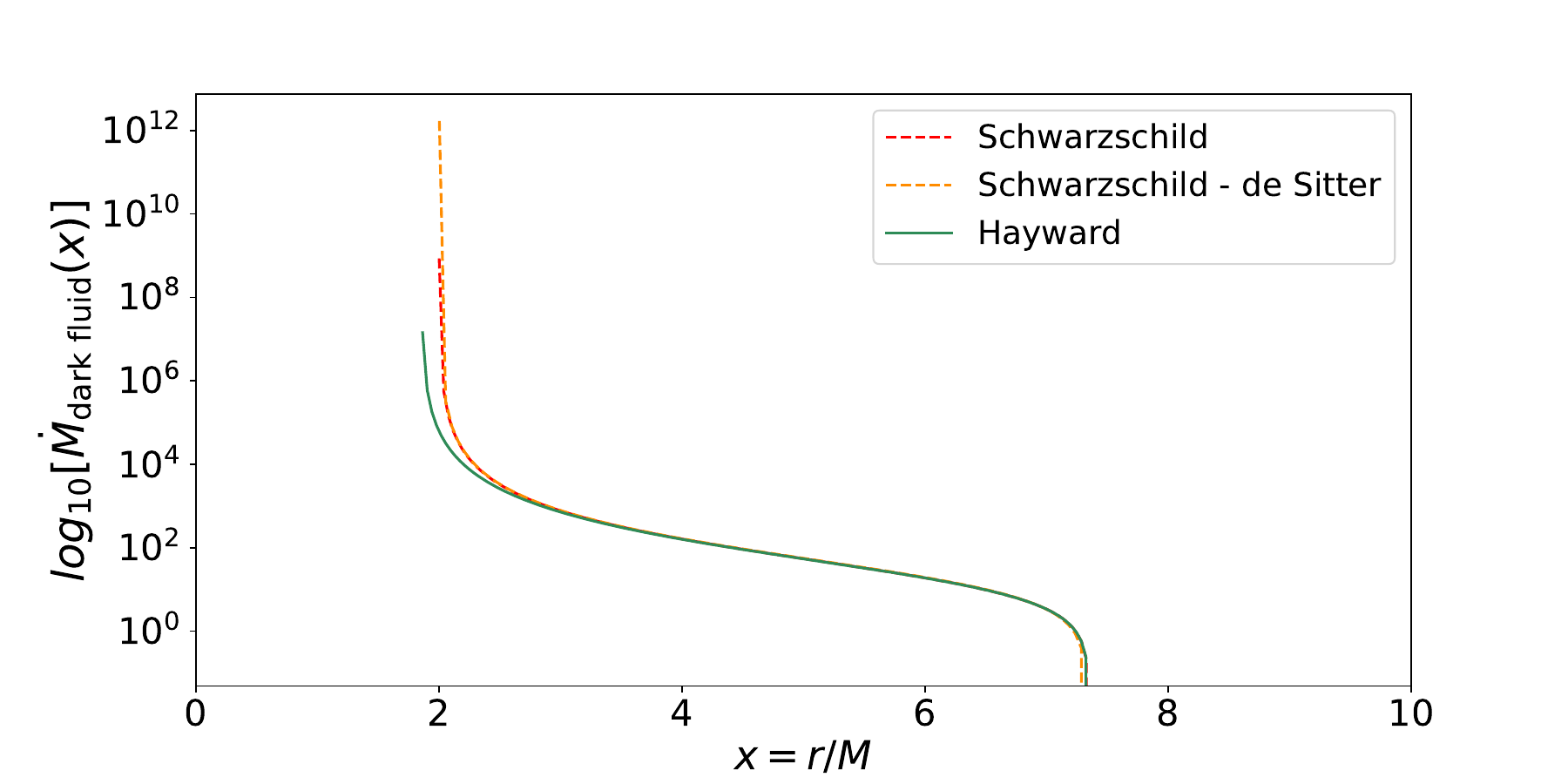}
\end{subfigure}
\hfill
\begin{subfigure}{0.5\textwidth}
\includegraphics[width=\linewidth]{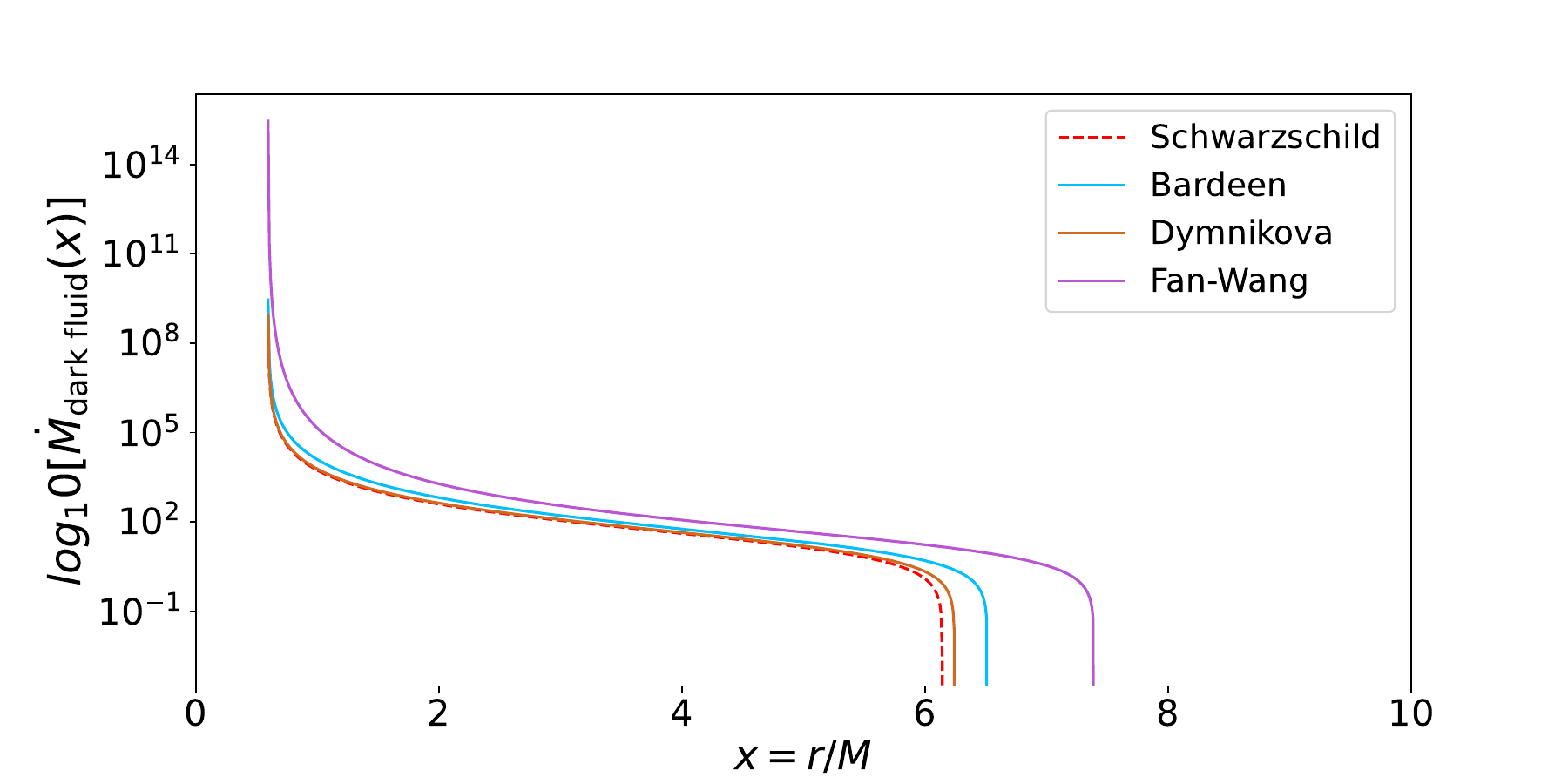}
\end{subfigure}
  \caption{\justifying Logarithm of the Mass accretion rate as a function of $x = r/M$ for different RBH solutions. {\bf Top:} Behavior of the vacuum solution compared with the Schwarzschild and Schwarzschild-de Sitter solutions. {\bf Bottom:} behavior of the topologically charged and Fan-Wang solutions compared with the Schwarzschild solution.}
\label{fig:Mdot_confront}
\end{figure}

\begin{figure}[h!]
\begin{subfigure}{0.5\textwidth}
\includegraphics[width=\linewidth]{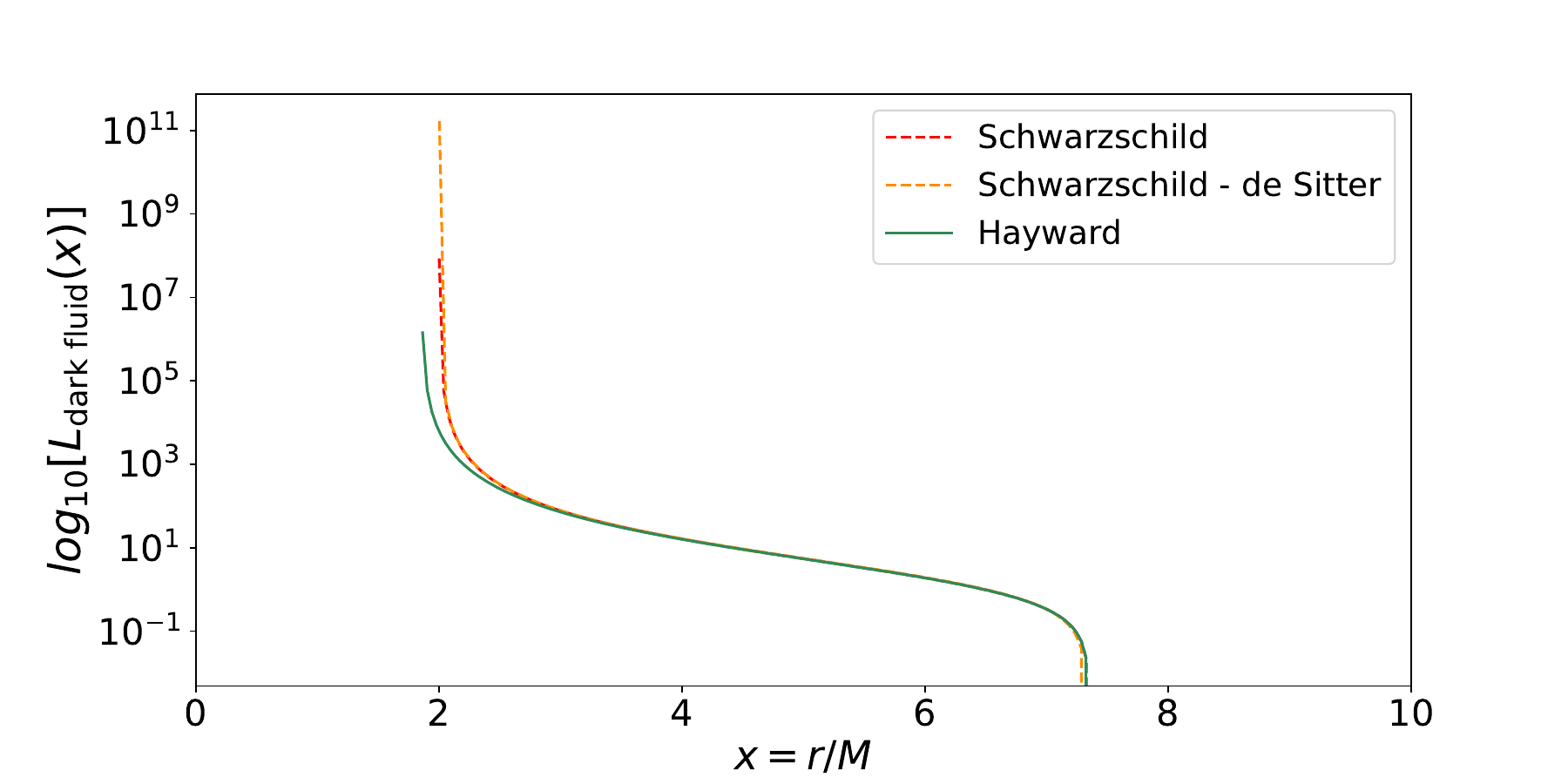}
\end{subfigure}
\hfill
\begin{subfigure}{0.5\textwidth}
\includegraphics[width=\linewidth]{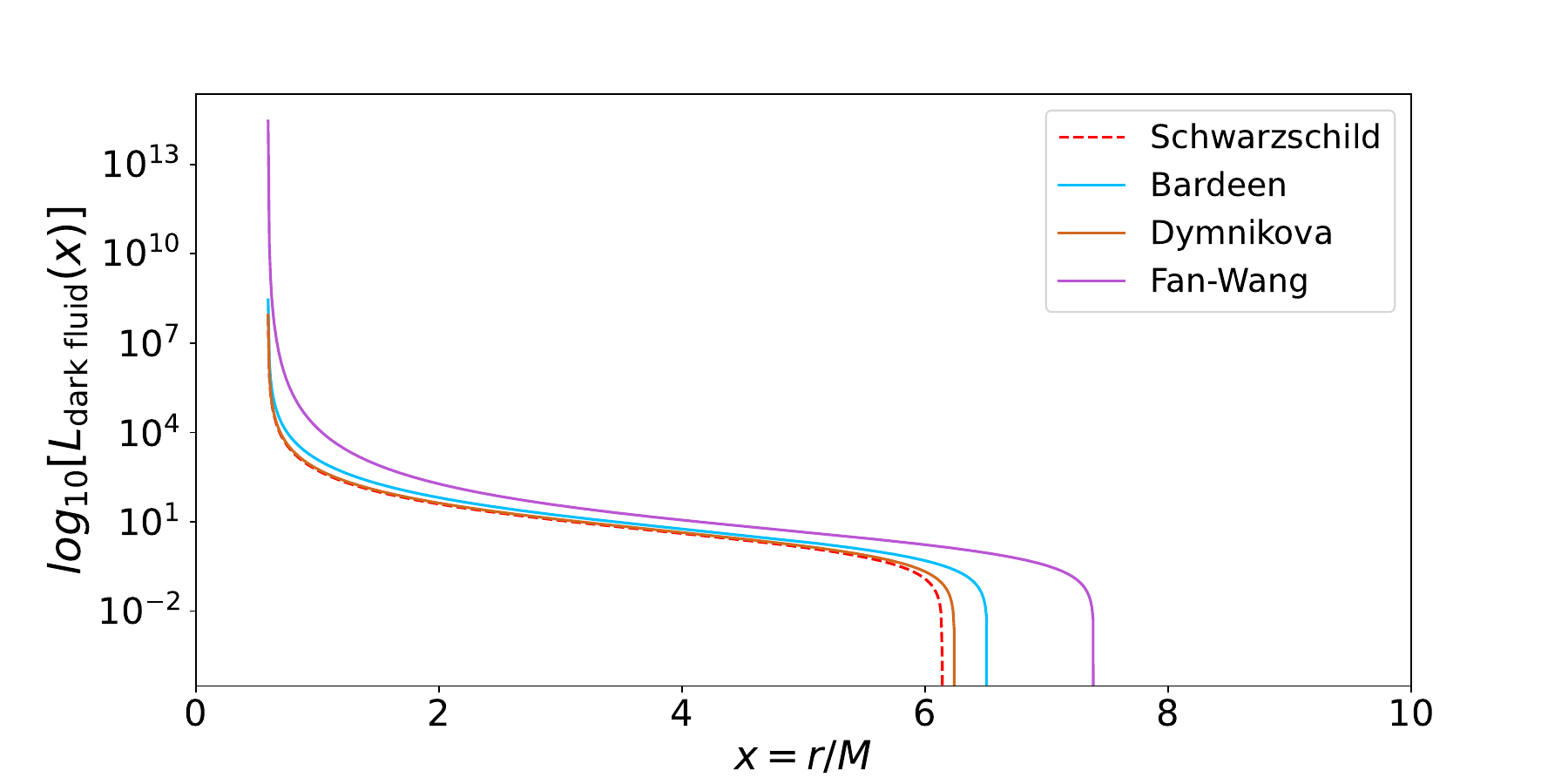}
\end{subfigure}
  \caption{\justifying Logarithm of the Bondi luminosity as a function of $x = r/M$ for different RBH solutions. {\bf Top:} Behavior of the vacuum solutions compared with Schwarzschild and Schwarzschild-de Sitter solutions. {\bf Bottom:} behavior of the topologically charged and Fan-Wang solutions compared with the Schwarzschild solution.}
\label{fig:L_confront}
\end{figure}

Next, we analyze the mass accretion rate in Fig.~\ref{fig:Mdot_confront}
and the luminosity, illustrated in Fig.~\ref{fig:L_confront}.
Let us start with the mass accretion rate. In the top panel, the Schwarzschild and Schwarzschild-de Sitter solutions show an overlapping trend and, like the Hayward solution, an increase in the inner region of the accretion disk. In particular, the Hayward BH has lower maximum values of the mass accretion rate and hence of the luminosity, but shows this growth in a more inner region of the disk, albeit slightly, than the comparison solutions. Both the mass accretion rate $\dot{M}$ and the luminosity $L$ approach zero as one moves away from the accretion disk, and the solutions show a completely superimposed behaviour in this outer region.

For the topologically charged and Fan-Wang solutions, the Dymnikova and Bardeen solutions almost overlap with the Schwarzschild solution at small $x$, while their decay starts at different values of $x$, with the Schwarzschild solution reaching zero first. The Fan-Wang one shows similar trends, but with relatively higher values of $\dot{M}$ and $L$ at large $x$ and therefore never overlap with the other solutions. As in the other cases, we observe the "collapse", again due to the logarithmic scale, that brings both $\dot{M}$ and $L$ to zero as we move away from the accretion disk.

\begin{figure}[h!]
\begin{subfigure}{0.5\textwidth}
\includegraphics[width=\linewidth]{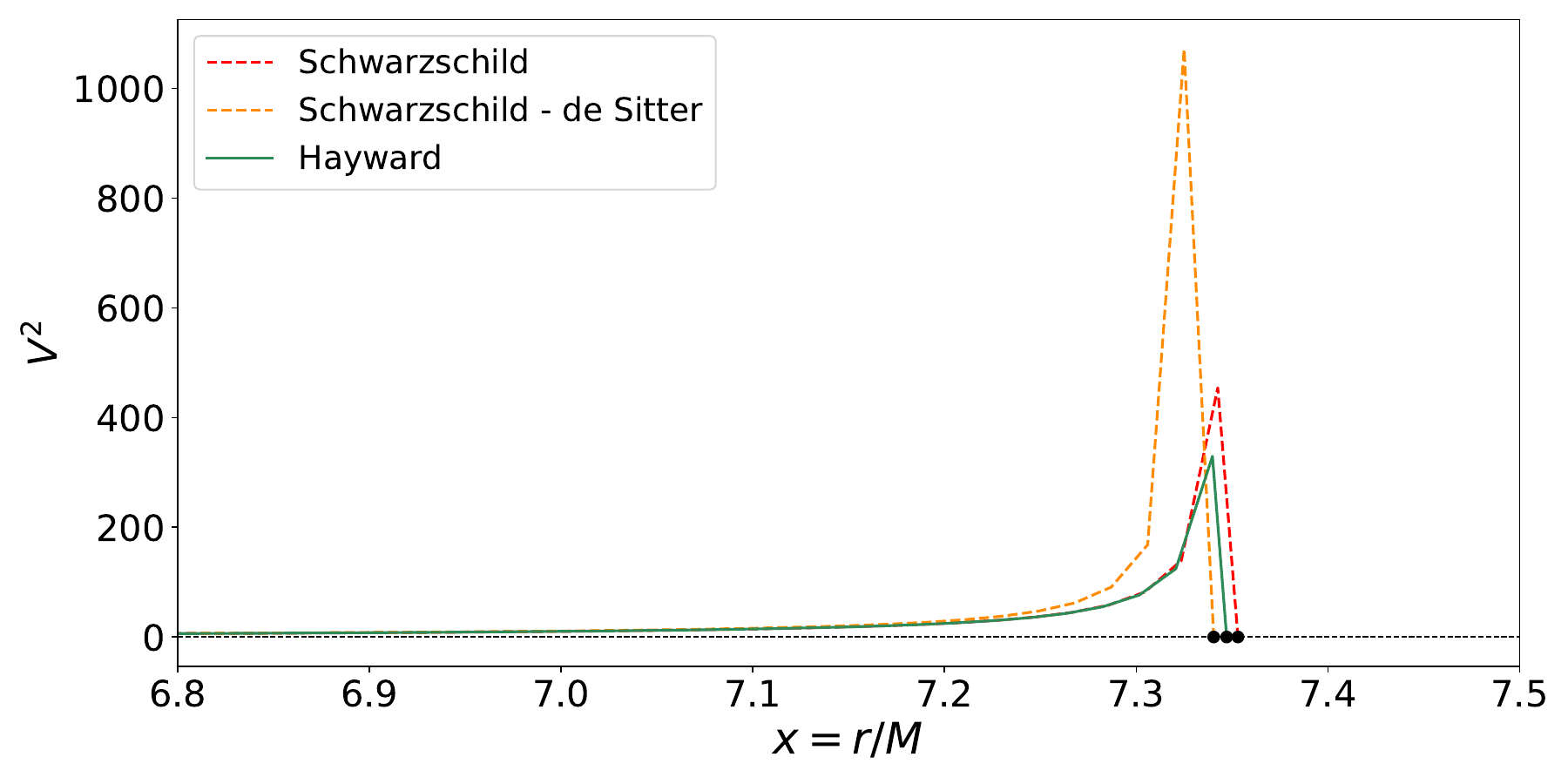}
\end{subfigure}

\hfill \begin{subfigure}{0.5\textwidth}
\includegraphics[width=\linewidth]{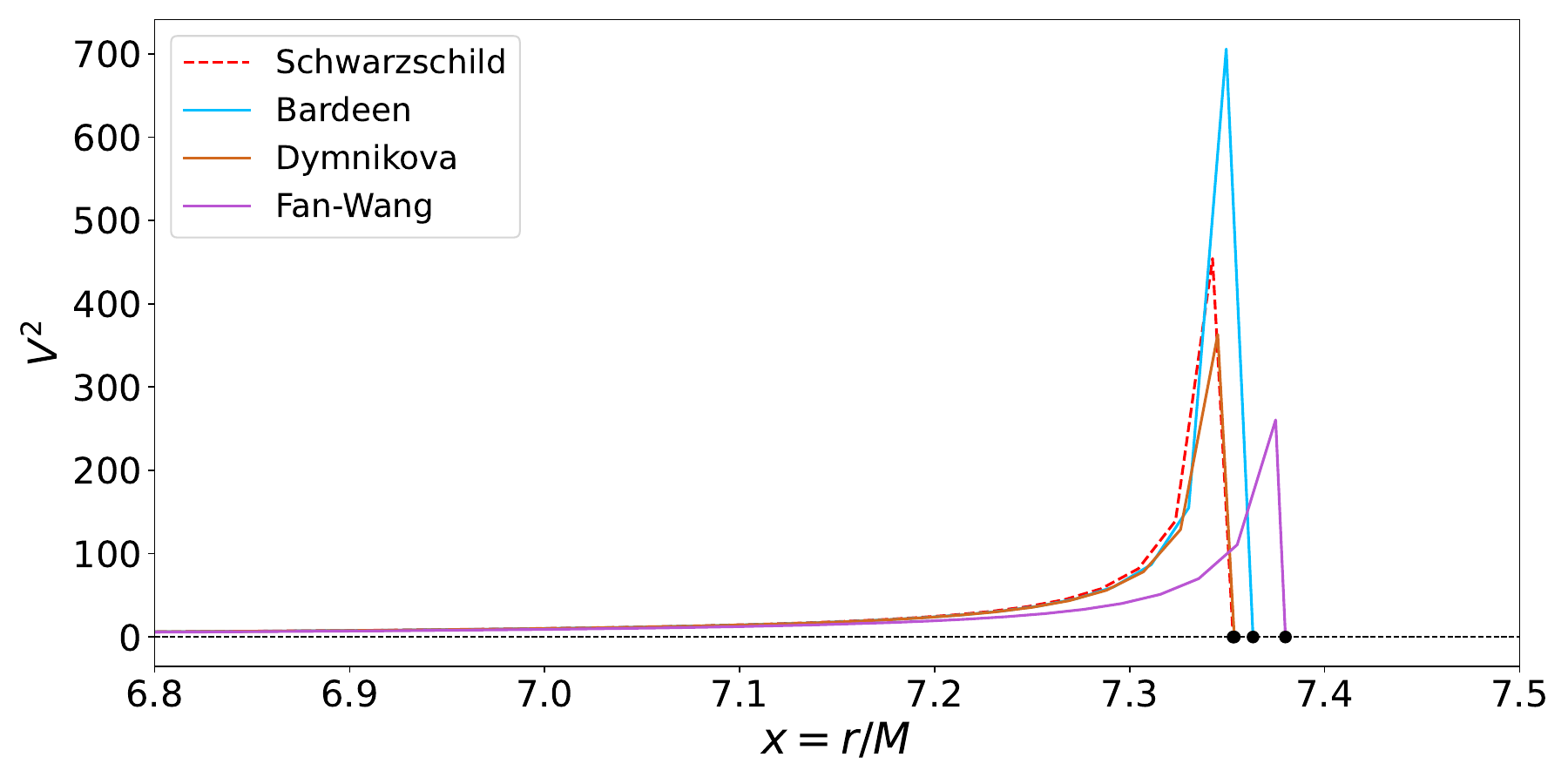}
\end{subfigure}
\caption{\justifying Positive range of values of $V^2$ as a function of $x=r/M$ for different RBH solutions. {\bf Top:} The behavior of the vacuum solution compared with the Schwarzschild and Schwarzschild-de Sitter solutions. {\bf Bottom:} The behavior of the topologically charged and Fan-Wang solutions compared with the Schwarzschild solution.}
\label{fig:V2_confront_positive}
\end{figure}

\begin{figure}[h!]
\begin{subfigure}{0.5\textwidth}
\includegraphics[width=\linewidth]{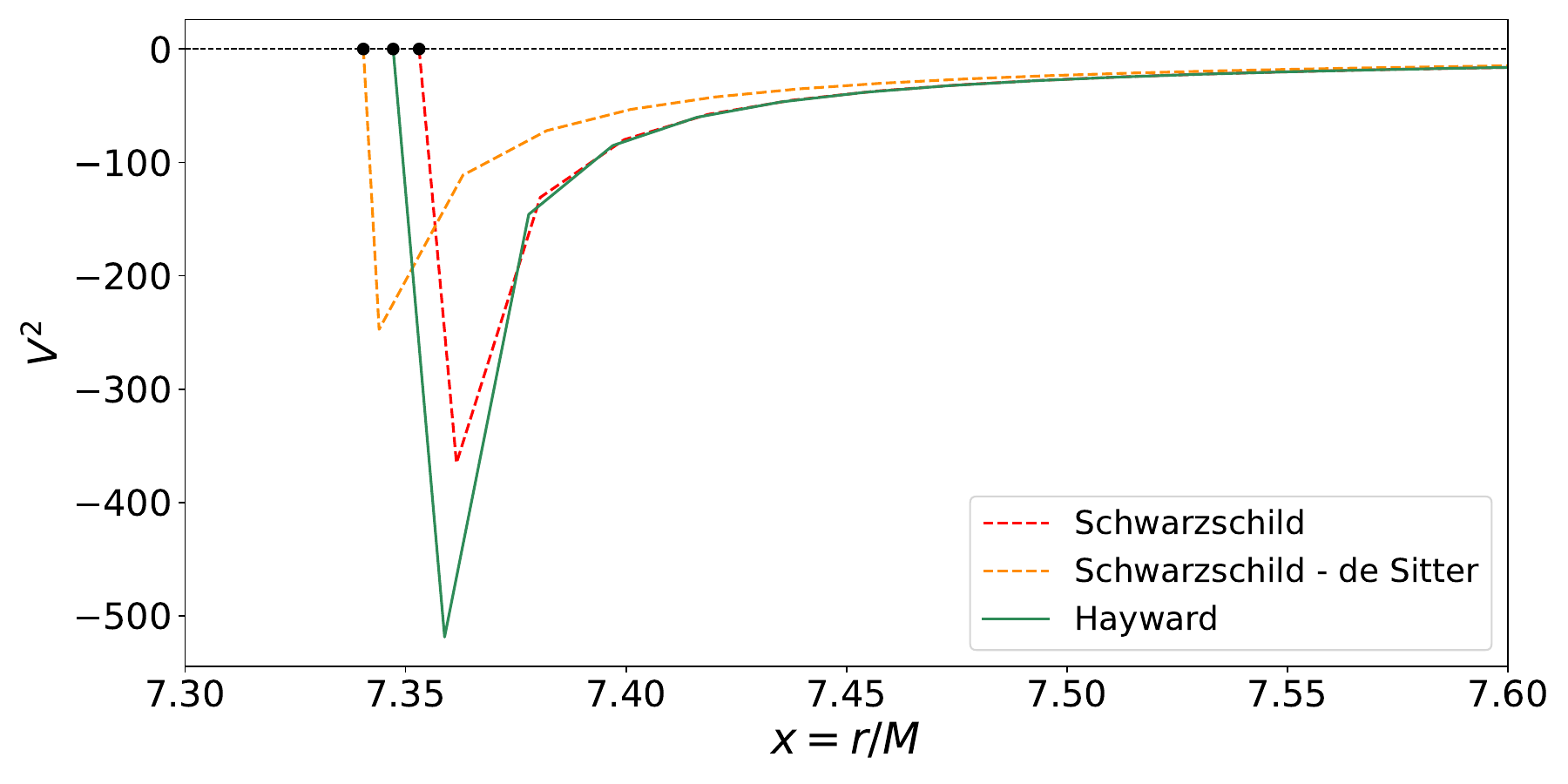}
\end{subfigure}
\hfill
\begin{subfigure}{0.5\textwidth}
\includegraphics[width=\linewidth]{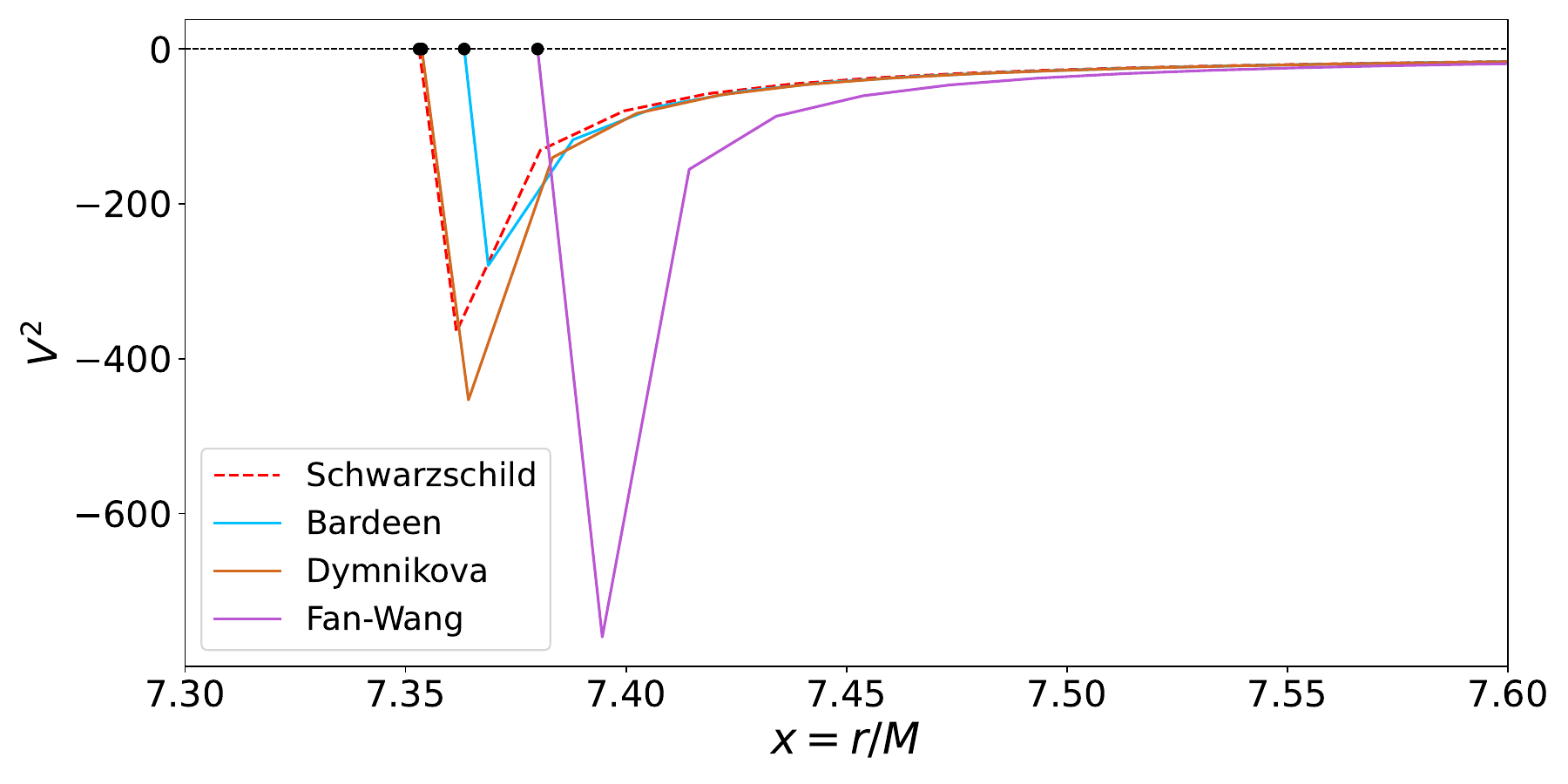}
\end{subfigure}
\caption{\justifying Negative range of values of $V^2$ as a function of $x=r/M$ for different RBH solutions. {\bf Top:} The behavior of the vacuum solution compared with the Schwarzschild and Schwarzschild-de Sitter solutions. {\bf Bottom:} The behavior of the topologically charged and Fan-Wang solutions compared with the Schwarzschild solution.}
\label{fig:V2_confront_negative}
\end{figure}

Finally, we compare the variable $V^2$ for the RBH solutions with the Schwarzschild and Schwarzschild-de Sitter solutions. To provide a clearer view of the behavior of all the solutions, we separate the positive and negative ranges of $V^2$ values, as was done in the previous section \ref{SEC:BONDI_df_results}. The positive and negative ranges are displayed in Fig.~\ref{fig:V2_confront_positive} and Fig.~\ref{fig:V2_confront_negative}, respectively.
Only the Fan-Wang solution in the lower panel has a critical point that deviates slightly from the other regular solutions, as well as from Schwarzschild BH. All other solutions, however, exhibit a critical point very similar to the comparison BHs, as can be seen in Tab. \ref{TAB:const_val_1st_method}.
%
%
%
\subsection{The accretion process with exponential density profile}
\label{SEC:BONDI_exp_results}
Let us now turn to the second case, which involves an exponential density profile. Our objective is to analyze the changes in the behavior of variables compared to the first case and to explore how this problem differs from previous studies, see, for example, \cite{Boshkayev:2020}, where an exponential density profile is used to model out dark matter within the framework of the Novikov-Thorne accretion model.

The goal remains the same as before: to study the behavior of the key variables involved in modeling accretion. However, the procedure differs slightly from the previous case. In this scenario, we analyze how the different RBH solutions respond, starting from the same sample density profile.

To achieve this, we employ \textit{critical point analysis}, which allows us to determine the critical radius and the corresponding critical velocity values. This analysis also provides the value of the integration constant $\mathcal{C}_3$, while the others are manually chosen as they cannot be constrained using the same approach.
%
%
%
\subsubsection{The numerical setup}
\label{SEC:BONDI_exp_rc}

The critical point, characterized by the sound transition, is analyzed using the variable $V^2$ as defined in Eq. \eqref{new_variable_V}. Contrary to the dark fluid case, $V^2$ remains positive throughout the domain, which simplifies its interpretation.

As we discussed before in Subsec. \ref{SEC:BONDI_exp_variables}, we construct and derive the equations for the radial velocity and the variable $V^2$ at the critical point, according to Eq. \eqref{critical_point_eq}. By solving this system numerically and imposing that the two equations might be satisfied simultaneously, we determine the critical radius $r_c$ and the critical velocity $u_c$.

Next, if we impose the equality between Eq. \eqref{variables_crit_point_uc2} and its general counterpart, Eq. \eqref{exponential_velocity_expression}, evaluated at the critical point as follows:
\begin{equation}
\label{condition_for_C3}
    u_c^2 = u^2_{|r=r_c} = \frac{\mathcal{C}_3}{\rho(r_c) r_c^2} \sqrt{\frac{B(r_c)}{A(r_c)}}\,,
\end{equation}
this leads to a constrained value for the constant $\mathcal{C}_3$. The results for all numerical calculations are presented in Tab. \ref{TAB:const_val_2nd_method}. However, we were unable to find a critical set of variables $(r_c, u_c)$ for some of the BH solutions considered. Specifically, for both the Schwarzschild and Schwarzschild-de Sitter metrics. For the latter, also despite testing various values of $\Lambda$ within the allowed range for positive vacuum energy, no critical point was found. This result holds even when considering different combinations of parameters, with the critical radius always being smaller than the event horizon.

This outcome is likely due to the density profile chosen for the study, since, with the "classical" polytropic description, the Schwarzschild solution has a well-defined critical point. The conclusion is that this specific setup of the problem is not well-suited for the analysis of these particular metrics.

The constants defining the density profile, $\rho_0$ and $r_0$, are determined by the reference values reported in \cite{Boshkayev:2020}, with the final parameters summarized in Tab. \ref{TAB:const_val_2nd_method}.

We now proceed to plot and study the variables of the problem. All variables are plotted starting from the event horizon for each metric, grouped consistently with the analysis in the first case.

\subsubsection{Variables profiles}
\label{SEC:BONDI_exp_plots}
\begin{figure}[h!]
\begin{subfigure}{0.5\textwidth}
\includegraphics[width=\linewidth]{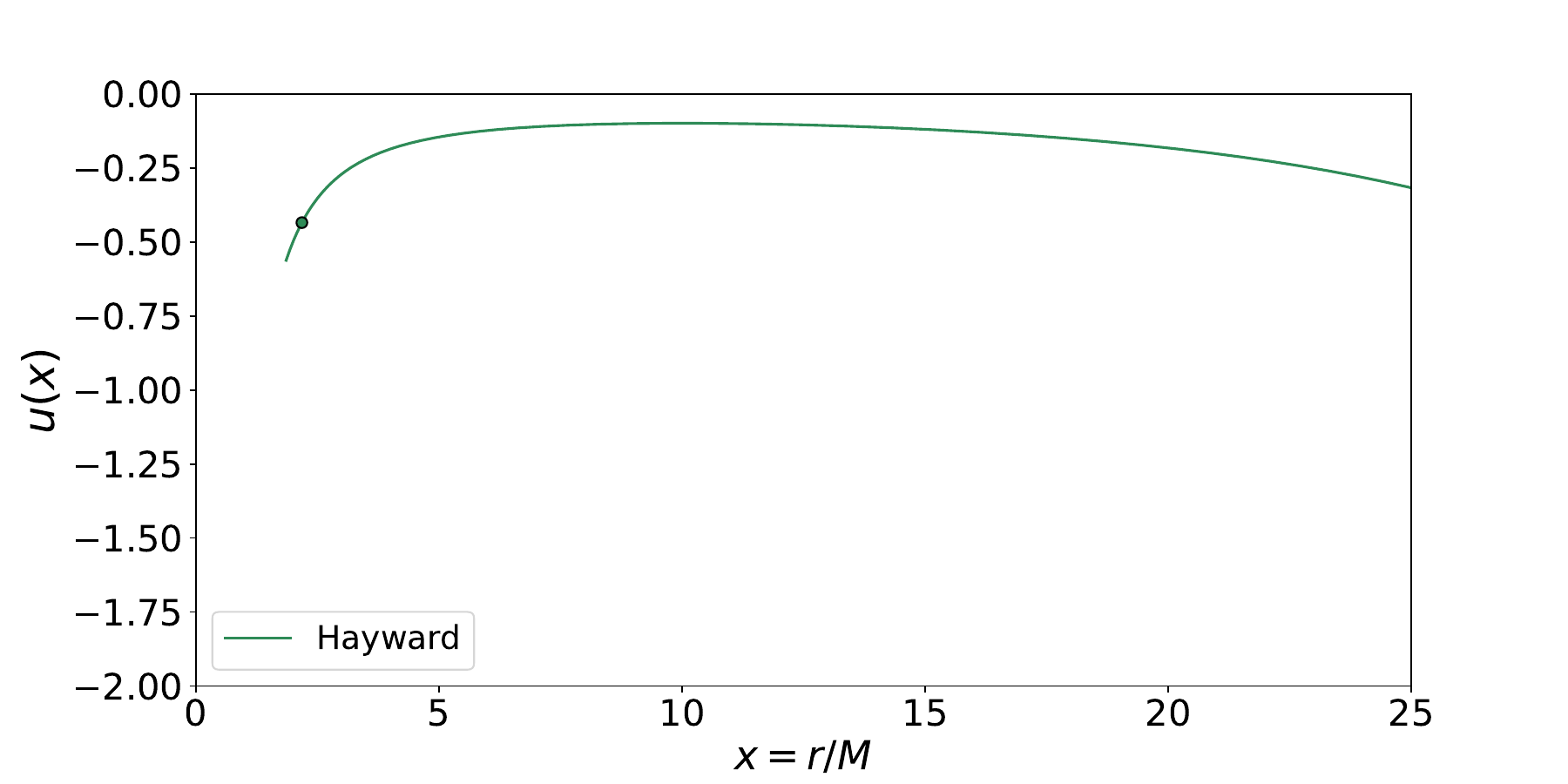}
\end{subfigure}
\hfill
\begin{subfigure}{0.5\textwidth}
\includegraphics[width=\linewidth]{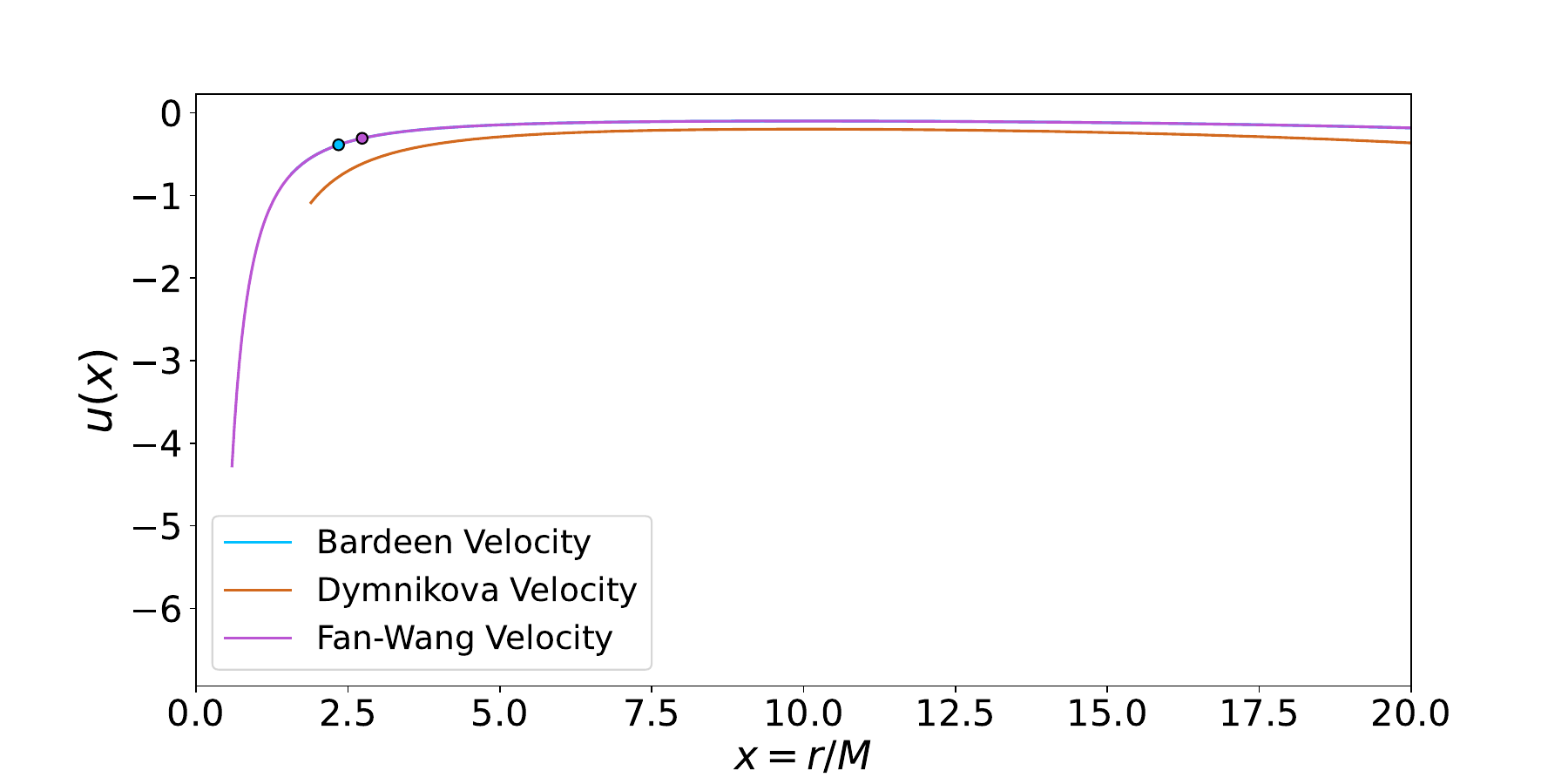}
\end{subfigure}
  \caption{\justifying Radial velocity profiles as a function of $x = r/M$ for different RBH solutions. \textbf{Top:} Behavior of the vacuum solution. \textbf{Bottom:} Behavior of the topologically charged and Fan-Wang solutions.
  The point on the curve, highlighted by the colour of the solution itself, correspond to the critical points of each metric.}
\label{fig:exponential_velocity}
\end{figure}

%
%
We start to analyze the radial velocity profiles in Fig.~\ref{fig:exponential_velocity}.

The radial velocity $u$, obtained from the conservation equations, remains consistently negative, indicating the inward motion of accreted matter. In the top panel, the Hayward solution exhibits a steep decline in velocity as $x$ decreases, becoming more pronounced near the event horizon, with the critical point located just outside it. Moving outward from the accretion disk, the velocity gradually decreases, though at a slower rate.

For topologically charged solutions and the Fan-Wang metric in the lower panel, the profiles overlap due to their common dependence on the density distribution and symmetry properties of the metrics such as $A(r) = B(r)$, with a slight decrease of the Dymnikova near the horizon. These trends indicate uniform accretion dynamics in these solutions, characterized by a consistent deceleration of matter with increasing radius. Within this range, the critical points of the different RBHs are observed, and it can be seen that the critical point of the Bardeen solution is much more internal than those of the Dymnikova and Fan-Wang solutions, which are located not only at large $x$ from the event horizon, but also extremely close to each other.

\begin{figure}[h!]
\begin{subfigure}{0.5\textwidth}
\includegraphics[width=\linewidth]{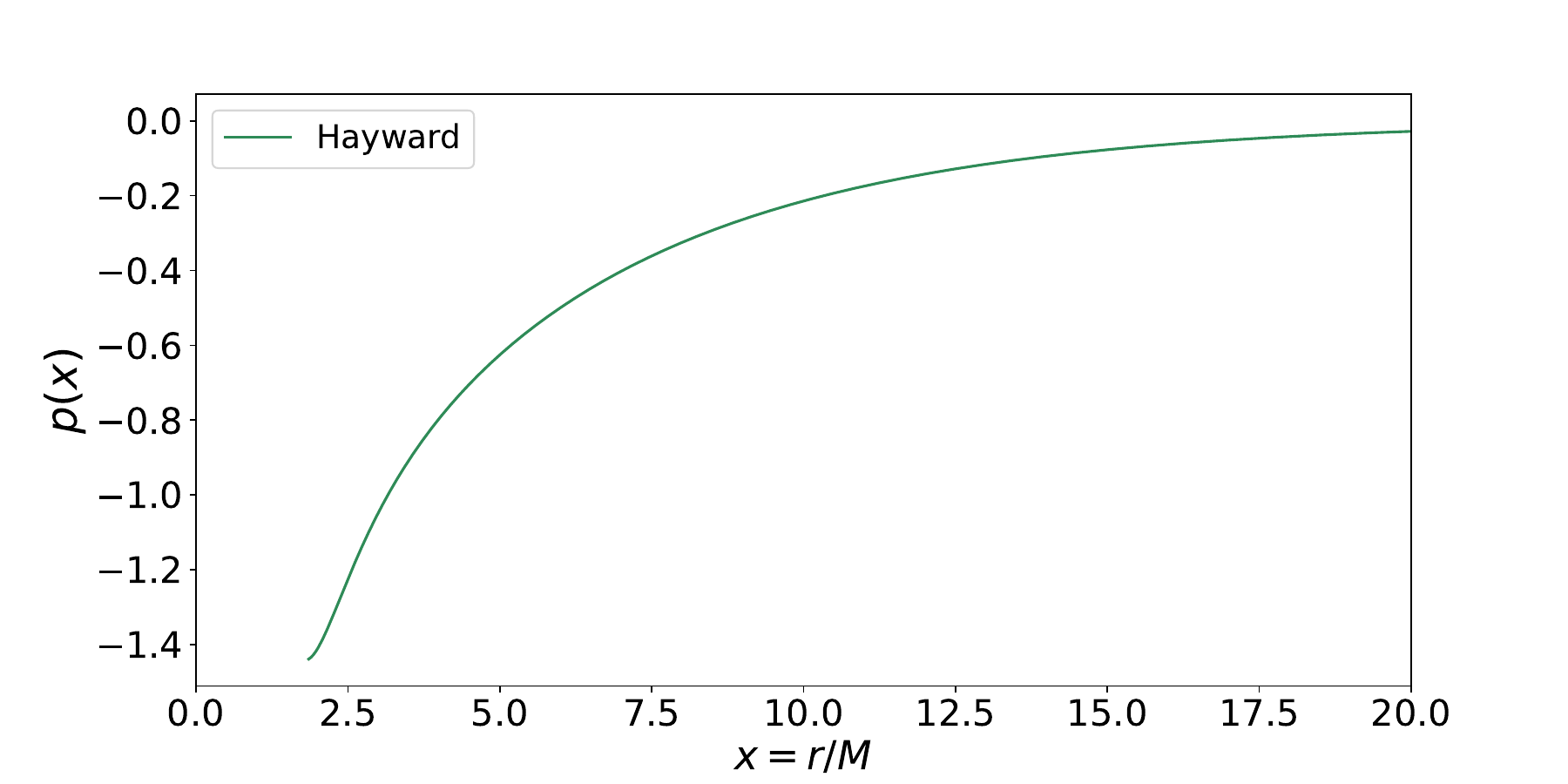}
\end{subfigure}
\hfill
\begin{subfigure}{0.5\textwidth}
\includegraphics[width=\linewidth]{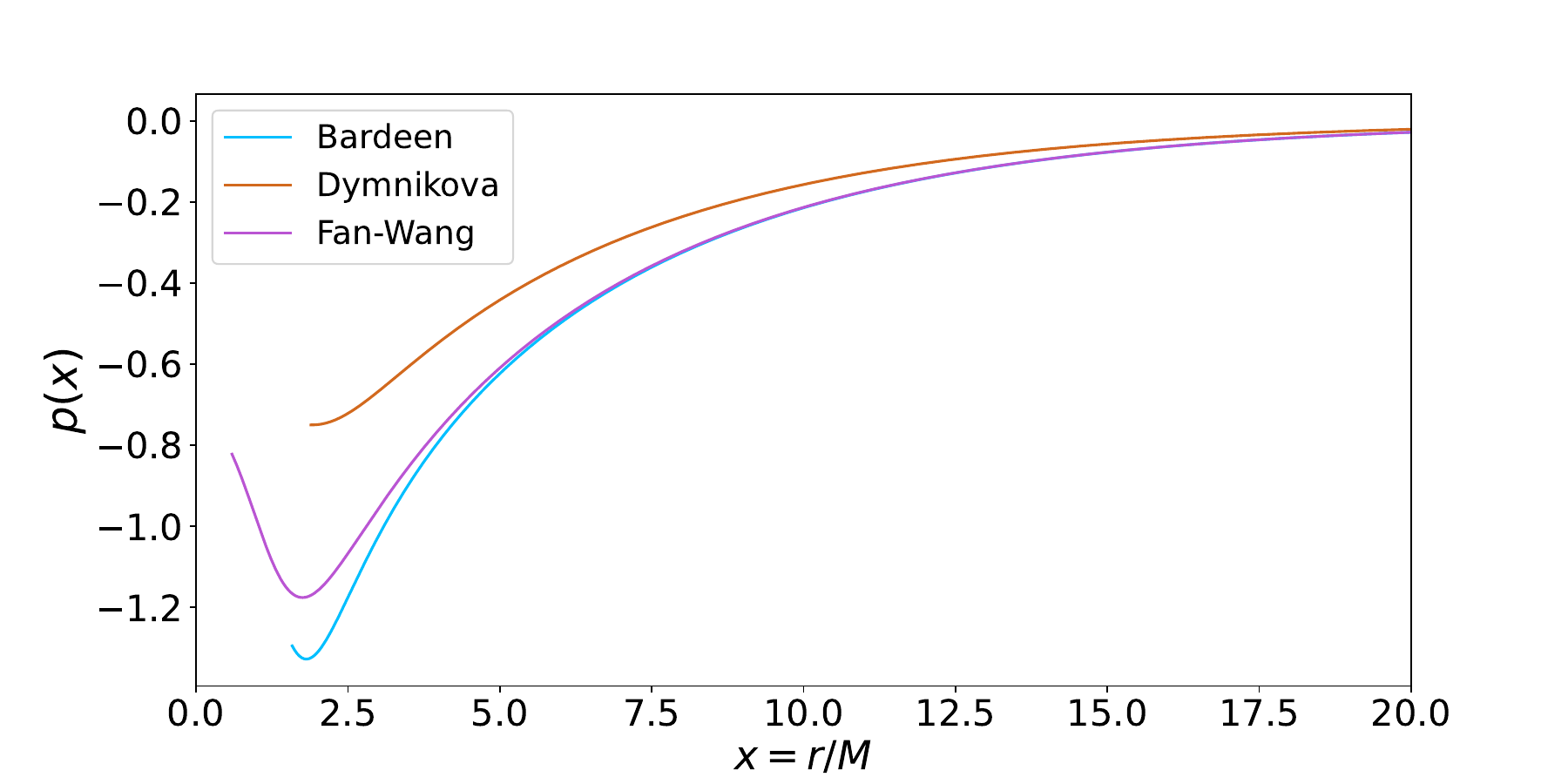}
\end{subfigure}
  \caption{\justifying Pressure profiles in function of $x = r/M$ for different RBH solutions. {\bf Top:} Behavior of the vacuum solution. {\bf Bottom:} behavior of the topologically charged and Fan-Wang solutions.
  }
\label{fig:exponential_pressure}
\end{figure}

%
%
Let us now study the pressure profiles. In this second case, we do not have a constant pressure but it is derived from the conservation equations, Eq. \eqref{exponential_pressure_expression}. Starting with the top panel of Fig.~\ref{fig:exponential_pressure},
Starting from the vacuum energy in the top panel, we observe a moderate decrease as we approach the event horizon and tends to zero for large $x$, moving further away from the event horizon.

For topologically charged solutions in the lower panel, we observe behavior similar to that of vacuum energy solutions at large distances from the event horizon, confirming the reduced dependence on the specific metric terms at high radii. In contrast, for small values of $x$, differences emerge: the curves diverge, showing a sharp decrease followed by a further increase as we approach the event horizon. In particular, the Dymnikova solution exhibits the less pronounced trend, with slightly higher pressure values compared to the Fan-Wang and Bardeen solutions, while the Fan-Wang solution, which has the smallest event horizon, shows an additional increase in pressure as the innermost region is approached.

\begin{figure}[h!]

\begin{subfigure}{0.5\textwidth}
\includegraphics[width=\linewidth]{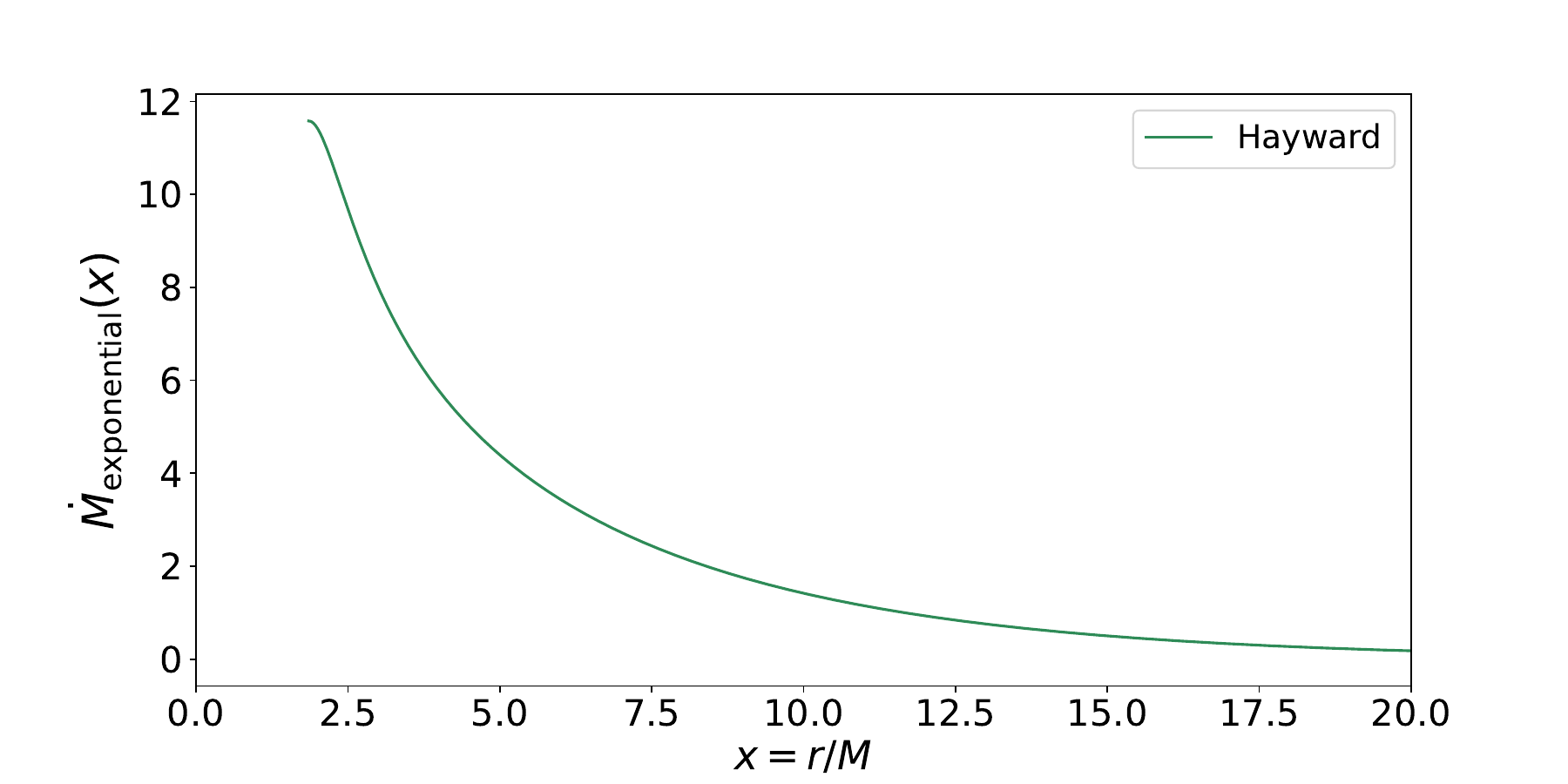}
\end{subfigure}
\hfill
\begin{subfigure}{0.5\textwidth}
\includegraphics[width=\linewidth]{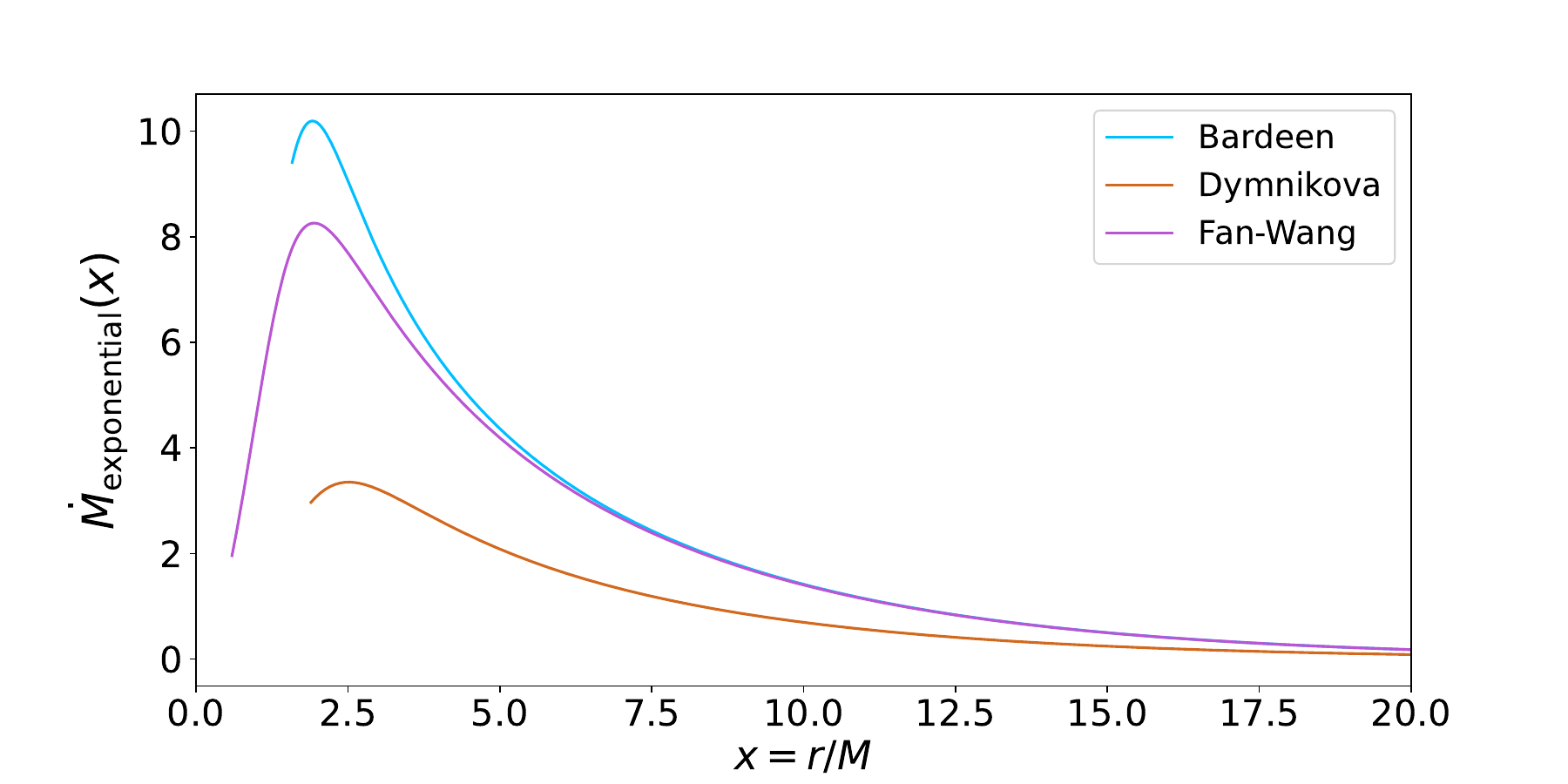}
\end{subfigure}
  \caption{\justifying Mass accretion rate in function of $x = r/M$ for different RBH solutions. {\bf Top:} Behavior of the vacuum solutions. {\bf Bottom:} Behavior of the topologically charged and Fan-Wang solutions.}
\label{fig:exponential_acc_rate}
\end{figure}

\begin{figure}[h!]
\begin{subfigure}{0.5\textwidth}
\includegraphics[width=\linewidth]{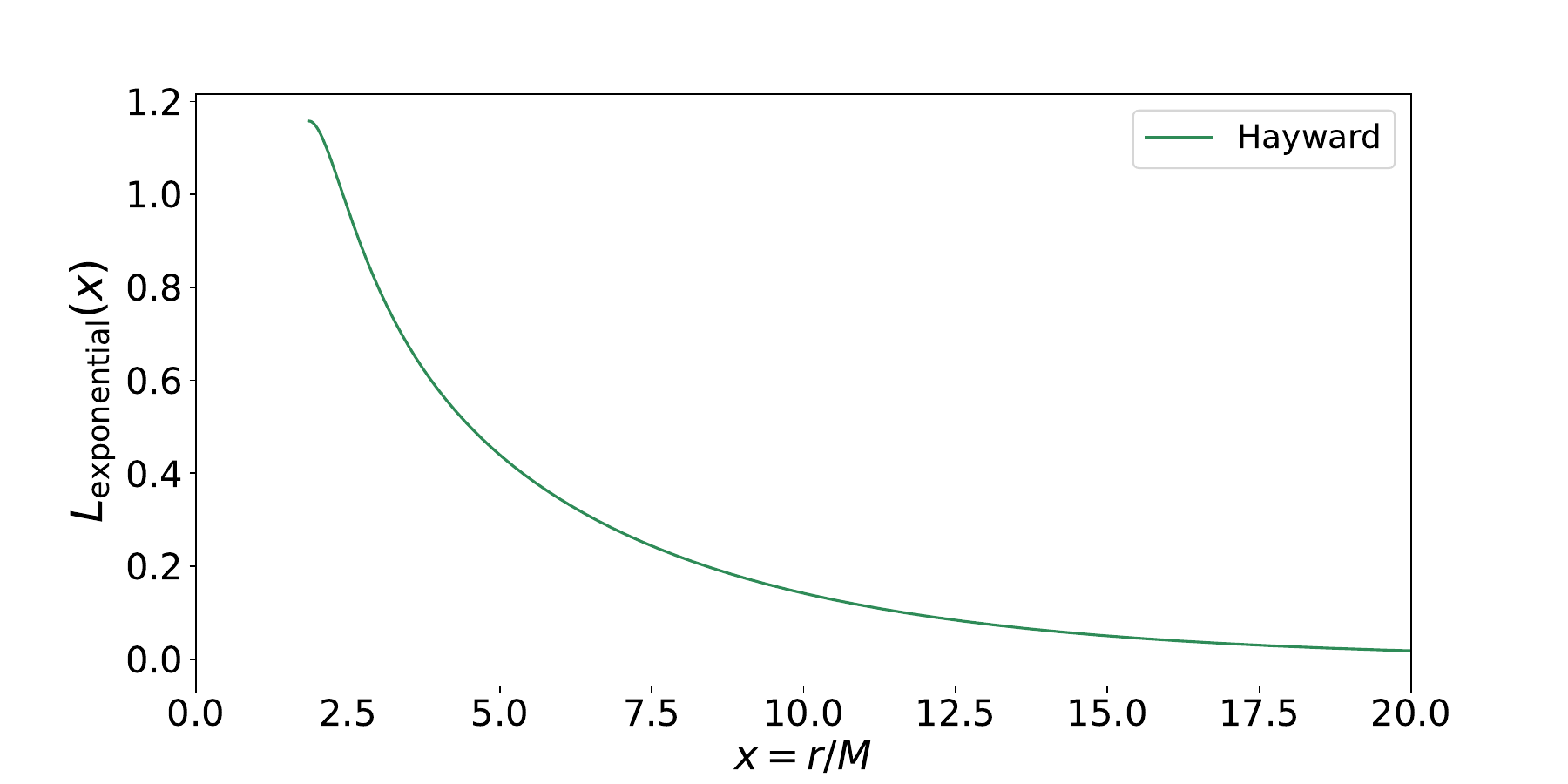}
\end{subfigure}
\hfill
\begin{subfigure}{0.5\textwidth}
\includegraphics[width=\linewidth]{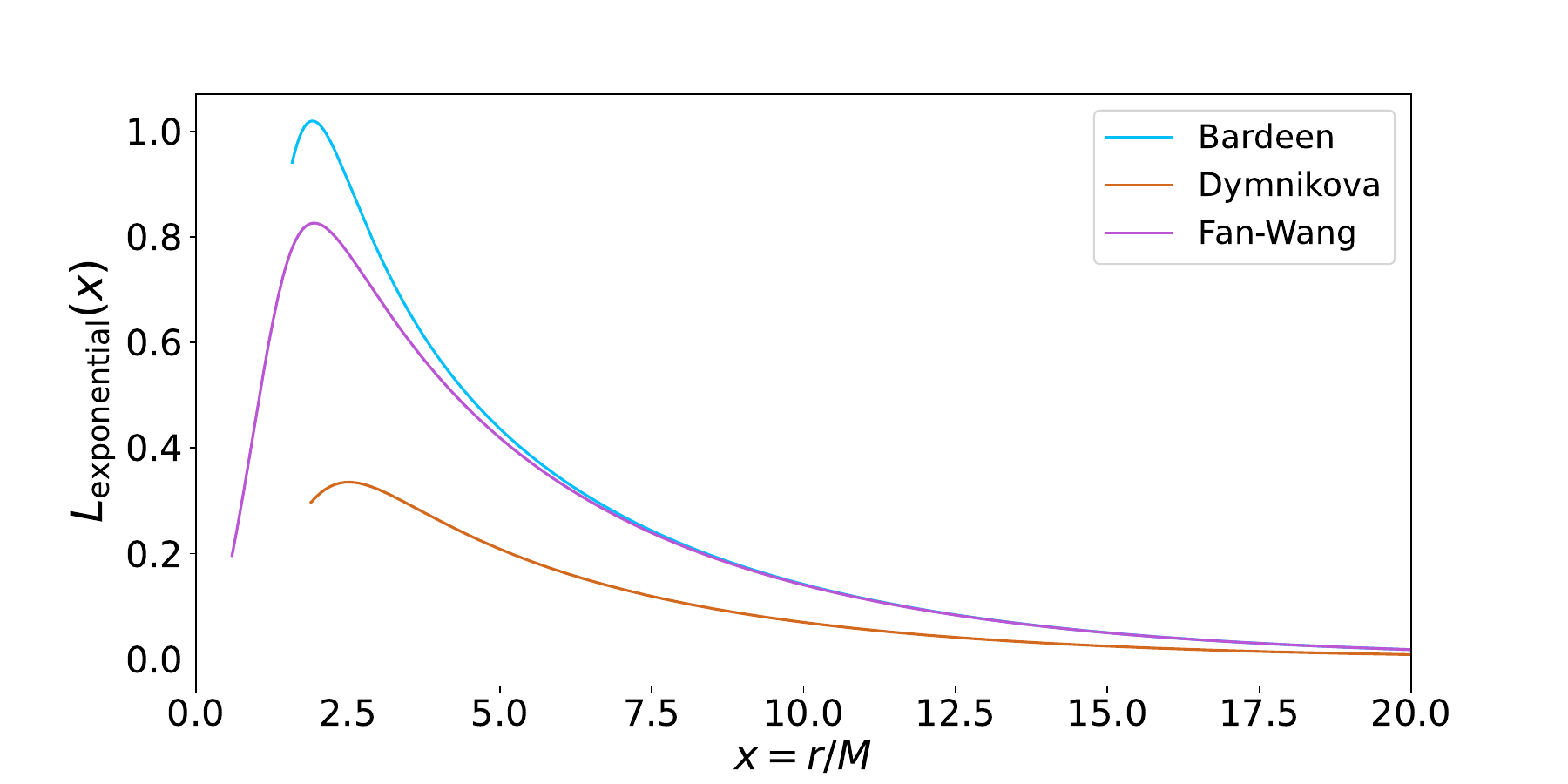}
\end{subfigure}
  \caption{\justifying Luminosity in function of $x = r/M$ for different RBH solutions. {\bf Top:} Behavior of the vacuum solution. {\bf Bottom:} Behavior of the topologically charged and Fan-Wang solutions.}
\label{fig:exponential_luminosity}
\end{figure}

%
%
Fig.~\ref{fig:exponential_acc_rate} and \ref{fig:exponential_luminosity} show the mass accretion rate $M\dot{M}$ and luminosity $L$, respectively.
The mass accretion rate, $\dot{M}$, increases as $x$ decreases toward the BH, with each solution exhibiting a distinct slope. At large radii, all accretion rates gradually approach zero.

In the top panel, corresponding to vacuum energy solutions, the accretion rate rises sharply near the event horizon. In particular, the Hayward solution exhibits a steeper growth, with no peak observed as the event horizon is approached.

In the bottom panel, which displays the behavior of topologically charged solutions and the Fan-Wang metric, the trends are more uniform. Among these, the Dymnikova solution consistently shows higher accretion rates across all radii.

The luminosity $L$, scaled by an efficiency factor $\eta_{\text{eff}}$, follows a similar trend. In the top panel, the Hayward solution maintains a steep profile and high luminosity values, mirroring its accretion rate behavior. Similarly, in the bottom panel, the Dymnikova solution exhibits the highest luminosities near the event horizon, with a rapid increase analogous to its accretion rate trend. Among the topologically charged solutions, the Dymnikova metric yields the highest luminosities, followed by the Bardeen and Fan-Wang solutions.
\begin{table}[!h]
    \centering
    \renewcommand{\arraystretch}{1.3}
    \begin{tabular}{|l|c|c|c|c|c|c|}
        \hline
        \hline
    \multicolumn{7}{|c|}{\textbf{2\textsuperscript{nd} case: exponential density profile}} \\ \hline\hline
    \textbf{Solution} & $ \mathcal{C}_1 $ & $ \mathcal{C}_2 $ & $ \mathcal{C}_3 $ & $ r_{\text{EH}} $ & $ r_{\text{crit}} $ & $ u_{\text{crit}} $ \\
        \hline\hline
        Bardeen         & 1     & -1 & -1      & 1.58          & 2.59  & -0.33 \\
        \hline
        Hayward         & 1     & -1 & -1      & 1.85          & 2.18  & -0.43 \\
        \hline
        Fan-Wang        & 1     & -1 & -5.12   & 0.59          & 15.09 & -0.10 \\
        \hline
        Dymnikova       & 1     & -1 & -0.99 & 1.89            & 14.60 & -0.11 \\
        \hline
        Schwarzschild   & 1     & -1 & -2      & 2             & --    & --    \\
        \hline
        Schwarzschild-de Sitter & 1 & -1 & -2.50  & 2 - 53.74  & --    & --    \\
        \hline
        \hline
    \end{tabular}
    \caption{\justifying The table presents the integration constants and the numerical results for the critical radius $r_{\text{crit}}$ and the event horizons $r_{\text{EH}}$, using the exponential profile. Constants used: $ M=1 $, $ a=0.5 $, $ q_B=0.65 $, $ q_D=0.452 $, $ \Lambda=0.001 $, $ l_{FW}=8/27 $, $\rho_0 = 0.75 $, $r_0=5$.}
    \label{TAB:const_val_2nd_method}
\end{table}
%
%
\subsubsection{Comparing Schwarszchild and Schwarszchild-de Sitter solutions}
\label{SEC:BONDI_exp_comparison}

\begin{figure}[h!]
\begin{subfigure}{0.5\textwidth}
\includegraphics[width=\linewidth]{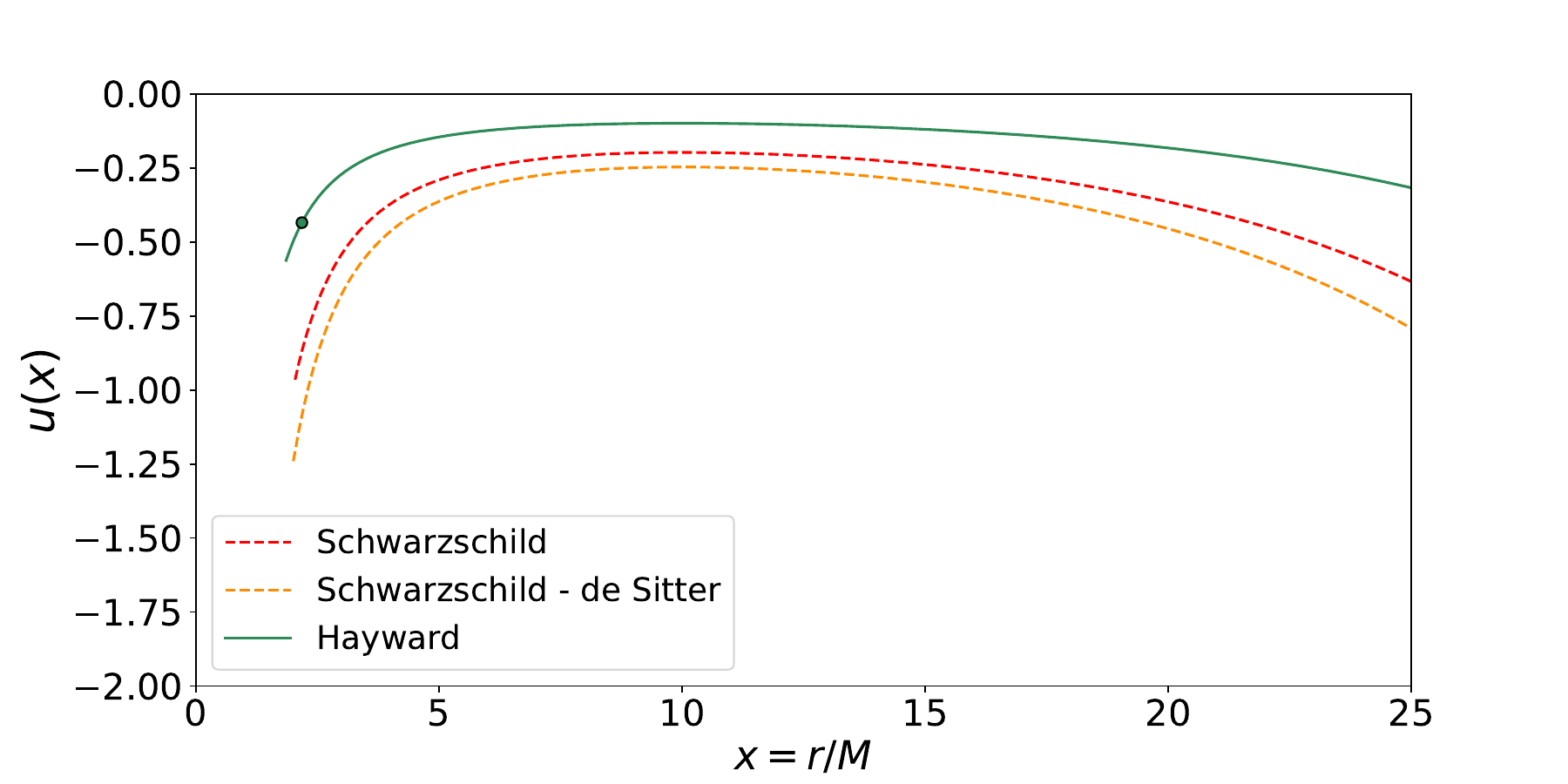}
\end{subfigure}
\hfill
\begin{subfigure}{0.5\textwidth}
\includegraphics[width=\linewidth]{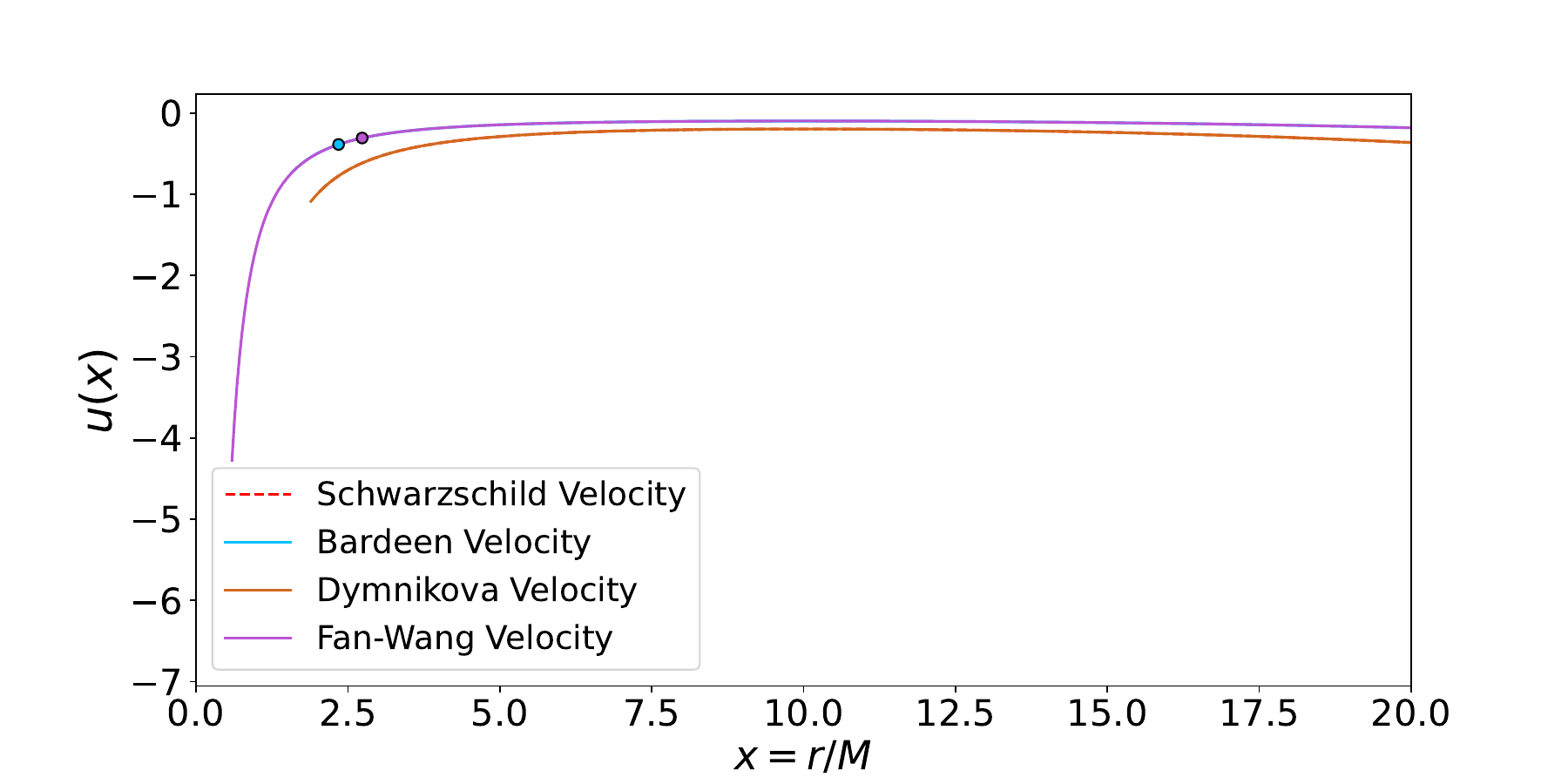}
\end{subfigure}
  \caption{\justifying Radial velocity profiles in function of $x = r/M$ for different RBH solutions. The points on the curve, highlighted by the colour of the solution itself, correspond to the critical points of each metric. {\bf Top:} Behavior of the vacuum solutions compared with Schwarzschild and Schwarzschild-de Sitter solutions. {\bf Bottom:} Behavior of the topologically charged and Fan-Wang solutions compared with the Schwarzschild solution.}
\label{fig:exponential_velocity_confronto}
\end{figure}

Let us now describe the comparison between the Schwarzschild and Schwarszchild-de Sitter solutions, as done in the previous case. Starting with the velocity profile shown in Fig.~\ref{fig:exponential_velocity_confronto},
we observe in the upper panel that the Hayward solution exhibits lower velocity values compared to the Schwarzschild and Schwarzschild-de Sitter solutions, with the latter having the smallest values in this group. However, all solutions follow a similar trend across the entire range of $x$. At larger radii, a gradual decrease in velocity is observed, and as the solutions move away from the event horizon, their velocity values remain distinct, though the overall decline in velocity becomes more pronounced.

The lower panel, showing the topologically charged and Fan-Wang solutions, shows that these solutions do indeed have the same trend as the Schwarzschild solution, but the latter has slightly more negative velocity values than the Bardeen and Fan-Wang BHs while overlapped with the Dymnikova solution.

We are not able to compare the critical point for the Scwarszchild and Scwarszchild-de Sitter solutions with the RBH solutions, since the critical point analysis in this framework does not give any results for these solutions.

\begin{figure}[h!]

\begin{subfigure}{0.5\textwidth}
\includegraphics[width=\linewidth]{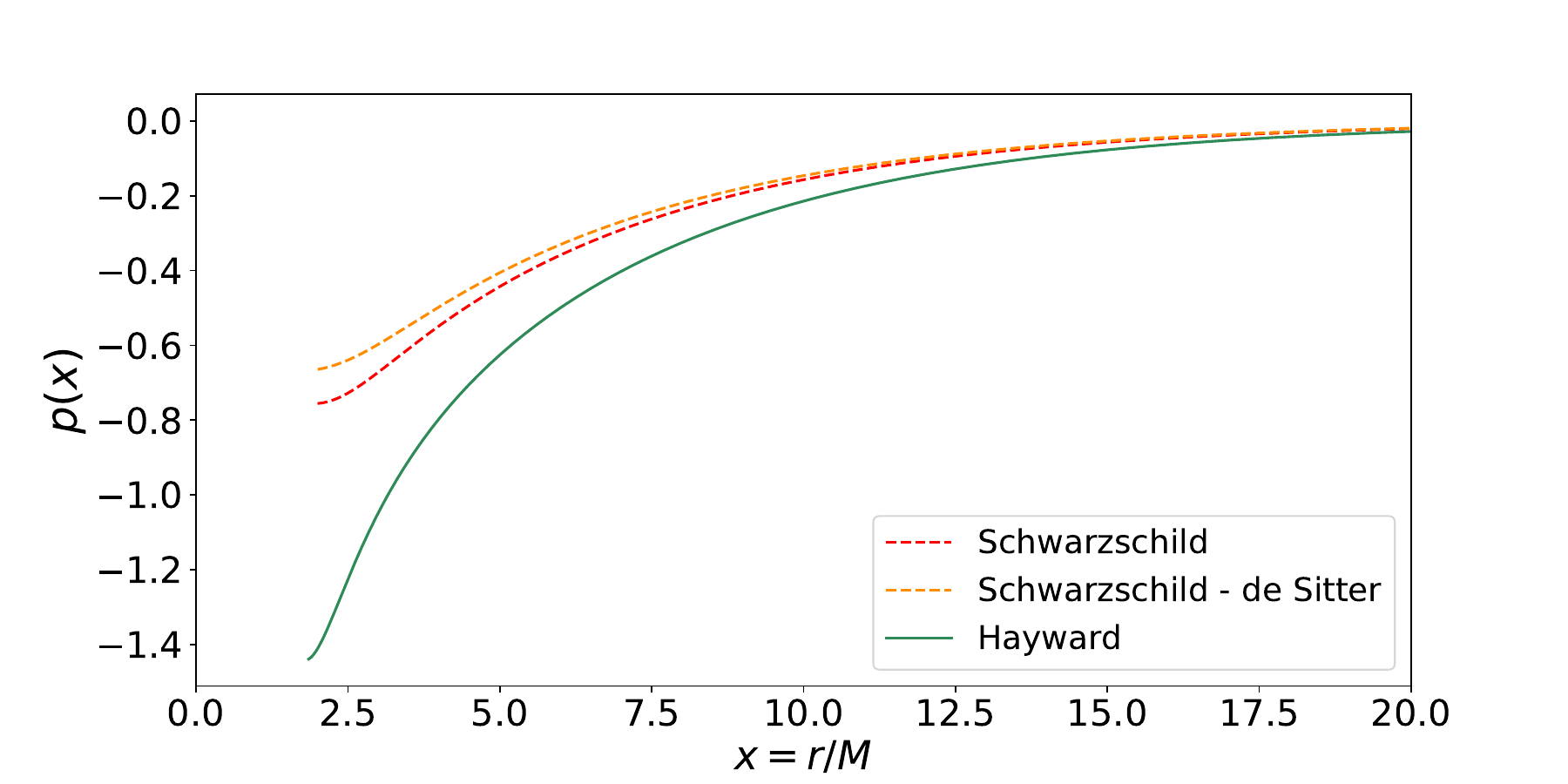}
\end{subfigure}
\hfill
\begin{subfigure}{0.5\textwidth}
\includegraphics[width=\linewidth]{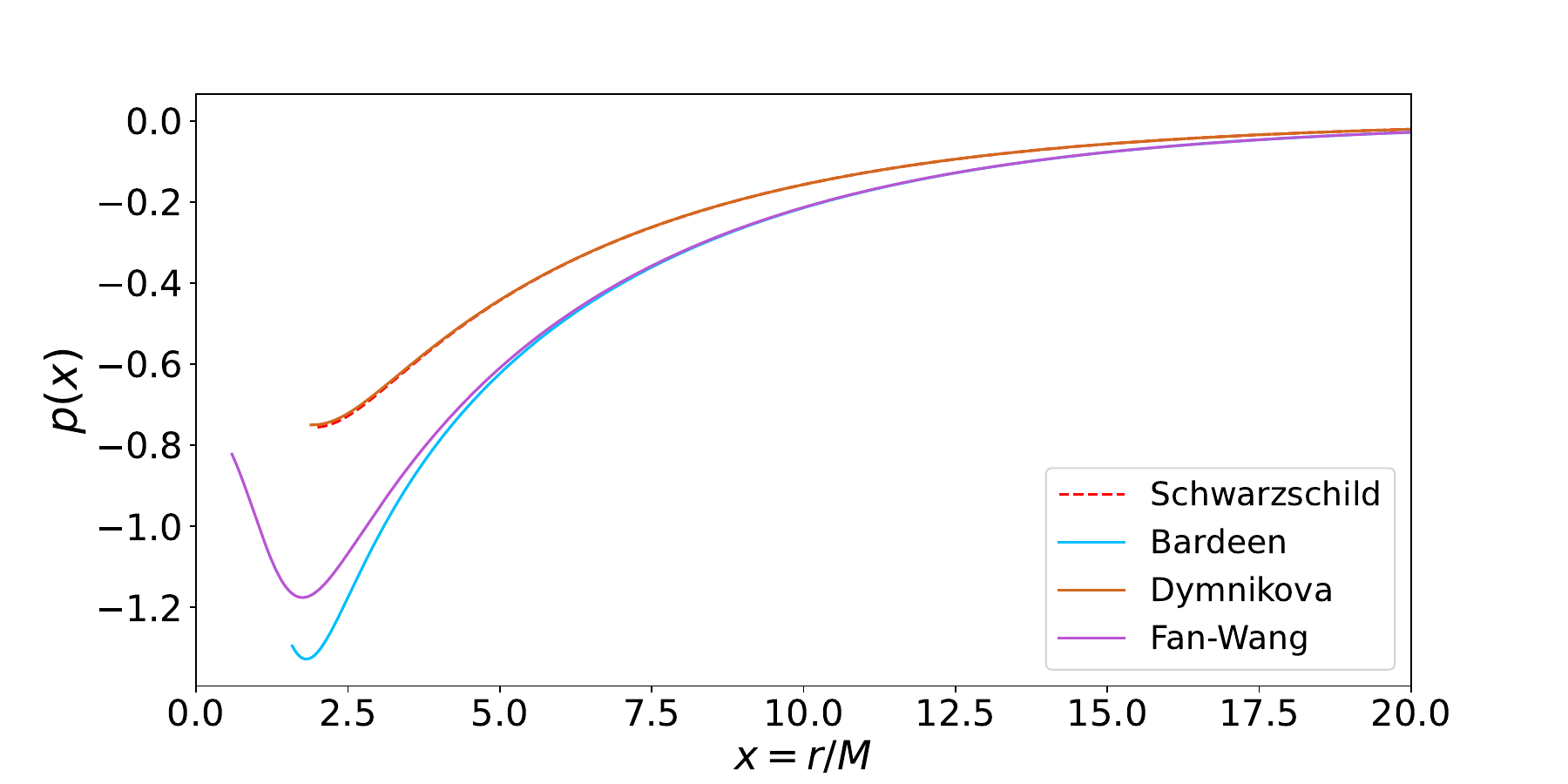}
\end{subfigure}
  \caption{\justifying Pressure profiles in function of $x = r/M$ for different RBH solutions. {\bf Top:} Behavior of the vacuum solutions compared with Schwarzschild and Schwarzschild-de Sitter solutions. {\bf Bottom:} Behavior of the topologically charged and Fan-Wang solutions compared with the Schwarzschild solution.}
\label{fig:exponential_pressure_confronto}
\end{figure}

In Fig.~\ref{fig:exponential_pressure_confronto},
we present the pressure profiles, with the vacuum solutions shown in the top panel. The Schwarzschild and Schwarzschild-de Sitter solutions exhibit nearly identical trends across all values of $x$, with only a slight deviation near the event horizon. This small difference was introduced by adjusting the integration constant $\mathcal{C}_3$ for the Schwarzschild-de Sitter solution to distinguish it from the Schwarzschild case, as they would otherwise be almost completely superimposed. However, this adjustment does not affect the critical point analysis, as both of them yielded an unconstrained result for this integration constant. As $x$ increases, all solutions gradually converge, eventually approaching zero.

For the topologically charged and Fan-Wang solutions, we note that the trends differ as we approach the event horizon, with the exception of the Dymnikova solution which is completely superposed to the Schwarzschild solution, having the lowest negative pressure values of all the other predicted solutions in the plot. As we move away from the disc towards larger radii, the trends also overlap with Schwarzschild, showing an increase in pressure asymptotically approaching zero.

\begin{figure}[h!]
\begin{subfigure}{0.5\textwidth}
\includegraphics[width=\linewidth]{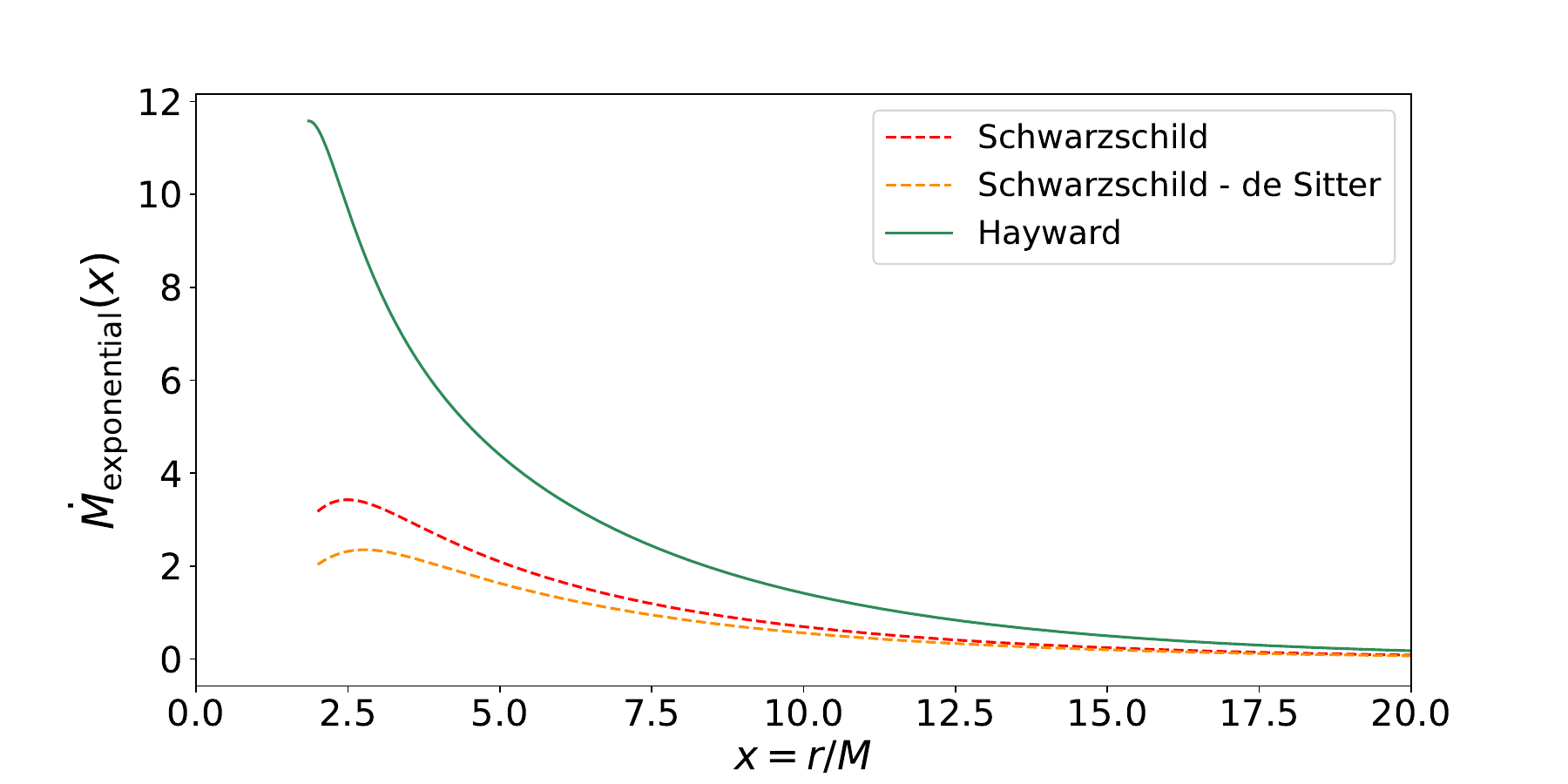}
\end{subfigure}
\hfill
\begin{subfigure}{0.5\textwidth}
\includegraphics[width=\linewidth]{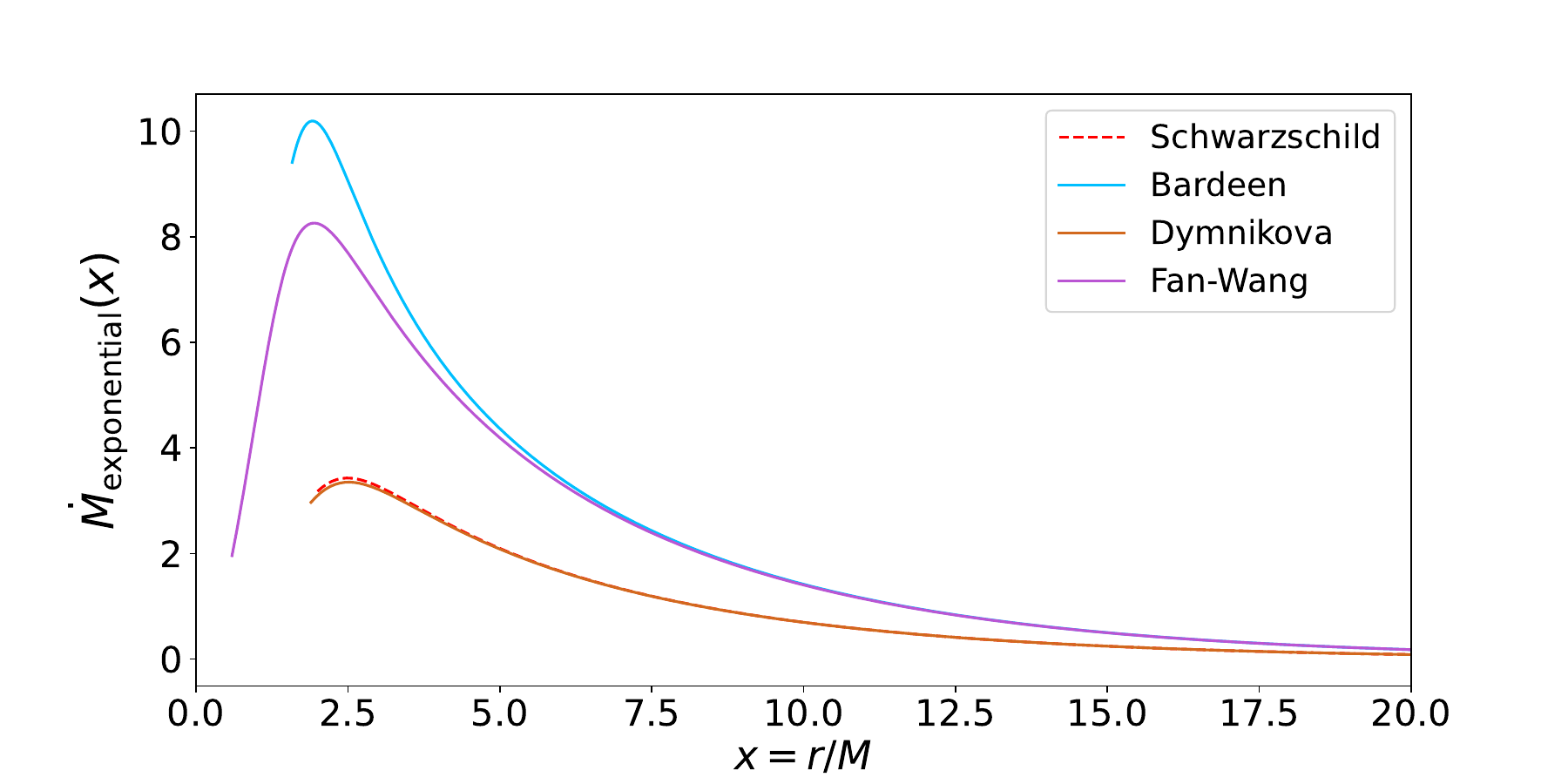}
\end{subfigure}
  \caption{\justifying Mass accretion rate in function of $x = r/M$ for different RBH solutions.
 {\bf Top:} Behavior of the vacuum solutions compared with Schwarzschild and Schwarzschild-de Sitter solutions. {\bf Bottom:} Behavior of the topologically charged and Fan-Wang solutions compared with the Schwarzschild solution.}
\label{fig:exponential_acc_rate_confronto}
\end{figure}

\begin{figure}[h!]
\begin{subfigure}{0.5\textwidth}
\includegraphics[width=\linewidth]{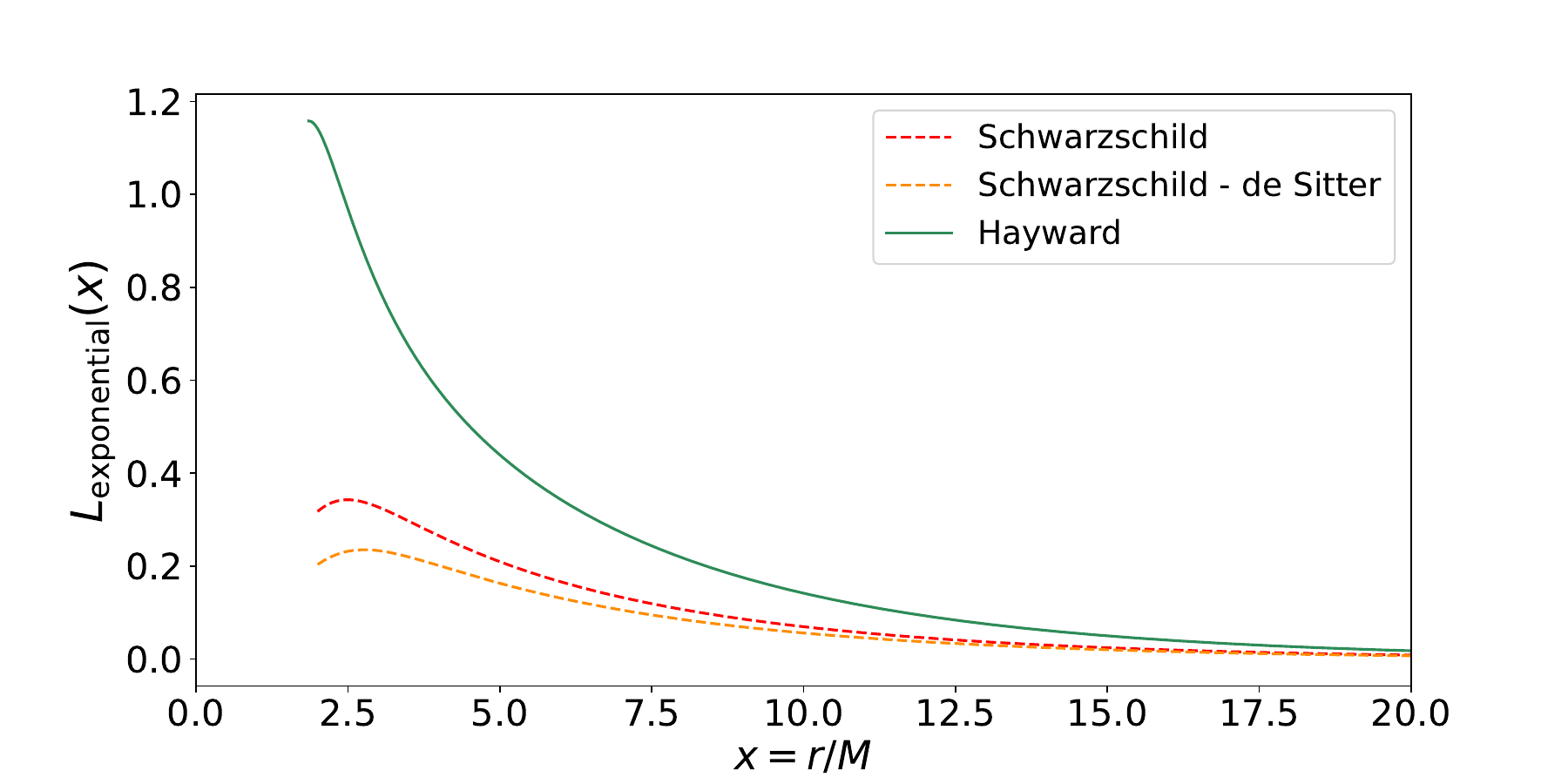}
\end{subfigure}
\hfill
\begin{subfigure}{0.5\textwidth}
\includegraphics[width=\linewidth]{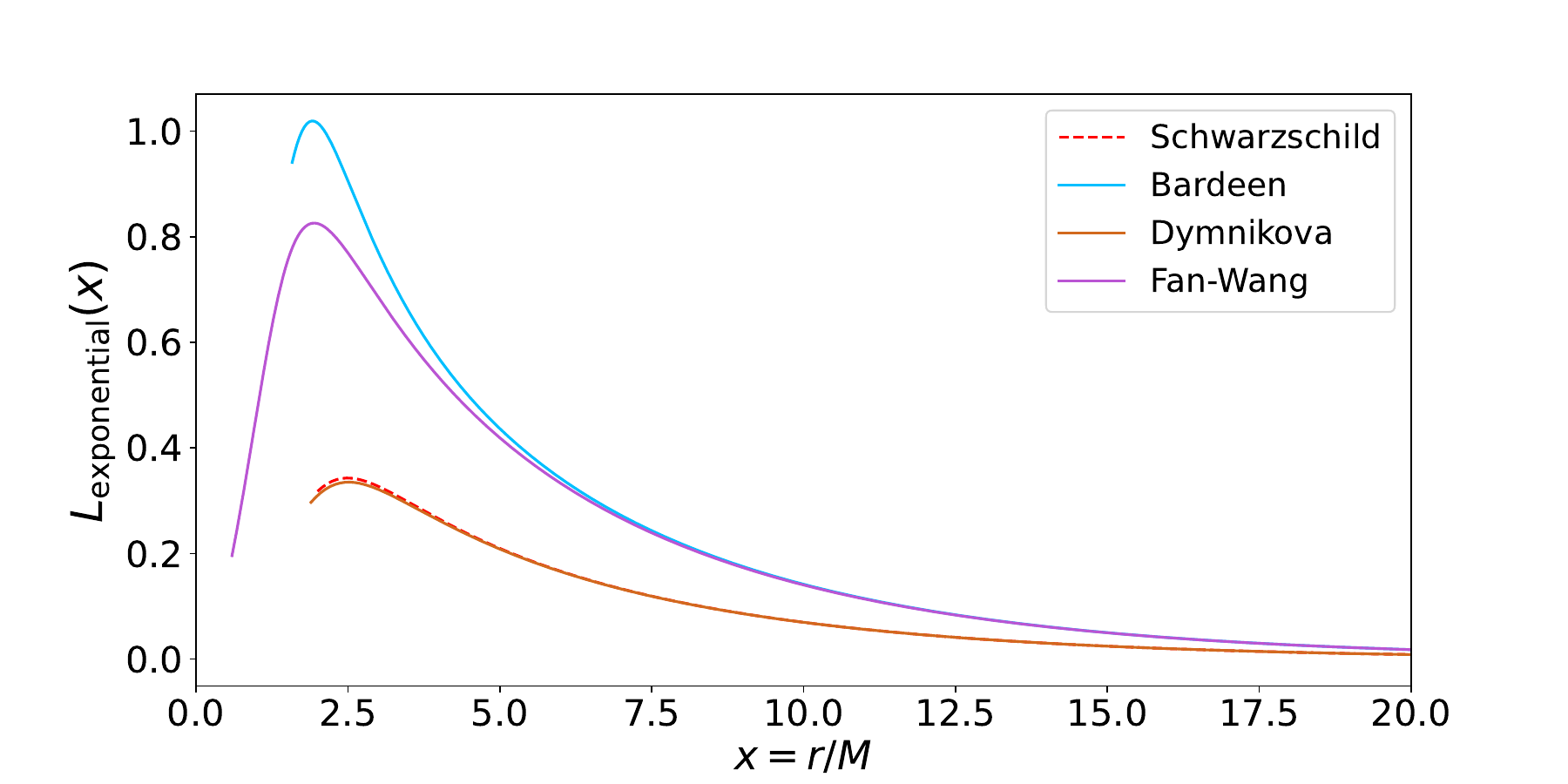}
\end{subfigure}
  \caption{\justifying  Luminosity in function of $x = r/M$ for different RBH solutions. {\bf Top:} Behavior of the vacuum solution compared with Schwarzschild and Schwarzschild-de Sitter solutions. {\bf Bottom:} Behavior of the topologically charged and Fan-Wang solutions compared with the Schwarzschild solution.}
\label{fig:exponential_luminosity_confronto}
\end{figure}

Next, we discuss the accretion rate in Fig.~\ref{fig:exponential_acc_rate_confronto} and the luminosity in Fig.~\ref{fig:exponential_luminosity_confronto}.

Starting from the upper panels of Figures \ref{fig:exponential_acc_rate_confronto}-\ref{fig:exponential_luminosity_confronto}, which illustrate the vacuum energy solution, we observe that for small $x$, the Hayward solution exhibits a significantly higher accretion rate—and consequently a higher luminosity—compared to the Schwarzschild and Schwarzschild-de Sitter solutions. Between these two, the Schwarzschild solution maintains slightly higher values than the Schwarzschild-de Sitter case, though the difference is not substantial. In the outer regions, all solutions follow a similar trend, gradually approaching zero.

For the topologically charged solutions and the Fan-Wang model, we observe that these solutions, along with the Schwarzschild solution, follow the same overall trend over the same range of $x$, all displaying a peak but with a distinct hierarchy in their values except the Dymnikova which is superposed to it. The Bardeen solution exhibits the highest values, while the Schwarzschild solution shows the lowest values of $\dot{M}$ and $L$ near the event horizon compared to the other BH solutions.
%
%
\section{The Analysis of the Novikov-Thorne Accretion Process}
\label{SEC:NT_results}
In this section, we analyze the Novikov-Thorne accretion model, extending our previous study of Bondi accretion. To maintain a consistent framework, we examine two distinct fluid models: a dark fluid accretion scenario and an accretion disk governed by an exponential density profile. The goal is to compare the behaviors of these fluids under the same astrophysical conditions, enabling a meaningful evaluation of their effects on disk properties.

We begin by studying the dark fluid case, where we present the emitted flux and luminosity for each RBH solution and compare these with both Schwarzschild and Schwarzschild-de Sitter solutions. The same approach is subsequently applied to the second case, where an exponential density profile is assumed.

In both cases, we maintain the same parameter choices as in the Bondi accretion model, with two exceptions: the Fan-Wang parameter is set to $l_{\rm FW}=4/27$, instead of $l_{\rm FW}=8/27$, and the cosmological constant for the Schwarzschild-de Sitter solution is adjusted to $\Lambda=10^{-5}$ instead of $\Lambda=10^{-4}$. These modifications were necessary to obtain physically meaningful results within the Novikov-Thorne framework.

The differential luminosity plots are presented as functions of the scaled radial coordinate $r/M_{\rm T}$, where $M_{\rm T}$ represents the total mass, which can either be $M(r_s)$ or $M_{\rm BH}$, depending on the chosen density profile. This scaling accounts for the mass dependence introduced by different fluid distributions.

\subsection{The Dark Fluid Accretion Process}
\label{SEC:NT_df_results}
Using the previously derived equations, we now investigate the first Novikov-Thorne accretion case for RBH solutions. To ensure a direct comparison with Bondi accretion results, we set the dark fluid pressure to $P = -0.75$.

Since pressure remains constant in this scenario, the sonic radius $r_{\rm s}$ is not determined using the TOV equation \eqref{TOV_eq}; instead, it is imposed as an external parameter. We choose $r_{\rm s} =35$ AU for all solutions, a value that aligns with those obtained in the exponential density profile case. The inner radius of the fluid envelope, $r_{\rm b}$, is set to a value smaller than the ISCO but sufficiently close to ensure consistency. The ISCO is determined by differentiating the angular momentum expression and solving for its critical points.

The three radii governing the accretion disk structure follow the established relation
\begin{equation}
   \label{radii_relation} r_{\rm b}\leq r_{\rm i}\leq r_{\rm s}\,,
\end{equation}
which ensures that a portion of the envelope extends beyond the ISCO while maintaining an inner edge larger than the event horizon, thus preserving physical validity. This approach diverges from vacuum BH cases, where $r_{\rm s} \leq r_{\rm i}$ would be realized.

The numerical values used in this case are listed in Table \ref{TAB:NT_DF}.

\begin{figure}[h!]
\begin{subfigure}{0.5\textwidth}
\includegraphics[width=\linewidth]{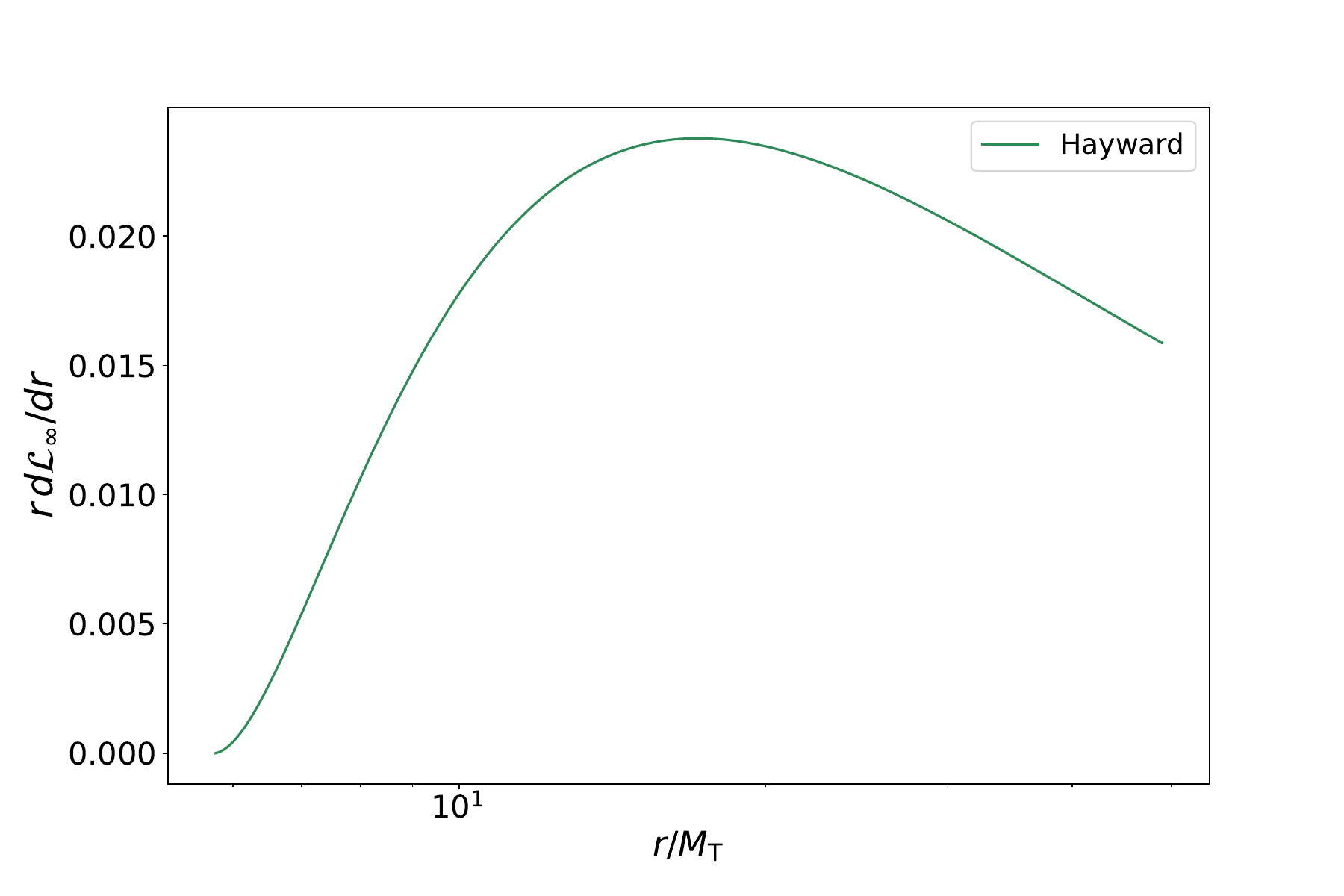}
\end{subfigure}
\hfill
\begin{subfigure}{0.5\textwidth}
\includegraphics[width=\linewidth]{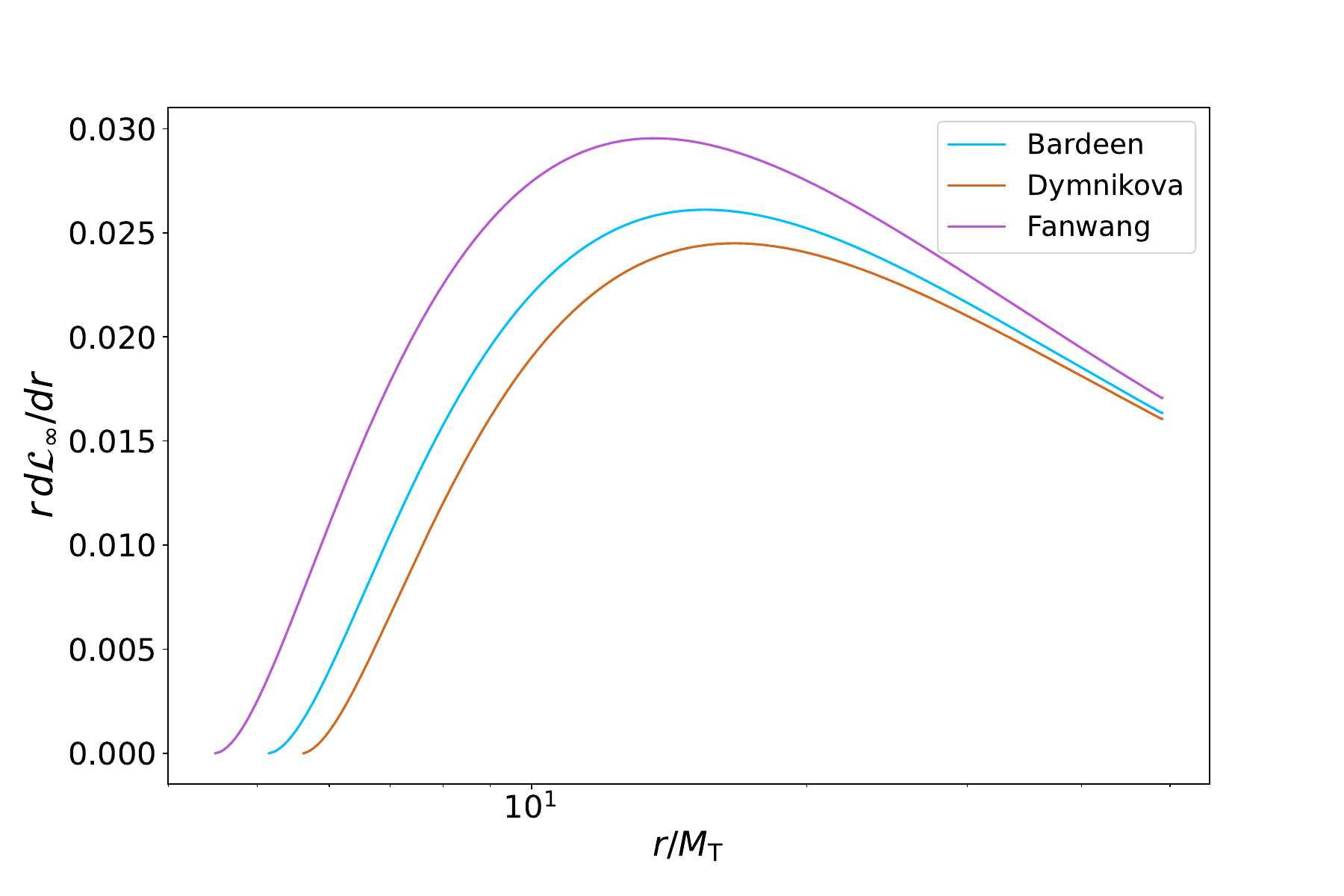}
\end{subfigure}
  \caption{\justifying Differential luminosity as a function of $r/M_{\rm T}$ (log scale) for different RBH solutions.
 {\bf Top:} Behavior of the vacuum solution. {\bf Bottom:} Behavior of the topologically charged and Fan-Wang solutions.}
\label{fig:NT_df_diff_lum}
\end{figure}

The differential luminosity is illustrated in Fig. \ref{fig:NT_df_diff_lum},
This quantity is directly derived from the flux using \eqref{differential_luminosity} and serves as the primary variable for comparison with Bondi accretion.

In the top panel, the vacuum Hayward solution shows an increase in luminosity, reaching a peak before gradually decreasing at larger radii. In the bottom panel, the charged and Fan-Wang solutions follow a similar trend, and their curves converge at large $r$, indicating asymptotic consistency among these models.

\subsubsection{Comparison with Schwarzschild and Schwarzschild-de Sitter}
We now compare the RBH models with the Schwarzschild and Schwarzschild-de Sitter BHs. A key difference from Bondi accretion is that the cosmological constant for Schwarzschild-de Sitter is set to $\Lambda = 10^{-6}$ instead of $\Lambda = 10^{-4}$. This adjustment was necessary to ensure physical ISCO and $r_{\rm s}$ values within the Novikov-Thorne framework. Table \ref{TAB:NT_DF} contains all reference values.

\begin{figure}[h!]
\begin{subfigure}{0.5\textwidth}
\includegraphics[width=\linewidth]{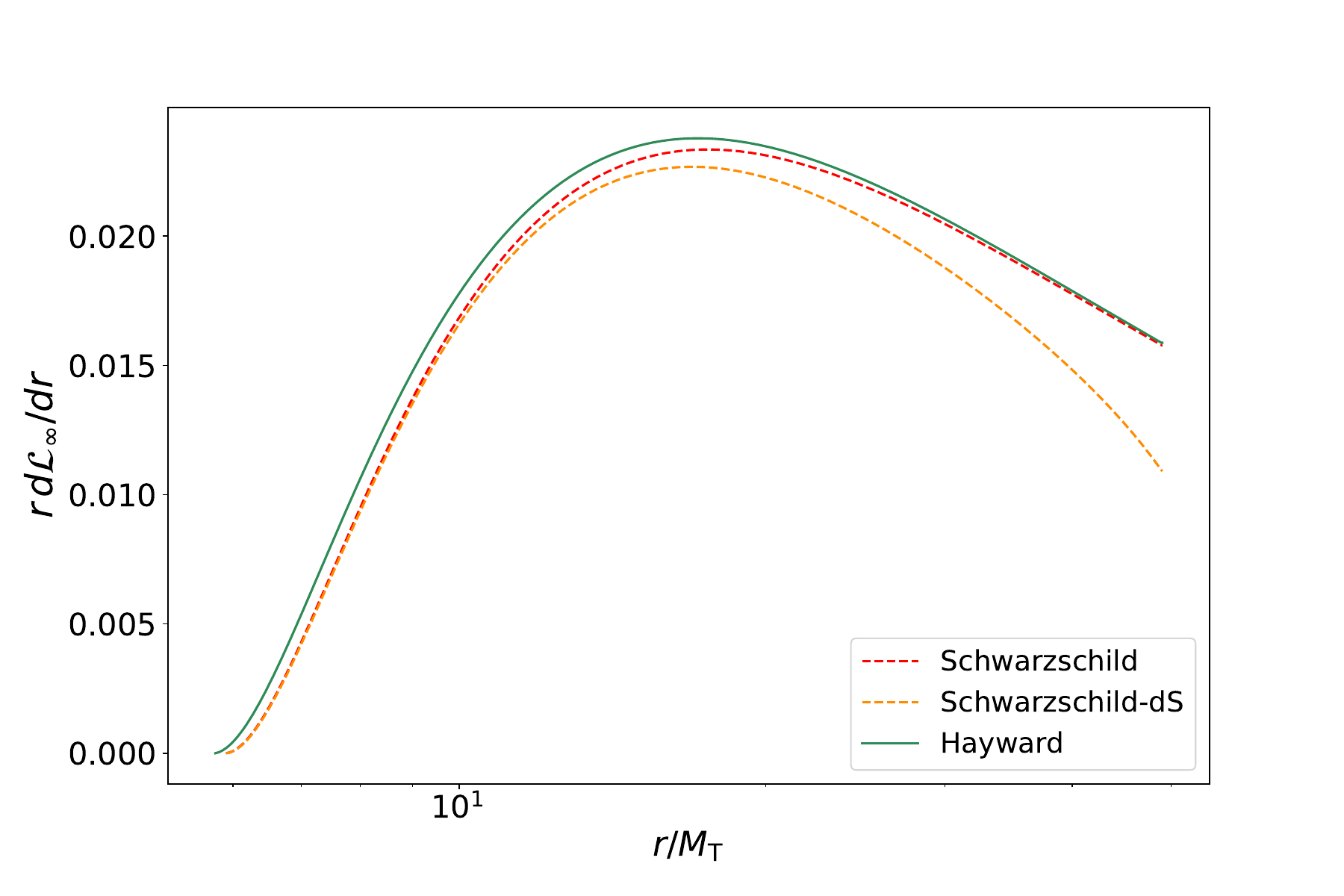}
\end{subfigure}
\hfill
\begin{subfigure}{0.5\textwidth}
\includegraphics[width=\linewidth]{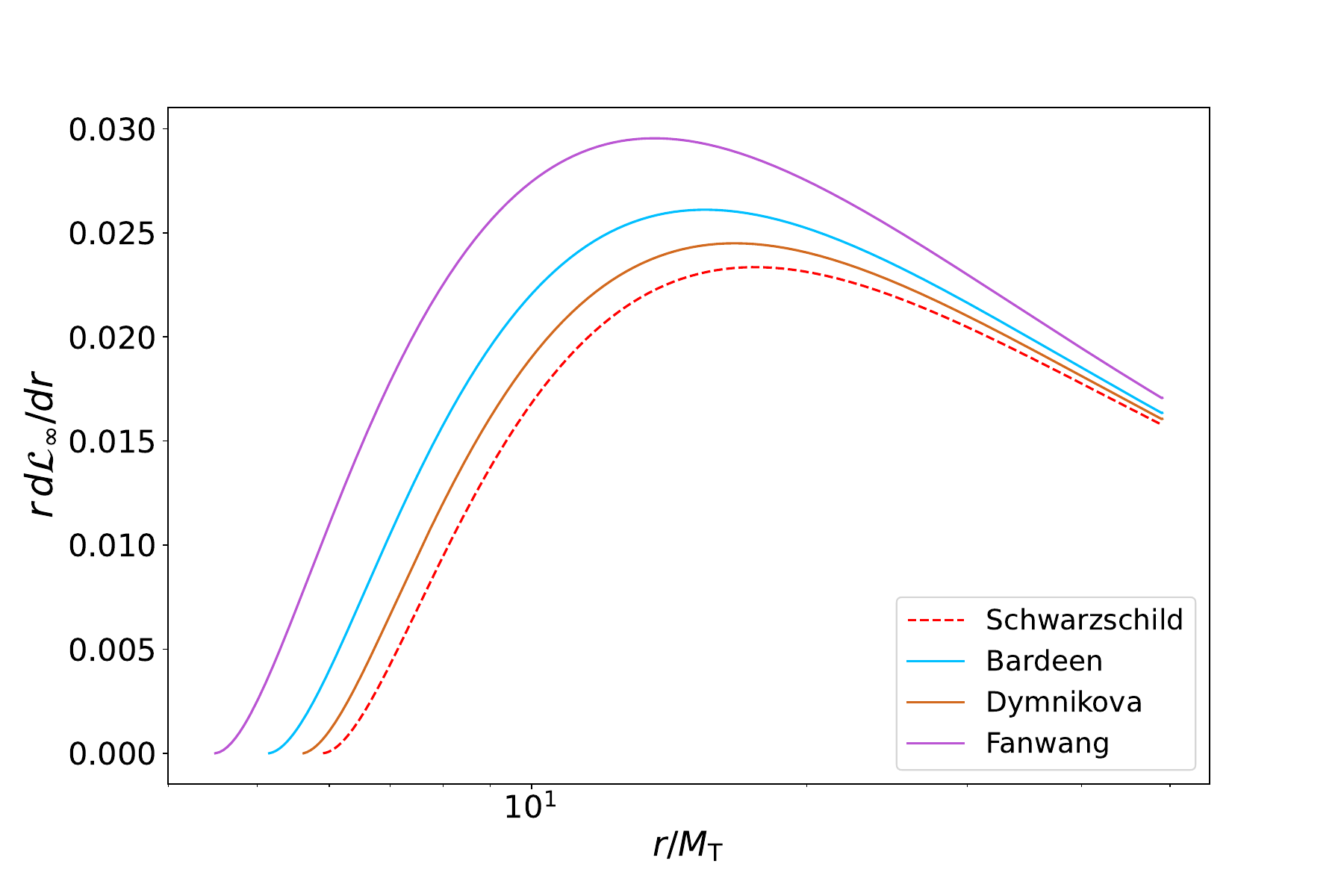}
\end{subfigure}
  \caption{\justifying Differential luminosity as a function of $r/M_{\rm T}$ (log scale) for different RBH solutions.
 {\bf Top:} Behavior of the vacuum solution compared with Schwarzschild and Schwarzschild-de Sitter solutions. {\bf Bottom:} Behavior of the topologically charged and Fan-Wang solutions compared with the Schwarzschild solution.}
\label{fig:NT_df_COMP_diff_lum}
\end{figure}

The differential luminosity comparison is shown in Fig. \ref{fig:NT_df_COMP_diff_lum}.
The upper panel shows that the Schwarzschild and Hayward solutions nearly overlap throughout the studied range, while the Schwarzschild-de Sitter solution deviates at larger $r$ by exhibiting a continual luminosity decrease. In the lower panel, the Fan-Wang solution has the highest luminosity values, followed by the Bardeen and Dymnikova solutions, with Schwarzschild showing the lowest values.
\begin{table}[!h]
    \centering
    \renewcommand{\arraystretch}{1.3}
    \begin{tabular}{|l|c|c|c|c|}
        \hline
        \hline
    \multicolumn{5}{|c|}{\textbf{2\textsuperscript{nd} Case: Exponential Density Profile}} \\ \hline\hline
    \textbf{Solution} & $r_{\text{EH}}$ & $r_{\rm b}$ & $r_{\rm i}$ & $r_{\rm s}$  \\
        \hline\hline
        Bardeen                 & 1.58   & 5    &   5.24    &   35      \\
        \hline
        Hayward                 & 1.85   & 5    &   5.85    &   35      \\
        \hline
        Fan-Wang                & 1.5    & 2.5  &   4.59    &    35     \\
        \hline
        Dymnikova               & 1.89   & 5    &   5.69    &     35    \\
        \hline
        Schwarzschild           & 2      & 5    &   6.01    &      35   \\
        \hline
        Schwarzschild-de Sitter & 2      & 4    &   6.02    &  35       \\
        \hline
        \hline
    \end{tabular}
    \caption{\justifying The table presents the numerical results for the inner and outer radii of the fluid envelope in the accretion disk, the event horizons, and the ISCO, using the exponential density profile. Constants used: $ M = 1 \, \text{AU} $, $ a = 0.5 $, $ q_B = 0.65 $, $ q_D = 0.452 $, $ \Lambda = 10^{-5} $, $ l_{FW} = 4/27 $, $\rho_0 = 0.75 \times 10^{-5} $, $r_0 = 10 \, \text{AU}$.}
    \label{TAB:NT_DF}
\end{table}
\subsection{The Accretion Process with Exponential Density Profile}
\label{SEC:NT_exp_results}
We now examine the second accretion model within the Novikov-Thorne framework, characterized by an exponential density profile. This model follows the formulation presented in Subsection \ref{SEC:NT_exp_variables}. The reference values and numerical results relevant to this case are summarized in Table \ref{TAB:NT_EXP}.

The pressure profile is determined through the numerical integration of the TOV equation. To ensure a smooth transition between the void and the inner boundary of the fluid envelope, we incorporate a Newtonian-limit pressure correction, increased by a relativistic factor, denoted as $f_{\rm P} = 1.5$.

The characteristic radii of the system are identified by first selecting the inner radius of the fluid envelope, $r_{\rm b}$, in accordance with the event horizon locations of each RBH solution. Notably, for this model, the BH mass deviates from unity, specifically taking the value $M_{\rm BH} = 4.933 , \text{AU}$, with all associated quantities and parameters scaled accordingly.

The outer boundary of the fluid envelope, $r_{\rm s}$, is determined by locating the radial coordinate at which the pressure vanishes, i.e., $P(r = r_{\rm s}) = 0$. This condition is numerically implemented by defining an appropriate threshold for near-zero pressure. The ISCO is then identified by the standard condition $dL/dr = 0$, consistent with the previous case and gives the same results of the dark fluid EoS given that we do not have dependence from density or pressure profiles.

\begin{figure}[h!]
\begin{subfigure}{0.5\textwidth}
\includegraphics[width=\linewidth]{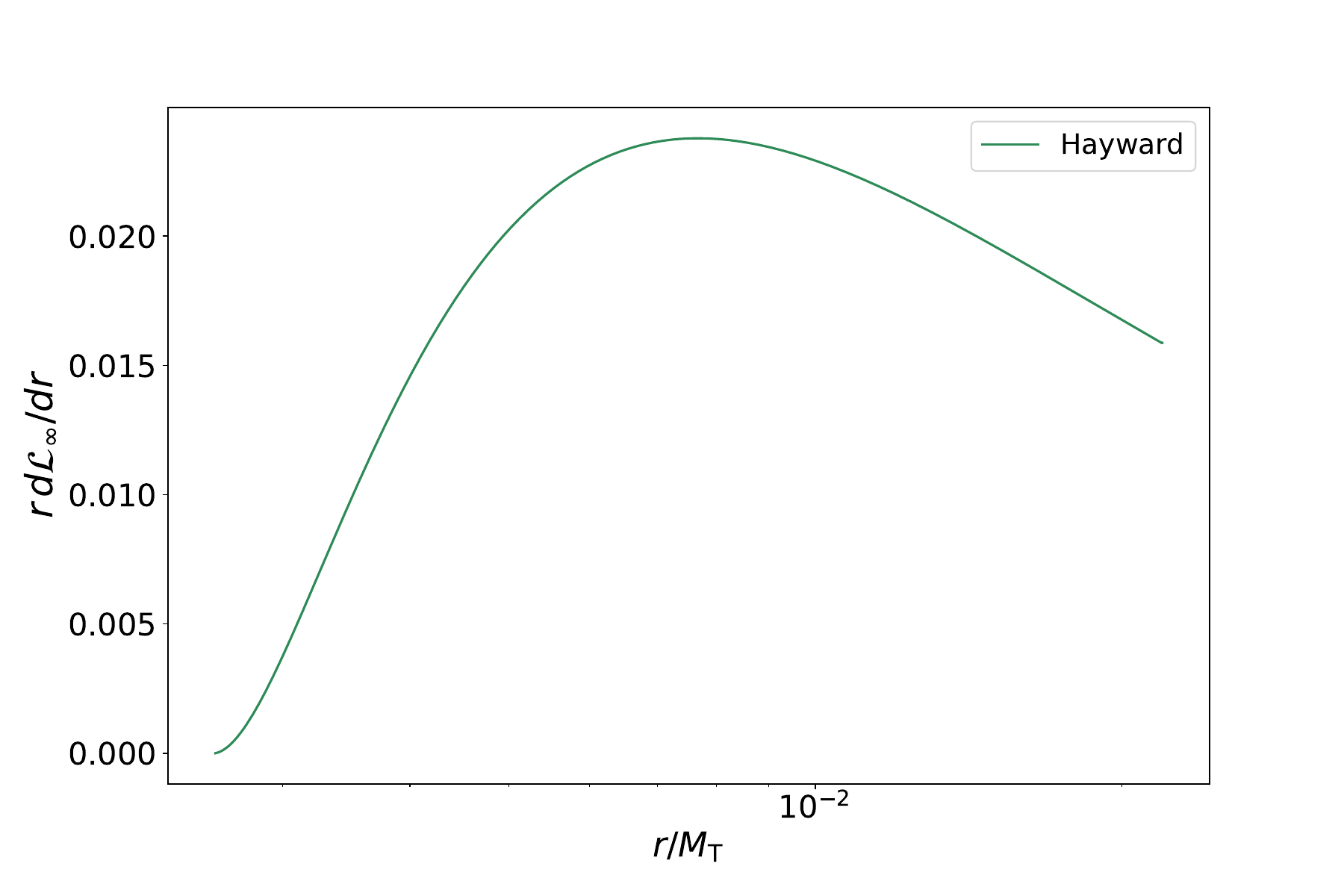}
\end{subfigure}
\hfill
\begin{subfigure}{0.5\textwidth}
\includegraphics[width=\linewidth]{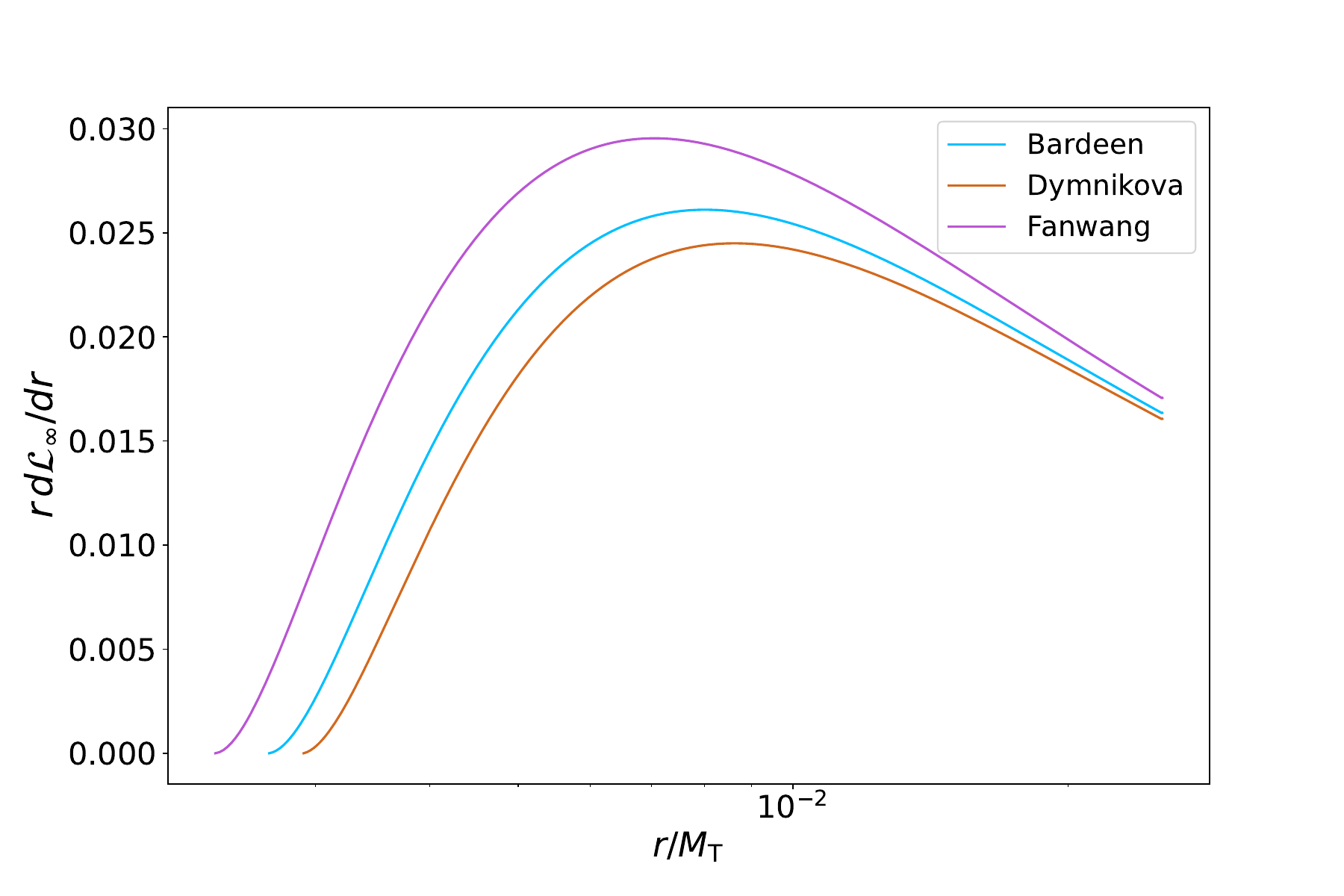}
\end{subfigure}
  \caption{\justifying Differential luminosity in function of $r/M_{\rm T}$ (in logarithmic scale) for different RBH solutions.
 {\bf Top:} Behavior of the vacuum solution. {\bf Bottom:} Behavior of the topologically charged and Fan-Wang solutions.}
\label{fig:NT_exp_diff_lum}
\end{figure}

We begin our analysis by examining the luminosity as observed at infinity, as shown in Fig. \ref{fig:NT_exp_diff_lum}.
In the upper panel, we present the differential luminosity for the Hayward vacuum solution, which exhibits a peak followed by a decline, forming a characteristic bump, similar to the dark fluid scenario. In the lower panel, we compare the topologically charged solutions, observing trends analogous to those of the Hayward case. However, the Fan-Wang solution consistently displays the highest luminosity across the entire range, followed by the Bardeen solution, while the Dymnikova solution exhibits the lowest luminosity values.

\subsubsection{Comparison with Schwarzschild and Schwarzschild-de Sitter}
We now contrast the results for RBHs with those of singular Schwarzschild and Schwarzschild-de Sitter solutions to highlight key differences.

\begin{figure}[h!]
\begin{subfigure}{0.5\textwidth}
\includegraphics[width=\linewidth]{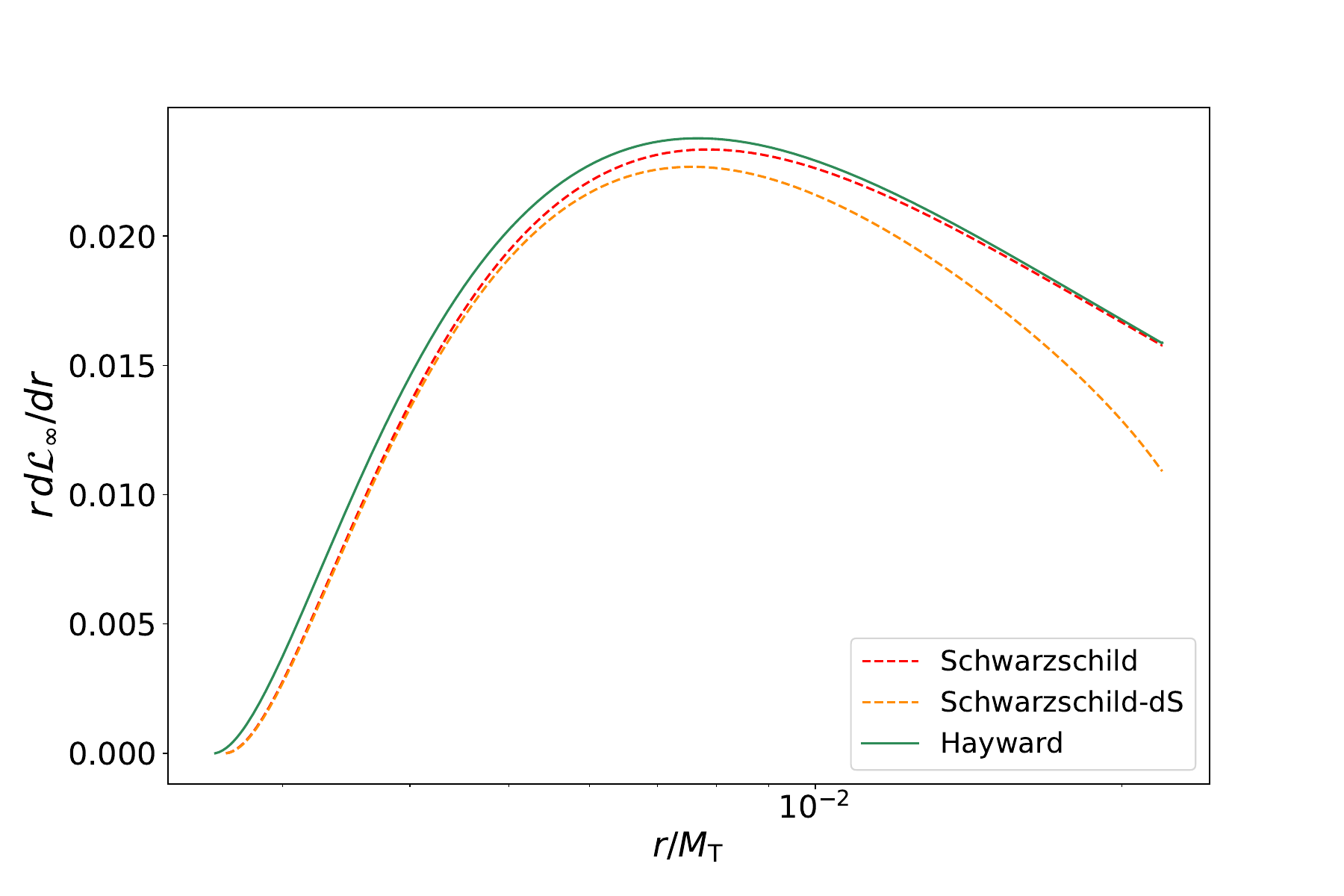}
\end{subfigure}
\hfill
\begin{subfigure}{0.5\textwidth}
\includegraphics[width=\linewidth]{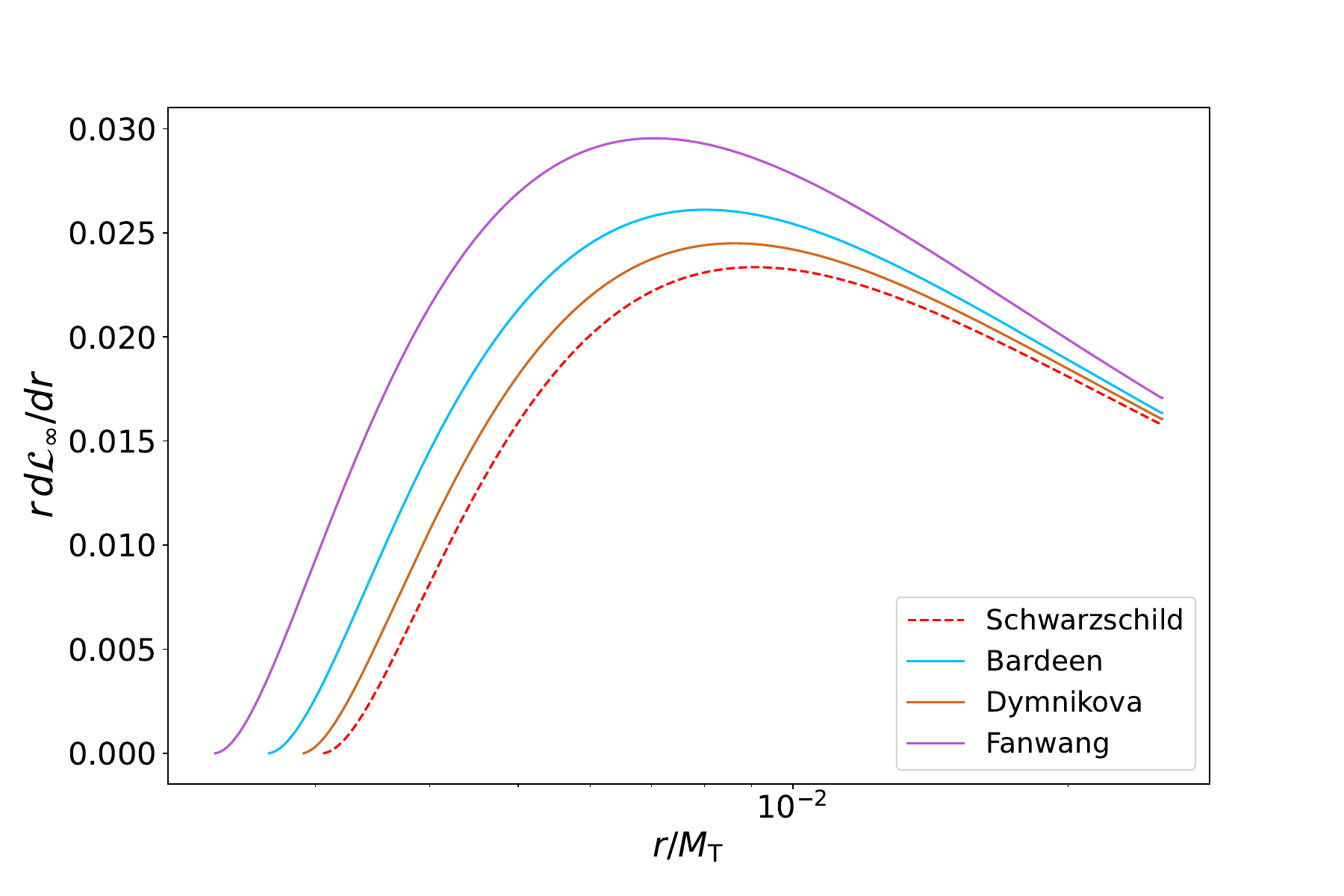}
\end{subfigure}
  \caption{\justifying Differential luminosity in function of $r/M_{\rm T}$ (in logarithmic scale) for different RBH solutions.
 {\bf Top:} Behavior of the vacuum solution compared with Schwarzschild and Schwarzschild-de Sitter solutions. {\bf Bottom:} Behavior of the topologically charged and Fan-Wang solutions compared with the Schwarzschild solution.}
\label{fig:NT_exp_COMP_diff_lum}
\end{figure}

In Fig. \ref{fig:NT_exp_COMP_diff_lum}
In the upper panel of Fig. \ref{fig:NT_exp_COMP_diff_lum}, the Hayward RBH solution closely follows the Schwarzschild behavior but exhibits slightly higher luminosity values at smaller radii. The Schwarzschild-de Sitter solution, on the other hand, decreases more rapidly at large radii while remaining nearly indistinguishable from the other solutions at small radii.

In the lower panel, the Schwarzschild solution consistently demonstrates the lowest luminosity values across the entire $r/M_{\rm T}$ range, although its general trend aligns with that of the regular solutions.
\begin{table}[!h]
    \centering
    \renewcommand{\arraystretch}{1.3}
    \begin{tabular}{|l|c|c|c|c|c|}
        \hline
        \hline
    \multicolumn{6}{|c|}{\textbf{2\textsuperscript{nd} Case: Exponential Density Profile}} \\ \hline\hline
    \textbf{Solution} & $r_{\text{EH}}$ & $r_{\rm b}$ & $r_{\rm i}$ & $r_{\rm s}$ & $M_{\rm T}/M_{\rm BH}$ \\
        \hline\hline
        Bardeen                 & 1.58 & 5 & 5.24 &  72.41    & 1.023         \\
        \hline
        Hayward                 & 1.85 & 5 & 5.85 &  37.53    & 1.023        \\
        \hline
        Fan-Wang                & 1.5  & 2.5& 4.59 &  23.49    & 1.019         \\
        \hline
        Dymnikova               & 1.89 &5  & 5.69 &  47.06    & 1.023          \\
        \hline
        Schwarzschild           & 2    & 5 & 6.01 &  36.48    &  1.023       \\
        \hline
        Schwarzschild-de Sitter & 2    & 4 & 6.02 &  27.70    &  1.021     \\
        \hline
        \hline
    \end{tabular}
    \caption{\justifying The table presents the numerical results for the inner and outer radii of the fluid envelope in the accretion disk, the event horizons, and the ISCO, using the exponential density profile. Constants used: $ M = 1 \, \text{AU} $, $ a = 0.5 $, $ q_B = 0.65 $, $ q_D = 0.452 $, $ \Lambda = 10^{-5} $, $ l_{FW} = 4/27 $, $\rho_0 = 0.75 \times 10^{-5} $, $r_0 = 10 \, \text{AU}$.}
    \label{TAB:NT_EXP}
\end{table}
\subsection{Considerations on the Mass Distribution}
From the analysis in the previous sections, it is evident that the results for the exponential density profile and the dark fluid case are nearly indistinguishable. Both exhibit the same trend and range of luminosity values. This similarity arises because the only distinguishing factor between these two cases, within the Novikov-Thorne framework, is the total mass. The density distribution itself does not directly affect the flux, and consequently, the differential luminosity remains largely unaffected by the different pressure profiles.

Some dependence on density and pressure can be recovered by considering the inner and outer radii of the accreting fluid envelope. However, since $ r_{\rm b} $ is an arbitrarily chosen parameter, and in the dark fluid case $ r_{\rm s} $ must be imposed due to the trivial integration of the TOV equation (which only constrains the density), their selection remains a free choice.

Under these conditions, studying the dark fluid case as rigorously as the exponential density profile becomes problematic. Furthermore, the matter distribution cannot be determined using the same procedure. If we incorporate our imposed constraints into Eq. \eqref{NT_mass}, the resulting mass distribution grows indefinitely or, practically, up to the chosen boundary of the envelope. Thus, using this relation directly may be unphysical. To maintain consistency, we have assigned a comparable mass in both cases to allow for a meaningful comparison.

However, in principle, one could choose a different mass. In the following, we analyze the effect of varying the mass in the dark fluid case and compare the results with the exponential density profile.

\begin{figure}[h!]
\begin{subfigure}{0.5\textwidth}
\includegraphics[width=\linewidth]{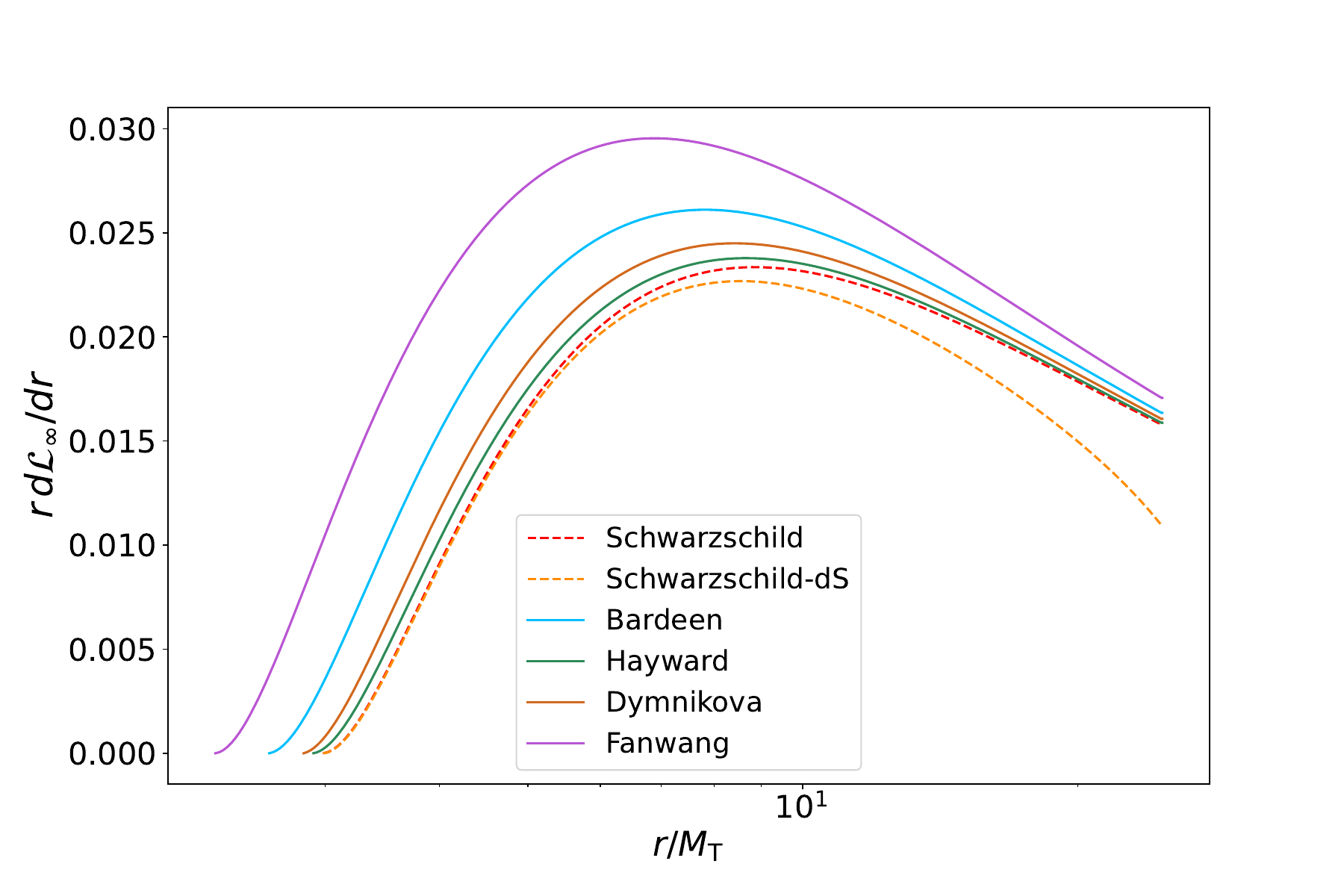}
\end{subfigure}
\hfill
\begin{subfigure}{0.5\textwidth}
\includegraphics[width=\linewidth]{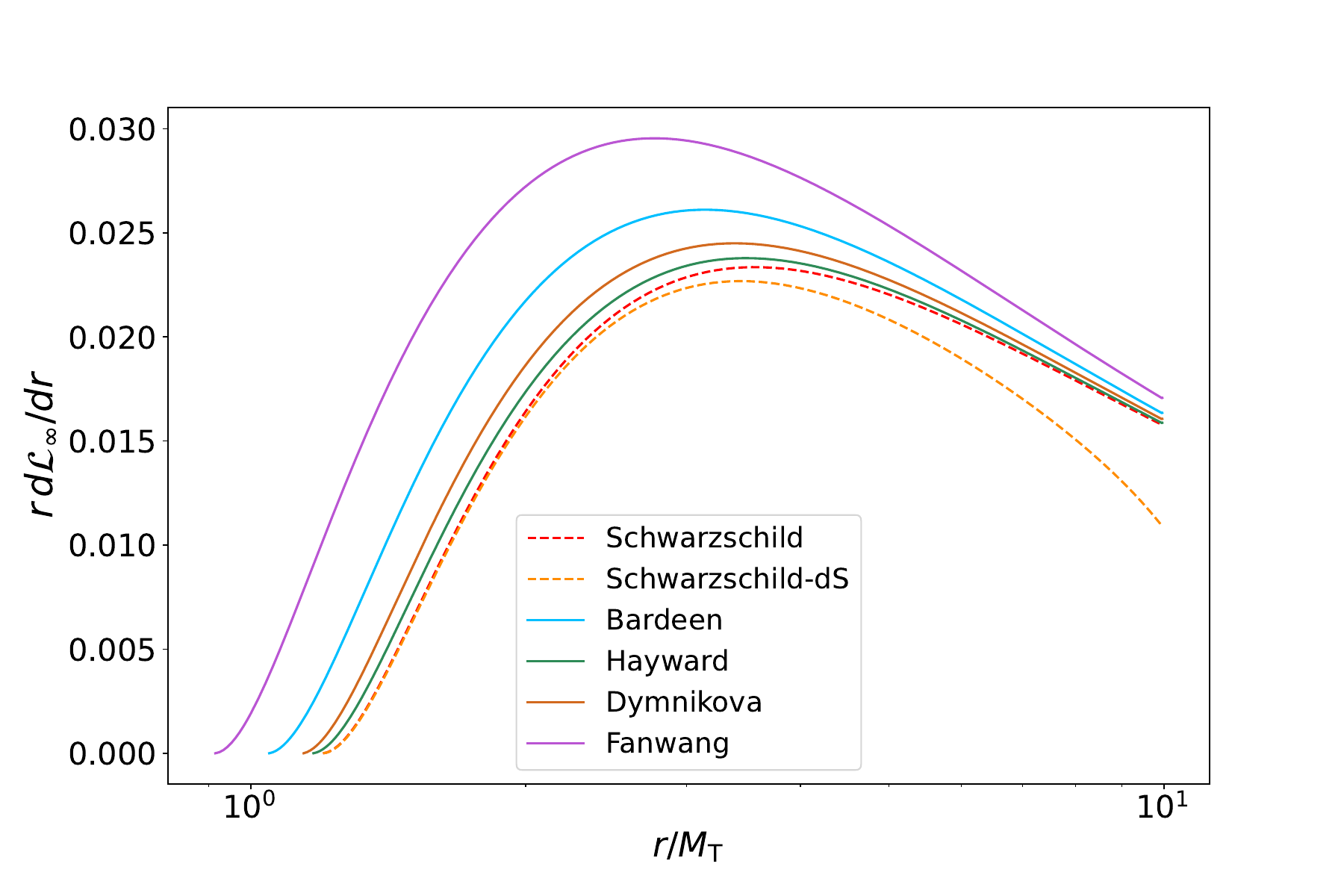}
\end{subfigure}
  \caption{\justifying Differential luminosity as a function of $ r/M_{\rm T} $ (in logarithmic scale) for all BH solutions, with different values of $ M_{\rm T} $.
  {\bf Top:} Differential luminosity with $ M_{\rm dark \ fluid} = 2.02 $.
  {\bf Bottom:} Differential luminosity with $ M_{\rm dark \ fluid} = 5.02 $.}
\label{fig:NT_df_different_mass}
\end{figure}

In Fig. \ref{fig:NT_df_different_mass}, we show the differential luminosity for different values of $ M_{\rm T} $.

In the upper panel, where the mass is increased to $ M_{\rm dark \ fluid} = 2.02 $, we observe that the differential luminosity reaches its peak values more quickly compared to the standard case. This behavior is a direct consequence of the reduced range of $ r/M_{\rm T} $ caused by the increased total mass. The same trend is even more pronounced in the bottom panel, where $ M_{\rm dark \ fluid} = 5.02 $. Here, the peak luminosity occurs at even smaller values of $ r/M_{\rm T} $, highlighting how the scaling of mass influences the observable luminosity profile.

Therefore, for this specific case, if we impose the same total mass for both fluid models, we find that the Novikov-Thorne accretion process does not reveal any differences between them. This is in contrast to the Bondi accretion case, where the choice of fluid significantly impacts the trends of the variables, including the velocity, density, and luminosity.

%
%
%
\section{Final remarks}
\label{SEC:Discussion}

In this paper, we analyzed and compared the Bondi and Novikov-Thorne accretion models. To this end, we explored a class of RBHs and compared their behavior with the simplest BH solution, the Schwarzschild metric. Additionally, for the vacuum solution case, we included a comparison with the Schwarzschild-de Sitter solution.

We characterized the problem by analyzing the key variables of the accretion processes. For Bondi this included the velocity profile, energy density, pressure, mass accretion rate, and luminosity. Instead, for the Novikov-Thorne accretion process the differential luminosity profile. Using relativistic conservation laws, we derived the equations governing these quantities and numerically solved them for different RBH solutions.

Our numerical results showed the different trends of the different RBH solutions under the two fluid models considered: the dark fluid EoS and the exponential density profile.

For the Bondi accretion, we started from the vacuum energy solution, the Hayward RBH, which exhibited a steady inflow velocity in both cases, with a small decrease near the event horizon and also a small decrease for the exponential profile in the external region of the disk. Also the density showed a similar behaviour in the two cases. In general, this metric behaved similar to both Schwarzschild and Schwarzschild-de Sitter while the major difference between the models was given by both the trends and the values of the accretion rate and luminosity.

The first examined topologically charged solution was the Bardeen RBH, which showed similar trends to Schwarzschild at larger radii, but had a slightly different critical radius in the case of dark fluid. The mass accretion rate and luminosity followed a similar pattern to the other solutions in the group, but with smaller variations near the event horizon.

The Dymnikova solution, also in the topologically charged category, showed the smallest increase in mass accretion rate and luminosity near the event horizon for the dark fluid case. The same trend is represented for the exponential density profile. Here indeed, it showed the smallest increase for all of the variables under study with almost a completely overlap with the Schwarzschild solution.

Instead, the Fan-Wang solution showed contrasting behaviour between the two cases. In the dark fluid scenario, it showed anomalies in the mass accretion rate and luminosity, with values significantly higher, even by orders of magnitude, compared to the other solutions. This suggested that this metric may not provide a physically meaningful accretion process with this EoS. However, in the exponential density case it followed a compatible trend, even with the smallest values of $\dot{M}$ and $L$, confirming that the anomalous behaviour in the dark fluid case was due to the chosen EoS rather than an intrinsic problem with the metric.

We considered the Schwarzschild and Schwarzschild-de Sitter solutions and compared them with the RBH metrics. The Schwarzschild BH, exhibited a well-defined critical point in the Bondi accretion, for the dark flui ase,  but failed to yield a consistent critical radius in the case of exponential density profile case. This suggested that this density distribution was not well-suited for this solution. The trends and values of all the variables were similar to most of the RBH solutions with the exception of the Fan-Wang and the Schwarzschild-de Sitter Sitter as mentioned before.

Similarly, the Schwarzschild-de Sitter BH, behaved like the Schwarzschild solution in the dark fluid case but did not produce a physically meaningful critical point in the exponential density profile case. This was an evidence of the limitations of applying this profile for certain solutions. The trends for the Schwarzschild-de Sitter solution was similar to those of the Schwarzschild solution, and therefore the similarities observed for the RBH solutions also apply to this metric.

The comparison between the two fluid models for the Bondi accretion showed important differences. While the velocity and density profiles were similar in both models, the mass accretion rate and luminosity showed different trends and also different ranges of values. In the case of dark fluid, these quantities followed a monotonically increasing pattern near the event horizon, while in the case of the exponential density profile, the curve showed a bump in the inner region, more similar to the Novikov-Thorne accretion.

The pressure behaved differently, as the dark fluid model assumed a constant pressure, while in the exponential density profile, the pressure is defined as a function of radius derived from conservation laws.

Another important consideration was the critical point: in the dark fluid model, the relativistic critical point analysis could not be implemented due to the peculiar EoS and led to a redefinition of certain variables: we recall in fact how the variable $V^2$ showed discontinuities when the dark fluid approach was used and, therefore, by imposing a constant negative pressure for the first case. Conversely, in the case of the exponential density profile, the critical point was well defined using the conventional formulation. However, the Schwarzschild and Schwarzschild-de Sitter solutions showed no physical critical points in the second model. This suggested that the assumed density profile was incompatible with these metrics.

The second accretion mechanism we investigated is the Novikov-Thorne model. In this case, the only variable we analyze is the differential luminosity, and the profiles obtained for the two types of fluids are identical.

We begin with the Hayward solution, which does not exhibit differences between the two fluid models. In comparison with the Schwarzschild and Schwarzschild-de Sitter solutions, the Hayward profile nearly overlaps with them throughout the entire plotted range, although it displays higher luminosity values than the Schwarzschild-de Sitter solution and a smoother decline.

Next, we consider the topologically charged and Fan--Wang solutions within this accretion model. The Bardeen solution follows the same trend as the Hayward solution but attains higher luminosity values. A comparison with the Schwarzschild solution shows that while both share a similar trend, the Bardeen solution reaches higher luminosity at the same radii.

The Dymnikova solution, following a similar overall trend as the other solutions, yields slightly lower luminosity than the Bardeen solution; however, it still does not overlap with the Schwarzschild BH, which consistently exhibits lower luminosity.

Finally, the Fan-Wang RBH, despite following the same trend as the others---shows luminosity values that are significantly higher than both the other topologically charged solutions and the Schwarzschild black hole.

For completeness, we also discuss the Schwarzschild and Schwarzschild-de Sitter solutions. The Schwarzschild solution follows the same trend as all the RBH models studied, but it reaches the smallest values for the asymptotic luminosity, $\mathcal{L}_\infty$. In contrast, the Schwarzschild-de Sitter solution shows a decrease, moving away from the inner region, a behavior not seen in any other solution. This unique decline can be attributed to the influence of the cosmological constant $\Lambda$, which is considered only for this case.

We now turn to the major differences encountered when modeling the two fluids. The exponential density profile follows a more classical derivation, using the same procedure as in \cite{Boshkayev:2020}, to model the various RBH solutions. This approach allowed us to numerically extrapolate the edge of the fluid envelope and determine the ISCO for each solution under study, imposing only the inner radius $r_{\rm b}$.

In contrast, the dark fluid model precludes the use of the usual TOV equation because its constant pressure renders the equation trivial, providing only information on the pressure profile. Consequently, both the edges of the fluid envelope and the mass distribution had to be chosen manually. With constant pressure, the conventional expression for the mass distribution is not available, so we imposed a constant mass for each solution. However, the ISCO could still be derived using the standard approach even with the dark fluid EoS.

The only notable difference when plotting the luminosity was the range of values for the variable $r/M_{\rm T}$. This range shows a direct dependence on the mass; by manually adjusting the mass of the dark fluid envelope, we observed differences compared to the exponential density profile. These differences appeared as shifts of the curves toward smaller or larger $r$ values, without changing the differential luminosity itself.

Finally, we compared the results with the Novikov–Thorne model. Although their formulations differ, as outlined in Subsubsec. \ref{SEC:NT_accretion}, there are similarities in both the luminosity values and trends between the Novikov–Thorne model (using both the dark fluid EoS and the exponential density profile) and the exponential density profile case of Bondi accretion. This agreement suggests that the exponential density profile, when applied to either our spherically symmetric setup or the thin-disc approximation of Novikov–Thorne, shares key physical properties regarding luminosity.

Each solution exhibited a similar trend: a bump at small radii followed by a decrease as we move further away from the accretion disk. However, the range of luminosity values differed by two orders of magnitude, with Bondi accretion displaying higher luminosities than the Novikov–Thorne model. Since we used the same parameters for each solution and for both fluid models, we conclude that the differences in magnitude are due to the distinct nature of the expressions derived to reproduce the luminosity behavior in the two accretion cases. Moreover, the mass used to scale the radius differed between the two models; in the Novikov–Thorne accretion, the entire mass within the chosen range was considered, including both the fluid and the bh mass.

We also observed that the various RBH solutions did not maintain the same hierarchy of luminosity values across the two models. Specifically, for Bondi accretion, the Hayward solution yielded the highest luminosity, followed by the Bardeen, Fan–Wang, and then the Dymnikova and the two rbh solutions. In contrast, for the Novikov–Thorne model, the Fan–Wang solution showed the highest values of $\mathcal{L}_\infty$, followed by the Hayward, Bardeen, Dymnikova, and finally the Schwarzschild and Schwarzschild–de Sitter cases.

Overall, we can summarize the results of these two different accreting fluids. For the dark fluid EoS, we recall how the critical point of the Bondi accretion process shifts compared to the standard polytropic case, as the conventional sonic point definition does not hold due to the vanishing sound speed in the dark fluid model. The variable $V^2$ showed discontinuities when the dark fluid approach was used, requiring a redefinition of certain variables and demonstrating that a constant negative pressure impacts the standard relativistic formulation.
The velocity profiles show a rapid decrease near the event horizon, with RBHs exhibiting a slightly different behavior than the Schwarzschild and Schwarzschild-de Sitter cases and the same can also be found in the trend of the mass accretion rate and luminosity which are generally higher for RBHs compared to Schwarzschild and Schwarzschild-de Sitter BHs. The same result is indeed confirmed in the Novikov-Thorne accretion.

Instead, with the exponential density profile, we showed how the critical point analysis is applicable only to RBHs, as no well-defined critical radius can be found for the Schwarzschild and Schwarzschild-de Sitter solutions. This suggests that the chosen density profile is unsuitable for describing accretion in these backgrounds. The velocity profiles are again characterized by a steep decline near the event horizon. For RBHs, the location of the critical point varies depending on the specific metric considered, with the Bardeen critical point positioned deeper within the gravitational well compared to Dymnikova and Fan-Wang solutions. The accretion rate and luminosity display a rapid growth towards the BH with the Dymnikova solution exhibiting the highest values. At large radii, all solutions approach zero accretion rate and luminosity, consistent with expected asymptotic behavior. The exponential density profile led to results that resemble the Novikov-Thorne accretion model, particularly in terms of luminosity trends and the formation of a characteristic bump in the inner accretion region.

Future works will focus on extending these analyses to rotating RBH solutions and exploring additional properties of BH solutions also possibly using extended Bondi accretion models. In addition, different dark matter scenarios will be investigated, remarking how the profile-dependence influences the outputs obtained from the background accretions. Last but not least, we will investigate whether the action of repulsive effects in gravity can be detected within the physics of accretion disks, see e.g. \cite{Luongo:2010we,Luongo:2014qoa,Luongo:2023aib,Luongo:2015zaa,Luongo:2023xaw,Giambo:2020jjo}.

\section*{Acknowledgements}
S.C. and S.G. acknowledge the support of {\it Istituto Nazionale di Fisica Nucleare} (INFN) sez. di Napoli, {\it iniziative specifiche} QGSKY and MOONLIGHT2. S.C. thanks the {\it Gruppo Nazionale di Fisica Matematica} of {\it Istituto Nazionale di Alta Matematica} for the support. O.L. acknowledges financial support from the {\it Fondazione} ICSC, Spoke 3 Astrophysics and Cosmos Observations. National Recovery and Resilience Plan ({\it Piano Nazionale di Ripresa e Resilienza}, PNRR) Project ID CN$\_$00000013 "Italian Research Center on High-Performance Computing, Big Data and Quantum Computing" funded by {\it MUR Missione 4 Componente 2 Investimento 1.4: Potenziamento strutture di ricerca e creazione di "campioni nazionali di R$\&$S (M4C2-19)}" - Next Generation EU (NGEU) GRAB-IT Project, PNRR Cascade Funding Call, Spoke 3,  Italian National Institute for Astrophysics (INAF), Project code CN00000013, Project Code (CUP): C53C22000350006, cost center STI442016.

\end{document}